\pdfoutput=1
\documentclass[aps,pr,twocolumn,superscriptaddress,preprintnumbers,showpacs,amsmath,amssymb,floatfix]{revtex4}

\usepackage{graphicx} 
\usepackage{dcolumn}  

\usepackage{epsfig}
\usepackage{epsbox}
\usepackage[T1]{fontenc}
\usepackage{wrapfloat}
\usepackage{subfigure}
\usepackage{here}
\usepackage{array}
\usepackage{amsmath, amssymb, hhline, multirow, hangcaption, subfigure}
\usepackage{graphics}
\usepackage{graphicx}
\usepackage{lscape}
\usepackage{fancyhdr}
\usepackage{longtable}
\usepackage{supertabular}
\usepackage{refmerge}
\usepackage{dcolumn}
\usepackage{color}

\def \vec #1{\mbox{{\boldmath $#1$}}}

\def \lr #1{\left( #1 \right)}

\def \B {{\cal B}}

\def \GeV {{\rm GeV}}
\def \MeV {{\rm MeV}}
\def \keV {{\rm keV}}

\def \vec #1{\mbox{{\boldmath $#1$}}}

\begin{document}
\hspace{100mm}
\preprint{\vbox{ \hbox{   }
                 \hbox{Belle Preprint 2015-15}
                  \hbox{KEK  Preprint 2015-24}
                 \hbox{Revised Feb.,2016}
}}
\title{ 
Study of $\pi^0$ pair production in single-tag
two-photon collisions\\
}

\begin{abstract}
We report a measurement of the differential cross section of $\pi^0$ 
pair production in single-tag two-photon collisions,
$\gamma^* \gamma \to \pi^0 \pi^0$, 
in $e^+ e^-$ scattering.
The cross section is measured for $Q^2$ up to $30~\GeV^2$,
where $Q^2$ is the negative of the invariant mass squared 
of the tagged photon, in the kinematic range
$0.5~\GeV < W < 2.1~\GeV$ and $|\cos \theta^*| < 1.0$ 
for the total energy and pion scattering angle,
respectively, 
in the $\gamma^* \gamma$ center-of-mass system.
The results are based on a data sample of 759~fb$^{-1}$ collected
with the Belle detector at the KEKB asymmetric-energy $e^+ e^-$ collider.
The transition form factor of the $f_0(980)$ 
and that of the $f_2(1270)$ with
the helicity-0, -1, and -2 components separately
are measured for the first time and 
are compared with theoretical calculations.
\end{abstract}


\normalsize

\affiliation{University of the Basque Country UPV/EHU, 48080 Bilbao}
\affiliation{Beihang University, Beijing 100191}
\affiliation{University of Bonn, 53115 Bonn}
\affiliation{Budker Institute of Nuclear Physics SB RAS, Novosibirsk 630090}
\affiliation{Faculty of Mathematics and Physics, Charles University, 121 16 Prague}
\affiliation{Chonnam National University, Kwangju 660-701}
\affiliation{University of Cincinnati, Cincinnati, Ohio 45221}
\affiliation{Deutsches Elektronen--Synchrotron, 22607 Hamburg}
\affiliation{Justus-Liebig-Universit\"at Gie\ss{}en, 35392 Gie\ss{}en}
\affiliation{Gifu University, Gifu 501-1193}
\affiliation{SOKENDAI (The Graduate University for Advanced Studies), Hayama 240-0193}
\affiliation{Gyeongsang National University, Chinju 660-701}
\affiliation{Hanyang University, Seoul 133-791}
\affiliation{University of Hawaii, Honolulu, Hawaii 96822}
\affiliation{High Energy Accelerator Research Organization (KEK), Tsukuba 305-0801}
\affiliation{IKERBASQUE, Basque Foundation for Science, 48013 Bilbao}
\affiliation{Indian Institute of Technology Bhubaneswar, Satya Nagar 751007}
\affiliation{Indian Institute of Technology Guwahati, Assam 781039}
\affiliation{Indian Institute of Technology Madras, Chennai 600036}
\affiliation{Indiana University, Bloomington, Indiana 47408}
\affiliation{Institute of High Energy Physics, Chinese Academy of Sciences, Beijing 100049}
\affiliation{Institute of High Energy Physics, Vienna 1050}
\affiliation{Institute for High Energy Physics, Protvino 142281}
\affiliation{INFN - Sezione di Torino, 10125 Torino}
\affiliation{J. Stefan Institute, 1000 Ljubljana}
\affiliation{Kanagawa University, Yokohama 221-8686}
\affiliation{Institut f\"ur Experimentelle Kernphysik, Karlsruher Institut f\"ur Technologie, 76131 Karlsruhe}
\affiliation{Kennesaw State University, Kennesaw, Georgia 30144}
\affiliation{King Abdulaziz City for Science and Technology, Riyadh 11442}
\affiliation{Department of Physics, Faculty of Science, King Abdulaziz University, Jeddah 21589}
\affiliation{Korea Institute of Science and Technology Information, Daejeon 305-806}
\affiliation{Korea University, Seoul 136-713}
\affiliation{Kyungpook National University, Daegu 702-701}
\affiliation{\'Ecole Polytechnique F\'ed\'erale de Lausanne (EPFL), Lausanne 1015}
\affiliation{Faculty of Mathematics and Physics, University of Ljubljana, 1000 Ljubljana}
\affiliation{Luther College, Decorah, Iowa 52101}
\affiliation{University of Maribor, 2000 Maribor}
\affiliation{Max-Planck-Institut f\"ur Physik, 80805 M\"unchen}
\affiliation{School of Physics, University of Melbourne, Victoria 3010}
\affiliation{Middle East Technical University, 06531 Ankara}
\affiliation{Moscow Physical Engineering Institute, Moscow 115409}
\affiliation{Moscow Institute of Physics and Technology, Moscow Region 141700}
\affiliation{Graduate School of Science, Nagoya University, Nagoya 464-8602}
\affiliation{Kobayashi-Maskawa Institute, Nagoya University, Nagoya 464-8602}
\affiliation{Nara Women's University, Nara 630-8506}
\affiliation{National Central University, Chung-li 32054}
\affiliation{National United University, Miao Li 36003}
\affiliation{Department of Physics, National Taiwan University, Taipei 10617}
\affiliation{H. Niewodniczanski Institute of Nuclear Physics, Krakow 31-342}
\affiliation{Nippon Dental University, Niigata 951-8580}
\affiliation{Niigata University, Niigata 950-2181}
\affiliation{Novosibirsk State University, Novosibirsk 630090}
\affiliation{Osaka City University, Osaka 558-8585}
\affiliation{Pacific Northwest National Laboratory, Richland, Washington 99352}
\affiliation{Peking University, Beijing 100871}
\affiliation{University of Pittsburgh, Pittsburgh, Pennsylvania 15260}
\affiliation{University of Science and Technology of China, Hefei 230026}
\affiliation{Seoul National University, Seoul 151-742}
\affiliation{Soongsil University, Seoul 156-743}
\affiliation{Sungkyunkwan University, Suwon 440-746}
\affiliation{School of Physics, University of Sydney, New South Wales 2006}
\affiliation{Department of Physics, Faculty of Science, University of Tabuk, Tabuk 71451}
\affiliation{Tata Institute of Fundamental Research, Mumbai 400005}
\affiliation{Excellence Cluster Universe, Technische Universit\"at M\"unchen, 85748 Garching}
\affiliation{Department of Physics, Technische Universit\"at M\"unchen, 85748 Garching}
\affiliation{Toho University, Funabashi 274-8510}
\affiliation{Tohoku University, Sendai 980-8578}
\affiliation{Earthquake Research Institute, University of Tokyo, Tokyo 113-0032} 
\affiliation{Department of Physics, University of Tokyo, Tokyo 113-0033}
\affiliation{Tokyo Institute of Technology, Tokyo 152-8550}
\affiliation{Tokyo Metropolitan University, Tokyo 192-0397}
\affiliation{University of Torino, 10124 Torino}
\affiliation{Utkal University, Bhubaneswar 751004}
\affiliation{CNP, Virginia Polytechnic Institute and State University, Blacksburg, Virginia 24061}
\affiliation{Wayne State University, Detroit, Michigan 48202}
\affiliation{Yamagata University, Yamagata 990-8560}
\affiliation{Yonsei University, Seoul 120-749}

\author{M.~Masuda}\affiliation{Earthquake Research Institute, University of Tokyo, Tokyo 113-0032} 
\author{S.~Uehara}\affiliation{High Energy Accelerator Research Organization (KEK), Tsukuba 305-0801}\affiliation{SOKENDAI (The Graduate University for Advanced Studies), Hayama 240-0193} 
\author{Y.~Watanabe}\affiliation{Kanagawa University, Yokohama 221-8686} 
\author{H.~Nakazawa}\affiliation{National Central University, Chung-li 32054} 

  \author{A.~Abdesselam}\affiliation{Department of Physics, Faculty of Science, University of Tabuk, Tabuk 71451} 
  \author{I.~Adachi}\affiliation{High Energy Accelerator Research Organization (KEK), Tsukuba 305-0801}\affiliation{SOKENDAI (The Graduate University for Advanced Studies), Hayama 240-0193} 
  \author{H.~Aihara}\affiliation{Department of Physics, University of Tokyo, Tokyo 113-0033} 
  \author{S.~Al~Said}\affiliation{Department of Physics, Faculty of Science, University of Tabuk, Tabuk 71451}\affiliation{Department of Physics, Faculty of Science, King Abdulaziz University, Jeddah 21589} 
  \author{D.~M.~Asner}\affiliation{Pacific Northwest National Laboratory, Richland, Washington 99352} 
  \author{H.~Atmacan}\affiliation{Middle East Technical University, 06531 Ankara} 
% ...

\collaboration{The Belle Collaboration}


\medskip


\maketitle

\tighten



\section{Introduction}
\label{sec:intro}
Single-tag two-photon production of a $C$-even hadronic system ($\cal M$),
$\gamma^* \gamma \to {\cal M}$,
is an important reaction to investigate the nature of
strong interactions 
in the low energy region, where perturbative Quantum
Chromodynamics (QCD) cannot be applied. 
It also provides valuable 
information on the $Q^2$ dependence of the transition form factor (TFF),
where $Q^2$ is the negative of the invariant mass squared 
of the tagged photon.
This reaction can be studied through the process of
$e^+ e^- \to e^\pm (e^\mp) {\cal M}$,
where $(e^\mp)$ indicates an undetected electron or positron, and results
of the measurement can be directly compared to QCD-based theoretical 
predictions. 
Diehl, Gousset and Pire considered this process at large $Q^2$ and
small $W$ ($<1$~GeV) in terms of constituent-hard scattering and
generalized distribution amplitudes and predicted a sizable
cross section at LEP and $B$ factories~\cite{Diehl}; it
is indeed the case at a $B$ factory as reported here. 
Based on this framework, Braun and Kivel pointed out that the
measurement of the TFF of the $f_2(1270)$ will be useful to
cleanly determine a gluon admixture in tensor mesons at large
enough $Q^2$~\cite{Braun}.
In addition, a data-driven dispersive approach was suggested 
recently~\cite{mug2}, allowing a more precise estimate of the hadronic 
light-by-light contribution to the anomalous magnetic moment of
the muon ($g-2$).

Schuler, Berends, and van Gulik, who had calculated meson TFFs based 
on a heavy quark approximation~\cite{schuler},
found that their calculations apply well to light
mesons as well with  only minor modifications.
The predicted $Q^2$ dependence of TFFs
for mesons with $J^{PC} = 0^{++}$ and $2^{++}$
is summarized in Table~\ref{tab:pred}, where 
the $\gamma^*\gamma$ center-of-mass (c.m.) energy $W$ 
is replaced by the resonance mass $M$ and $\lambda$ represents the total 
helicity of the two incident photons.
Note that the helicity-1 ($\lambda = 1$) state is allowed
when a photon is off the mass shell.
According to Table~\ref{tab:pred}, TFFs for the helicity-0 and -1
components of a tensor meson grow with $Q^2$, 
a prediction which is amenable to investigation.

\begin{center}
\begin{table}
\caption{Predicted $Q^2$ dependence of transition form factors of 
mesons for various helicities $\lambda$ of two incident photons~\cite{schuler}.
Each term has a common divisor of
$(1 + Q^2/M^2)^2$. 
}
\label{tab:pred}

\begin{tabular}{c|ccc} \hline \hline
&&&\\[-10pt]
$J^{PC}$ & \multicolumn{3}{c}{$Q^2$ dependence 
$\left( \div \left( 1 + \frac{Q^2}{M^2} \right)^2 \right)$ } 
\\ \cline{2-4}
&&&\\[-10pt]
& $\lambda=0$ & $\lambda=1$ & $\lambda=2$ \\ \hline\hline
&&&\\[-10pt]
$0^{++}$ & $\left( 1 + \frac{Q^2}{3 M^2} \right)$
& -- & -- \\ 
&&&\\[-10pt]
\hline
&&&\\[-10pt]
$2^{++}$ & $\frac{Q^2}{\sqrt{6} M^2}$ & 
$\frac{\sqrt{Q^2}}{\sqrt{2}M}$ & 1 \\
&&&\\[-10pt]
 \hline \hline
\end{tabular}
\end{table}
\end{center}
Recently, Pascalutsa, Pauk, and Vanderhaeghen have 
formulated sum rules for $\gamma^* \gamma^*$ fusion cross sections,
finding several new exact superconvergence relations that are
integrated to zero~\cite{ppv}.
They derive two predictions for the helicity-2 TFF of the $f_2(1270)$
from two sum rules in the case of one virtual and one real photon
under the assumption that the sum rules
are saturated by low mass resonances including the $f_2(1270)$.
In one sum rule, the integrand has 
contributions from pseudoscalar mesons and tensor mesons.
In the other, axial-vector mesons and tensor mesons contribute to
its integrand.
The first (second) sum rule gives the helicity-2 TFF of the $f_2(1270)$ in 
terms of TFF information of the $\eta$ and $\eta'$ 
($f_1(1285)$ and $f_1(1420)$).
With pseudoscalar ($P$) mesons, the helicity-2 TFF of the
$f_2(1270)$ is given by
\begin{equation}
F_{f_2}(Q^2) = \sqrt{ \frac{f}{\left( 1 + \frac{Q^2}{\Lambda^2_{\eta}} \right)^2}
+ \frac{1-f}{\left( 1 + \frac{Q^2}{\Lambda^2_{\eta^\prime}} \right)^2} \;},
\label{eqn:pseudo}
\end{equation}
where $\Lambda_{\eta}$ and $\Lambda_{\eta^\prime}$ are the pole masses 
and $f = c_{\eta} / (c_{\eta} + c_{\eta^\prime})$, with 
$c_P = \Gamma_{\gamma \gamma} (P)/m_P^3$.
The relevant parameters are summarized in Table~\ref{tab:ppv}.
In another sum rule for axial-vector ($A$) mesons,
the helicity-2 TFF of the $f_2(1270)$ is given by
\begin{eqnarray}
F_{f_2}(Q^2) &=& \left(1 + \frac{Q^2}{m^2_{f_2}} \right)^{\frac{1}{2}}
\times \nonumber \\
&& \sqrt{ \frac{f}{\left( 1 + \frac{Q^2}{\Lambda^2_{f_1}} \right)^4}
+ \frac{1-f}{\left( 1 + \frac{Q^2}{\Lambda^2_{f^\prime_1 }} \right)^4} } \;,
\label{eqn:axial}
\end{eqnarray}
where $\Lambda_{f_1}$ and $\Lambda_{f^\prime_1}$ are the pole masses 
and $f = c_{f_1} / (c_{f_1} + c_{f^\prime_1})$, with 
$c_A = 3 \tilde{\Gamma}_{\gamma \gamma} (A)/m_A^5$.
The effective two-photon width of the $A$ resonance is defined as
\begin{equation}
\tilde{\Gamma}_{\gamma \gamma} (A) = 
\lim_{Q^2 \to 0} \frac{M_A^2}{Q^2} 
\Gamma(A \to \gamma^*_L \gamma_T) ,
\label{eqn:tilde}
\end{equation}
where $\Gamma(A \to \gamma^*_L \gamma_T)$ is the parameter
of the axial-vector meson $A$ decaying into a virtual longitudinal photon
and a real transverse photon.
The relevant parameters are also summarized in Table~\ref{tab:ppv}.

\begin{center}
\begin{table}
\caption{Parameters of the $\eta$, $\eta^\prime$~\cite{cleo1}, 
$f_1(1285)$, and $f_1(1420)$~\cite{l31, l32}.
$\Gamma_{\gamma \gamma}$ for an axial-vector 
meson shall read $\tilde{\Gamma}_{\gamma \gamma}$ defined in 
Eq.~(\ref{eqn:tilde}).
}
\label{tab:ppv}
\begin{tabular}{c|ccc} \hline \hline
Meson & $M_M$(MeV/$c^2$) & $\Gamma_{\gamma \gamma}$(keV)
& $\Lambda_M$(MeV/$c^2$) \\
\hline
$\eta$ & $547.853 \pm 0.024$ & $0.510 \pm 0.026$
& $774 \pm 29$ \\
$\eta '$ & $957.78 \pm 0.06$ & $4.29 \pm 0.14$
& $859 \pm 28$ \\ \hline
$f_1(1285)$ & $1281.8 \pm 0.6$ & $3.5 \pm 0.8$
& $1040 \pm 78$ \\
$f_1(1420)$ & $1426.4 \pm 0.9$ & $3.2 \pm 0.9$
& $926 \pm 78$ \\ \hline \hline
\end{tabular}
\end{table}
\end{center}

Experimentally, for pseudoscalar mesons,
the TFF of the $\pi^0$ meson has been measured recently by BaBar~\cite{babar1} 
and by Belle~\cite{pi0tff}, and those of
$\eta$ and $\eta^\prime$~\cite{babar2}
and $\eta_c$~\cite{babar3} by BaBar for $Q^2 \le 40~\GeV^2$.

Two-photon production of axial-vector mesons, $f_1(1285)$ and 
$f_1(1420)$, which is interpreted as 
a two-photon fusion of a longitudinal (helicity-0) and a real photon,
was studied by the L3 collaboration, who measured the 
parameters listed in Table~\ref{tab:ppv}~\cite{l31, l32}.
For scalar or tensor mesons, 
no significant data for the high-$Q^2$ region
beyond the $\rho$-meson mass scale
exist to be compared with QCD predictions; 
only yields consistent with zero 
were reported for $\gamma^* \gamma \to f_2(1270) \to \pi^+\pi^-$ at 
$Q^2 > 1.0$~GeV$^2$ by the
TPC/Two-Gamma collaboration~\cite{tpc2g}.

\begin{figure}
 \centering
   {\epsfig{file=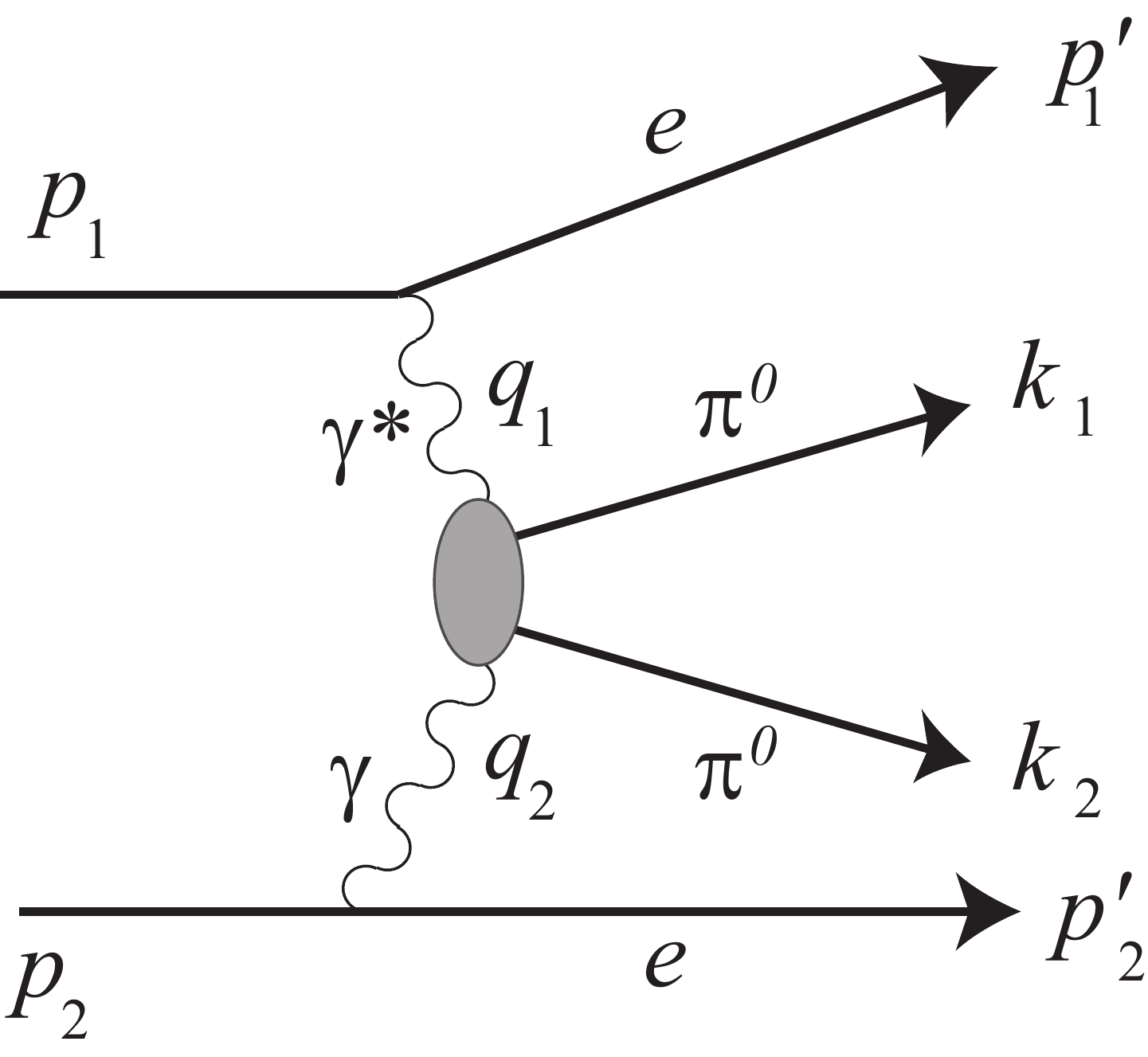,width=40mm}}
 \caption{Feynman diagram for the process 
$e^+ e^- \to e (e) \pi^0 \pi^0$.
$p_1$, $p'_1$, $p_2$, and $p'_2$ are the four-momenta of the 
incident and scattered electron or positron,
$q_1$, $q_2$ are those of 
the virtual and real photons
and $k_1$, $k_2$ are those of the produced $\pi^0$ mesons. 
}
\label{fig:eepp}
\end{figure}

\begin{figure}
 \centering
   {\epsfig{file=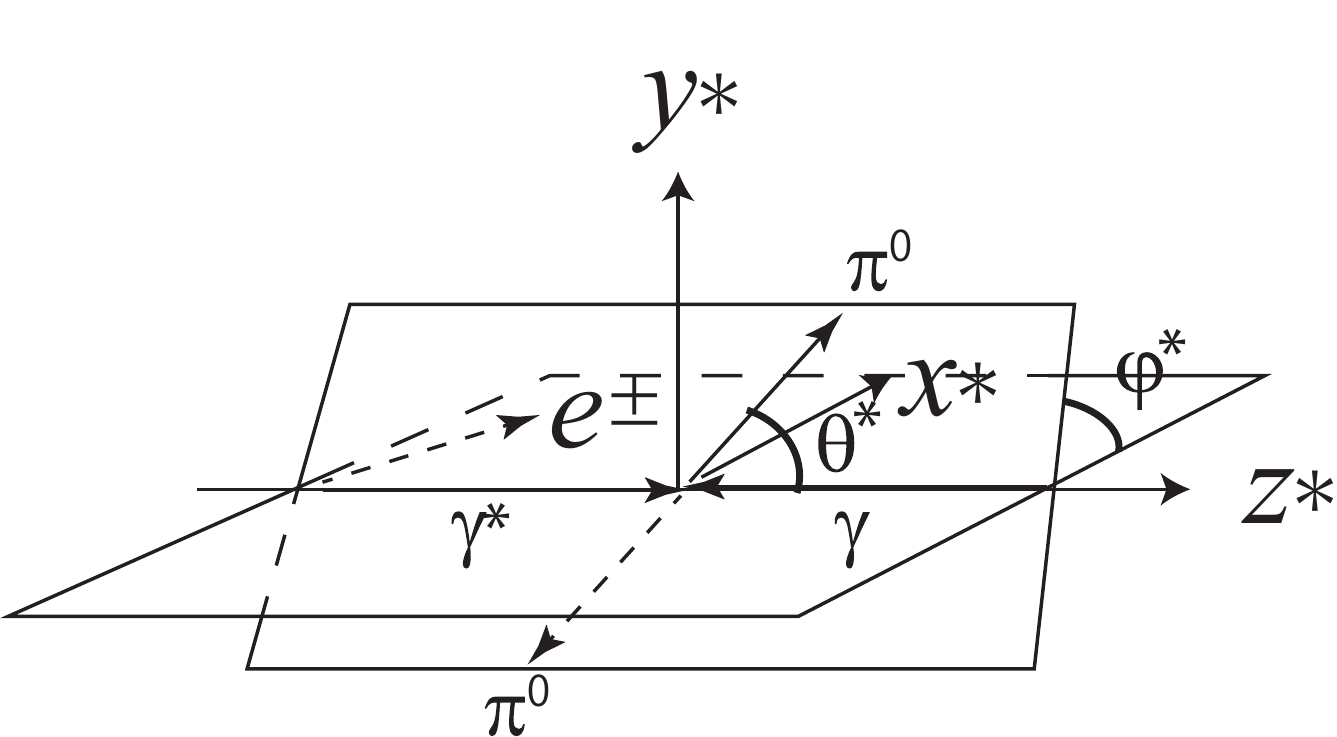,width=80mm}}
 \caption{Definition of the coordinate system
for $\gamma^* \gamma \to \pi^0 \pi^0$.
The incident $\gamma^*$ and $\gamma$ are along the $z^*$
axis, the tagged $e^{\pm}$ is in the $x^* z^*$ plane with
$p_{x^* {\rm tag}} > 0$, 
and a $\pi^0$ is produced at angles $(\theta^*, \varphi^*)$.
}
\label{fig:coord}
\end{figure}

We report a measurement of
the process $e^+ e^- \to e (e) \pi^0 \pi^0$,
where one of the $e^\pm$ is detected together with $\pi^0 \pi^0$
while the other $e^\mp$ is scattered in the forward direction
and undetected.
The Feynman diagram for the process is shown in Fig.~\ref{fig:eepp},
where the four-momenta of the particles are defined.
We consider the process  $\gamma^* \gamma \to \pi^0 \pi^0$
in the c.m. system of the $\gamma^* \gamma$.
We define the $x^* y^* z^*$-coordinate system as shown in 
Fig.~\ref{fig:coord} at fixed values of $W$ and $Q^2$; the
asterisks here denote the coordinate system that is 
used for angular variables for the
differential cross sections.
One of the $\pi^0$ mesons is scattered at angles ($\theta^*$, $\varphi^*$).
Because of the identical particles in the final state
and $P$ symmetry in the reaction, 
only the region where 
$\theta^* \le \pi/2$ and $0 \le \varphi^* \le \pi$ is meaningful.
The $z^*$-axis is along the incident $\gamma^*$ and the $x^* z^*$ plane 
is defined by the tagged $e^\pm$ such that 
$p_{x^* \; \rm tag} > 0$, where $\vec{p}_{\rm tag}$ is the 3-momentum
of the tagged $e^\pm$.

The differential cross section
for $\gamma^* \gamma \to \pi^0 \pi^0$ is given by~\cite{gss}
\begin{equation}
\frac{d \sigma(\gamma^* \gamma \to \pi^0 \pi^0)}{d \Omega} = 
\sum_{n=0}^2 t_n \cos (n \varphi^*),
\label{eqn:dsdo}
\end{equation}
with
\begin{eqnarray}
t_0 &=& 
|M_{++}|^2 + |M_{+-}|^2
+ 2 \epsilon_0 |M_{0+}|^2 
, \label{eqn:t0}\\
t_1 &=& 2 \epsilon_1 \Re \left( (M_{+-}^* - M_{++}^*) M_{0+} \right),
\label{eqn:t1}\\
t_2 &=& - 2 \epsilon_0 \Re (M_{+-}^* M_{++}) .
\label{eqn:t2}
\end{eqnarray}
Here, $M_{++}$, $M_{0+}$, and $M_{+-}$ are helicity amplitudes whose subscripts
$+$, $-$, and $0$ indicate the helicity state of the incident virtual photon
along, opposite, and transverse 
to the quantization axis, respectively,
and $\epsilon_0$ and $\epsilon_1$ are given by
\begin{eqnarray}
\epsilon_0 &=& \frac{1-x}{1 -  x + \frac{x^2}{2}} ,
\label{eqn:eps0}\\
\epsilon_1 &=& \frac{(2 - x)\sqrt{\frac{(1-x)}{2}}}{1 -  x + \frac{x^2}{2}} .
\label{eqn:eps1}
\end{eqnarray}
In turn, $x$ is defined as
\begin{equation}
x = \frac{(q_1 \cdot q_2)}{(p_1 \cdot q_2)} ,
\label{eqn:defx}
\end{equation}
where $q_1, q_2, p_1$, and $p_2$
are the four-momenta of the virtual and real photons and 
the incident electron and positron, respectively, as defined
in Fig.~\ref{fig:eepp}.
When Eq.~(\ref{eqn:dsdo}) is integrated over $\varphi^*$, we obtain
\begin{equation}
\frac{d \sigma(\gamma^* \gamma \to \pi^0 \pi^0)}{4\pi d |\cos \theta^*|} 
= |M_{++}|^2 + |M_{+-}|^2 + 2 \epsilon_0 |M_{0+}|^2 .
\label{eqn:dsdcos}
\end{equation}
The total cross section is obtained by integrating Eq.~(\ref{eqn:dsdcos}) 
over $\cos \theta^*$. 
It can be written as
\begin{equation}
\sigma_{\rm tot}(\gamma^* \gamma \to \pi^0 \pi^0)
= \sigma_{TT} +  \epsilon_0 \sigma_{LT},
\label{eqn:tots}
\end{equation}
where $\sigma_{TT}$ ($\sigma_{LT}$) corresponds to the total cross
section of two photons, both of which are transversely polarized 
(one transversely and the other longitudinally polarized);
as $Q^2 \to 0$, the second term vanishes and $\sigma_{TT}$ 
approaches the total cross section of real photon-photon
scattering.

Neutral-pion pair production in the final state $e(e)\pi^0\pi^0$
is different from the corresponding charged-pair 
process, $e(e)\pi^+\pi^-$: the $\pi^0 \pi^0$ is
a pure $C$-even state, whereas the $\pi^+ \pi^-$ is
a mixture of $C$-even and $C$-odd states.
Thus, the $\pi^0 \pi^0$ state has no contribution
from single-photon production (``bremsstrahlung process''),
whose effect must be considered in two-photon production 
of $\pi^+ \pi^-$.

In this paper, we report for the first time a measurement of the 
cross section for the process $\gamma^* \gamma \to \pi^0 \pi^0$
up to $Q^2 = 30~\GeV^2$, from which we extract 
the TFF of the $f_0(980)$ and helicity-0, -1, and -2 TFFs
of the $f_2(1270)$.

This article is organized as follows.
Section~\ref{sec:exper} briefly describes the Belle 
detector and the data sample used in this measurement.
The Monte Carlo (MC) program used for producing simulated events and for
efficiency determination is described in Sec.~\ref{sec:mc}.
Selection of events and comparison with MC data are 
explained in Sec.~\ref{sec:evsel}.
Section~\ref{sec:bkgest} is devoted to estimation of 
possible backgrounds.
The differential cross section is derived and its systematic
uncertainties are estimated in Sec.~\ref{sec:dcs}.
In Sec.~\ref{sec:meastff},
the cross section is parameterized and fitted to extract the TFFs 
of the $f_0(980)$ and the helicity-0, -1, and -2 components of the 
$f_2(1270)$ as a function of $Q^2$, which are compared to theoretical 
predictions.
Finally, Sec.~\ref{sec:summary} provides the summary and conclusion.

\section{Experimental apparatus and Data Sample}
\label{sec:exper}
In this section, we briefly describe the Belle 
detector and the data sample.
We use a 759~fb$^{-1}$ data sample recorded with
the Belle detector~\cite{belle, belleptep} at the KEKB asymmetric-energy 
$e^+e^-$ collider~\cite{kekb}.
We combine data samples collected at several beam energies: at
the $\Upsilon(4S)$ resonance $(\sqrt{s} = 10.58~{\rm GeV})$,
where the beam energy for the electron (positron) beam is
8 GeV (3.5 GeV), and 
60~MeV below it (637~fb$^{-1}$ in total); 
at the $\Upsilon(3S)$ resonance $(\sqrt{s} = 10.36~\GeV$, 
3.2~fb$^{-1}$); and near the $\Upsilon(5S)$ resonance 
$(\sqrt{s} = 10.88~\GeV$, 119~fb$^{-1}$).
Correspondingly, when combining the data, the slight dependence of the 
two-photon cross section on beam energy is taken into account
as described in Sec.~\ref{sec:dcs}.

This analysis is performed in the ``single-tag'' mode, where  
either the recoil electron or positron (hereinafter referred to as an electron)
alone is detected. 
As described in Sec.~\ref{sec:evsel} in more detail,
we restrict the virtuality ($Q^2$) of the untagged-side photon to be small 
by imposing a strict transverse-momentum balance
between the tagged electron and the final-state
neutral pion pair in the $e^+e^-$ c.m. frame with respect to
the beam axis.
In this article, we refer to events tagged by an $e^+$ or an $e^-$ 
as ``p-tag'' (positron-tag) or ``e-tag'' (electron-tag), 
respectively.

\subsection{Belle detector}
\label{sub:belle}
A comprehensive description of the Belle detector is
given elsewhere~\cite{belle, belleptep}. 
In the following, we describe only the
detector components essential for this measurement.
Charged tracks are reconstructed from the drift-time information in a central
drift chamber (CDC) located in a uniform 1.5~T solenoidal magnetic field.
The $z$ axis of the detector and the solenoid is along the positron beam
direction, with the positron beam pointing in the $-z$ direction.  
The electron-beam direction at the collision point is 
22 mrad from the $z$ axis.
The CDC measures the longitudinal and transverse momentum components, 
{\it i.e.}, along the $z$ axis and in the $r\varphi$ plane perpendicular to
the beam, respectively.
Track trajectory coordinates near the collision point are provided by a
silicon vertex detector.  
Photon detection and energy measurements are performed with a CsI(Tl) 
electromagnetic calorimeter (ECL)
by clustering of the energy deposits in the crystals of the electromagnetic 
shower from an energetic photon, where a possible connection of a
charged track to the cluster is examined by extrapolating the track
trajectory.
Electron identification (ID) is based on  $E/p$, the ratio of the 
calorimeter energy to the track momentum.

\subsection{Data sample}
\label{sub:sample}
To be recorded, events of interest here must satisfy 
one of the two ECL-based triggers: 
the HiE (High-energy threshold) 
trigger and the Clst4 (four-energy-cluster) trigger~\cite{ecltrig}. 

The HiE trigger requires that the sum of the energies measured 
by the ECL in an event exceed
1.15~GeV but that the event be not Bhabha-like;
the latter requirement is enforced by the absence of the CsiBB trigger 
(``Bhabha-veto''),
which is designed to identify
back-to-back Bhabha events~\cite{ecltrig}.
For the purpose of monitoring trigger performance, we record one in 50 
events that satisfy the CsiBB trigger
({\it i.e.}, prescaled by a factor of 50).

The Clst4 trigger requires at least four energy clusters
in the ECL with each cluster energy larger than 0.11~GeV.
This trigger is not vetoed by the CsiBB because the
Bhabha-event rate is manageable for the Clst4 trigger sample. 
Five clusters are expected in total in the 
signal events of interest if all the final-state particles 
are detected in
the region where the ECL trigger is sensitive.

We do not use information from the charged-track triggers because they
require two or more charged tracks whereas our signal
has only one.

\section{Signal Monte Carlo Code}
\label{sec:mc}
\subsection{Signal Monte Carlo, TREPSBSS}
\label{sub:treps}
We use the signal Monte Carlo (MC) generator TREPSBSS,
which has been developed to calculate the efficiency for
single-tag two-photon events, $e^+ e^- \to e (e) X$,
as well as the two-photon luminosity function for
$\gamma^* \gamma$ collisions at an $e^+e^-$ collider.

TREPSBSS implements Eqs.~(2.16) to (2.20) 
of Ref.~\cite{tpbw} and is based on the MC code in Ref.~\cite{treps},
which was modified to match the single-tag configuration. 
We regard Eq.~(\ref{eqn:tots})
as the total cross section of the 
$\gamma^* \gamma$ collisions, according to 
Eq.~(2.16) of Ref.~\cite{tpbw}, although $\epsilon_0$ is a 
variable depending on experimental conditions.
It is possible to estimate $\epsilon_0$ for specific experimental conditions by 
taking the average value of $\epsilon_0$ calculated for the selected
signal events in MC or experimental data. 
Under our experimental conditions, $\epsilon_0$ ranges from
0.7 to 0.9.
However,  $\sigma_{\rm LT}$ cannot be separated from $\sigma_{\rm TT}$ 
based on this information only.

In TREPSBSS, the following kinematical variables
are used for characterizing 
$\gamma^* \gamma$ collisions of the generated events 
and for an integration to calculate the two-photon
luminosity function: $Q^2_1$ and $Q^2_2$ (the absolute
value of momentum transfer squared of the highly virtual and the less
virtual incident photons, respectively), $W$ (c.m.
energy of the incident $\gamma^* \gamma$ system),
$\omega_2$ (the energy of the photon with the smaller  
virtuality), and $\Delta \varphi$ (the azimuthal-angle
difference between the two virtual photons). 
The choice of kinematical variables is discussed in Ref.~\cite{schuler2}.

We always retain the condition $Q^2_1 > Q^2_2$ for the virtuality of the two
colliding photons by requiring $Q^2_1 > 3.0$~GeV$^2$
and $Q^2_2 < 1.0$~GeV$^2$ for the ranges of
integration, event generation, and selection. 
This $Q^2_2$ range is sufficient to generate signal events over
the kinematical region for the $|\Sigma \vec{p}_t^*|$
selection criterion.

The form-factor effect for the photon with the smaller  virtuality is
assumed to follow the $1/(1+Q^2_2/m_{\rho}^2)^2$ dependence,
where $m_{\rho}$ is the $\rho$-meson mass, 0.77~GeV/$c^2$.
The cross section defined below is that extrapolated to $Q^2_2=0$,
assuming this $Q^2_2$ dependence.

The distribution of $Q^2_1$ ($\equiv Q^2$) in the MC sample is 
arbitrary 
and should not affect the final result. 
Conventionally, we choose the distribution corresponding to the flat
form factor in order to retain high statistics of the signal
MC in the high-$Q^2$ region
but we weight the MC sample by an additional $1/Q^2$ factor to 
model a more realistic distribution for the efficiency derivation 
in each $Q^2$ region discussed in this analysis, and for comparison 
of distributions between the signal-MC samples and the 
experimental data.

The luminosity function, which
is a conversion factor between the
$e^+e^-$-based cross section $\sigma_{ee}$ and the $\gamma^* \gamma$-based
cross section  $\sigma_{\gamma^* \gamma}$, is calculated in the same code.
It is defined by
\begin{equation}
\frac{d^2 \sigma_{ee}}{dQ^2 dW} = 2\frac{d^2 L_{\gamma^* \gamma}}{dQ^2 dW}
\sigma_{\gamma^* \gamma}(Q^2, W) ,
\end {equation}  
where $2L_{\gamma^* \gamma}$ corresponds to 
the value of $e^+ e^-$-based integrated cross section per unit 
$\gamma^* \gamma$-based cross section
in Eq.~(2.16) of Ref.~\cite{tpbw}, and the differential
luminosity function $d^2 L_{\gamma^* \gamma}/dQ^2 dW$ is calculated
by performing a three-dimensional numerical integration over
$Q^2_2$, $\omega_2$, and $\Delta \varphi$. 
The factor of two incorporates for the contributions from the e-tag and p-tag
processes included in the entire $e^+e^-$-based cross section.

\subsection{Event generation}
\label{sub:evtgen}
The event generation is performed using the same
integrand but including initial-state radiation (ISR)
effects from the tag-side electron. 
Inclusion of ISR changes the kinematics and $Q^2_1$ significantly. 
Meanwhile, ISR from the untagged side has little effect
because an ISR photon is nearly parallel not only
to the initial-state electron but also to the final-state
untagged electron. 
We use an exponentiation technique~\cite{expo}
for the photon emission based on the parameter 
$\eta = (2\alpha/\pi)(\log(Q^2_1/m_e^2)-1)$ and
the probability density for the photon energy distribution,
$dP(r_k) \propto r_k^{\eta-1} dr_k$, where 
\begin{equation}
r_k \equiv \frac{E^*_{\rm ISR}}{E_{\rm beam}^*}.
\label{eqn:rk1}
\end{equation}
As an approximation, the photon is always emitted along the incident 
electron direction on the tagged side.
We limit the fractional energy of radiation to below $r_k^{\rm max}=0.25$
in the MC generation.

In this configuration, the correction factor $1+\delta$ to the tree-level 
cross section is close to unity~\cite{radcor1}.
Most of the events with large $r_k$, typically $r_k > 0.1$,
are rejected 
by the selection criterion that uses  $E_{\rm ratio}$
(the definition and criterion being described in Sec.~\ref{sub:selec})
by requiring energy-momentum conservation between the initial- and
final-state particle systems without radiation.
This effect is accounted for as a loss of efficiency for events
with $r_k \leq r_k^{\rm max}$.

We generate events with a virtuality of 
the tagged-side photon $Q^2_1$ distributed with a constant form factor
over its continuous range $Q^2 > 3.0$~GeV$^2$.
The $Q^2$ value of each event is modified by ISR. 
We use the momentum of the ISR photon to determine
the true $Q^2$ value in signal-MC events
and to study the $Q^2$ dependence of the detection efficiency.
We correct the experimental $Q^2$ dependence for the ISR effect
using factors obtained from the signal MC (see Sec.~\ref{sub:radcor}).
We choose 15 different $W$ points between 0.4~GeV and 2.5~GeV
for the calculation of the luminosity function and event generation.

We use a GEANT3-based detector simulation~\cite{geant}
to study the propagation of the generated particles through the
detector; the same code as for the experimental 
data is used for reconstruction and selection of the MC simulated events.
We thus obtain the selection efficiency as functions
of $Q^2$, $W$, and $|\cos \theta^*|$ for a flat $\varphi^*$
dependence of the differential cross section. 
A correction for the observed $\varphi^*$ dependence
is discussed in Sec.~\ref{sub:effcor}.

\section{Event Selection}
\label{sec:evsel}
In this section, we describe the event selection
and present some raw distributions to compare with those
from MC.

\subsection{Selection criteria for signal candidate events}
\label{sub:selec}
A signal event contains an energetic electron and four photons.
The kinematical variables are calculated in the laboratory system 
unless otherwise noted; those in the $e^+e^-$ or $\gamma^* \gamma$
c.m. frame are identified with an asterisk in this section.
We require exactly one track that satisfies $p_t > 0.5~\GeV/c$,
$-0.8660 < \cos \theta <0.9563$, $dr < 1$~cm, and $|dz| < 5$~cm.
There must be no other tracks that satisfy $p_t > 0.1$~GeV/$c$, 
$dr < 5$~cm, and $|dz| < 5$~cm in the above angular range. 
Here, $p_t$ is the transverse
momentum in the laboratory frame with respect to the positron beam axis,
$\theta$ is the polar angle of the momentum
direction with respect to the $z$ axis,
and ($dr$, $dz$) are the cylindrical coordinates of the point of closest 
approach of the track to the beam axis. 
We also require one or more neutral clusters in the ECL, whose
energy sum is greater than 0.5~GeV,
as a pre-selection criterion for the experimental samples.
These conditions are efficient in selecting a signal process
within the kinematical regions 
of  $e^+ e^- \to e (e) \pi^0 \pi^0$ in which one electron escapes
detection at small forward angles.

For electron ID, we require $E/p > 0.8$ for the candidate
electron track.
The absolute value of the momentum of the electron must be 
greater than 1.0~GeV/$c$,
where the electron energy is corrected for photon radiation or 
bremsstrahlung in the following way.
In a $3^\circ$ cone around the track, we collect all photons 
in the range 0.1~GeV$< E_\gamma < p_e c/3$,
where $p_e$ is the measured absolute momentum of the electron track. 
The absolute momentum of the electron 
is replaced by $p_e + \Sigma E_\gamma$.
The cosine of the polar angle for the electron ($\theta_e$)
must be within the range $-0.6235 < \cos \theta_e < 0.9481$,
which is the sensitive region for the HiE and Clst4 triggers. 

We search for a $\pi^0$ candidate 
reconstructed from a photon pair with each photon having an
energy above 0.1~GeV
and a polar angle in the range $-0.6235 < \cos \theta_\gamma < 0.9481$.
We constrain the polar angle of the photons from 
at least one $\pi^0$ in the sensitive region of the ECL triggers
by the latter condition,
in order to reduce the systematic uncertainty of the trigger efficiency.
The two-photon invariant mass is required to satisfy the criterion
0.115~GeV/$c^2 < M_{\gamma\gamma} < 0.150$~GeV/$c^2$.
This $\pi^0$ candidate is referred to as ``$\pi 1$''.

In parallel, we search for $\pi^0$ candidates with the $\pi^0$-mass-constrained
fit among all pairwise combinations of photons in the entire ECL region. 
We select only combinations of photon pairs whose
goodness of fit satisfies $\chi^2<16$. 
If $\pi 1$ is also selected by the mass-constrained
fit, we replace the four-momentum of $\pi 1$ by that of the result
of the fit.

We require that only one more pion be found among the $\pi^0$ candidates from
the mass-constrained fit which does not share any photons with $\pi 1$.
We refer to the second $\pi^0$ as ``$\pi 2$''.

If there are two or more possible assignments of
$\pi 1$ and $\pi 2$, we choose the one with the highest-energy 
photon to construct $\pi 1$. 
If there are still two or more
combinations that share the highest-energy photon in $\pi 1$,
we choose the one in which the other photon in $\pi 1$ has the higher 
energy.

\begin{figure}
\centering
\includegraphics[width=7cm]{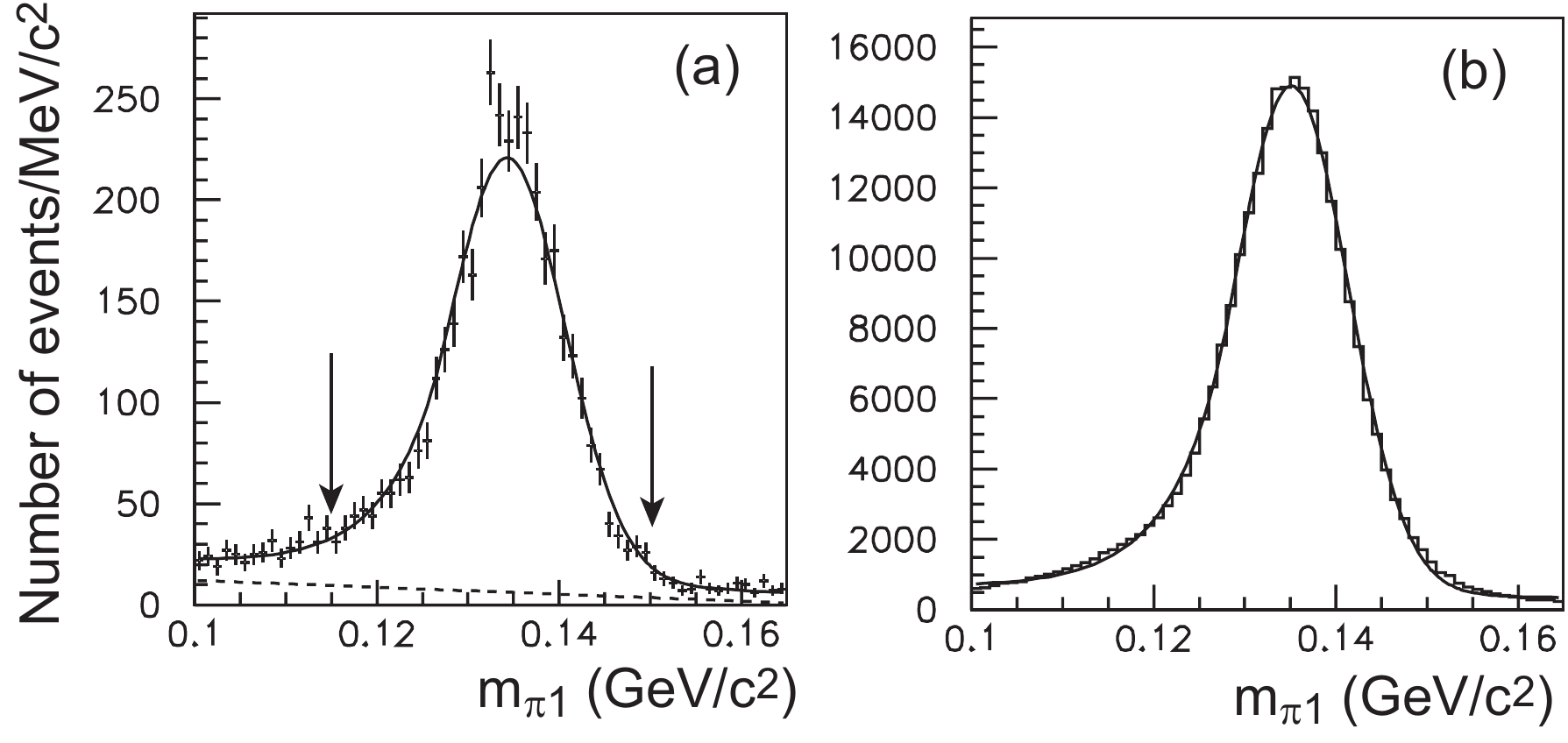}
\centering
\caption{
(a) The experimental 
distributions for the $\gamma \gamma$
invariant mass forming $\pi 1$. The arrows indicate the
selection range. (b) The distribution from the signal MC for
the same variable.
Statistics of the MC figure are arbitrary.
The fit for the comparison is shown by a solid curve 
in each of the distributions (see the text). 
The background component in the experimental data from the fit
is shown by the dashed line.
}
\label{fig:widepi0}
\end{figure}

Figure~\ref{fig:widepi0} shows the $\gamma \gamma$ invariant-mass
distribution when constructing $\pi 1$ 
with a looser criterion of the two-photon invariant mass.
The experimental data are compared with the distribution from the
signal-MC sample using fits described below.
The signal-MC distribution is fitted
by the sum of a Crystal Ball function~\cite{crybal} and a 
linear function (signal component). 
Then, the experimental distribution is
fitted by the sum of the signal component with the determined shape 
parameters and an additional linear function (background 
component), where the normalization and horizontal position 
for the signal component are allowed to float (resulting in a shift 
of the peak position by $-0.8$~MeV/$c^2$).
The sole purpose of this fit is to compare the figures,
which indicates a reasonable agreement and provides an
estimate of the background contamination.

We apply additional selection criteria for $\pi 1$ and $\pi 2$
to reduce contamination from low-energy background photons.
The energy asymmetry for the two daughters
($\gamma 1$ and $\gamma 2$) of either pion, defined as 
\begin{equation}
E_{\rm asym} = \frac{|E_{\gamma 1}- E_{\gamma 2}|}{E_{\gamma 1}+ E_{\gamma 2}} ,
\label{eqn:easym}
\end{equation}
must satisfy $E_{\rm asym} <0.8$.
We require that the $\pi^0$ energies and transverse
momenta satisfy
$E_{\pi1} > 0.4$~GeV,  $E_{\pi2} > 0.3$~GeV,
$E_{\pi1} + E_{\pi 2} > 1$~GeV,
$p_{t,\pi1} > 0.15$~GeV/$c$, and $p_{t,\pi2} > 0.15$~GeV/$c$, respectively.
We require that the polar angle of $\pi 2$ in the laboratory frame 
satisfy $-0.8660 < \cos \theta <0.9563$.

We reject events with a back-to-back configuration of an electron
and $\pi 1$ in the
$e^+e^-$ c.m. frame, to suppress Bhabha events
in which a track is not reconstructed; we require
$\zeta^*(e, \pi 1) < 177^\circ$,
where $\zeta^*(e, \pi 1)$ is the opening angle between the
electron and the $\pi 1$ system. 

We require the tagged lepton to have the correct charge sign 
(``right-sign'') with respect to the beam from which it originates
in the $e^+ e^-$ c.m. frame:
\begin{equation}
q_{\rm tag} \times (p^*_{z,e} + p^*_{z,\pi 1} + p^*_{z,\pi 2})< 0 ,
\label{eqn:tagchg}
\end{equation}
where $q_{\rm tag}$ is the tagged lepton charge.

We apply a kinematical selection of 
$0.85 < E_{\rm ratio} <1.1$, where $E_{\rm ratio}$
is defined as
\begin{equation}
E_{\rm ratio} = \frac{E^{* {\rm measured}}_{\pi^0\pi^0}}
{E^{* {\rm expected}}_{\pi^0\pi^0}} 
\label{eqn:eratio}
\end{equation}
and $E^{* {\rm measured}}_{\pi^0\pi^0}$ 
($E^{* {\rm expected}}_{\pi^0\pi^0}$) is the $e^+ e^-$ c.m.
energy of the $\pi^0 \pi^0$ system measured directly 
(expected by kinematics without radiation). 
This requirement is motivated by
a three-body kinematical calculation for  $e^+ e^- \to e (e) R$ 
that is to be followed by $R \to \pi^0 \pi^0$, where
$R$ need not be a physical 
resonance because this is a kinematical calculation.
We impose a four-momentum conservation condition
$p_{\rm initial}(e^+e^-) = p_{\rm final}(e (e) R)$ 
wherein the direction of the $R$ momentum is taken to be 
parallel to that of the observed $\pi^0 \pi^0$ system in 
the $e^+e^-$ c.m. frame.
The expected energy of the $R(=\pi^0 \pi^0)$ system,
$E_{\pi^0\pi^0}^{\rm *expected}$, is obtained
by assigning the measured $\pi^0\pi^0$ invariant mass 
to the $R$ system.

\begin{figure}
\centering
\includegraphics[width=6cm]{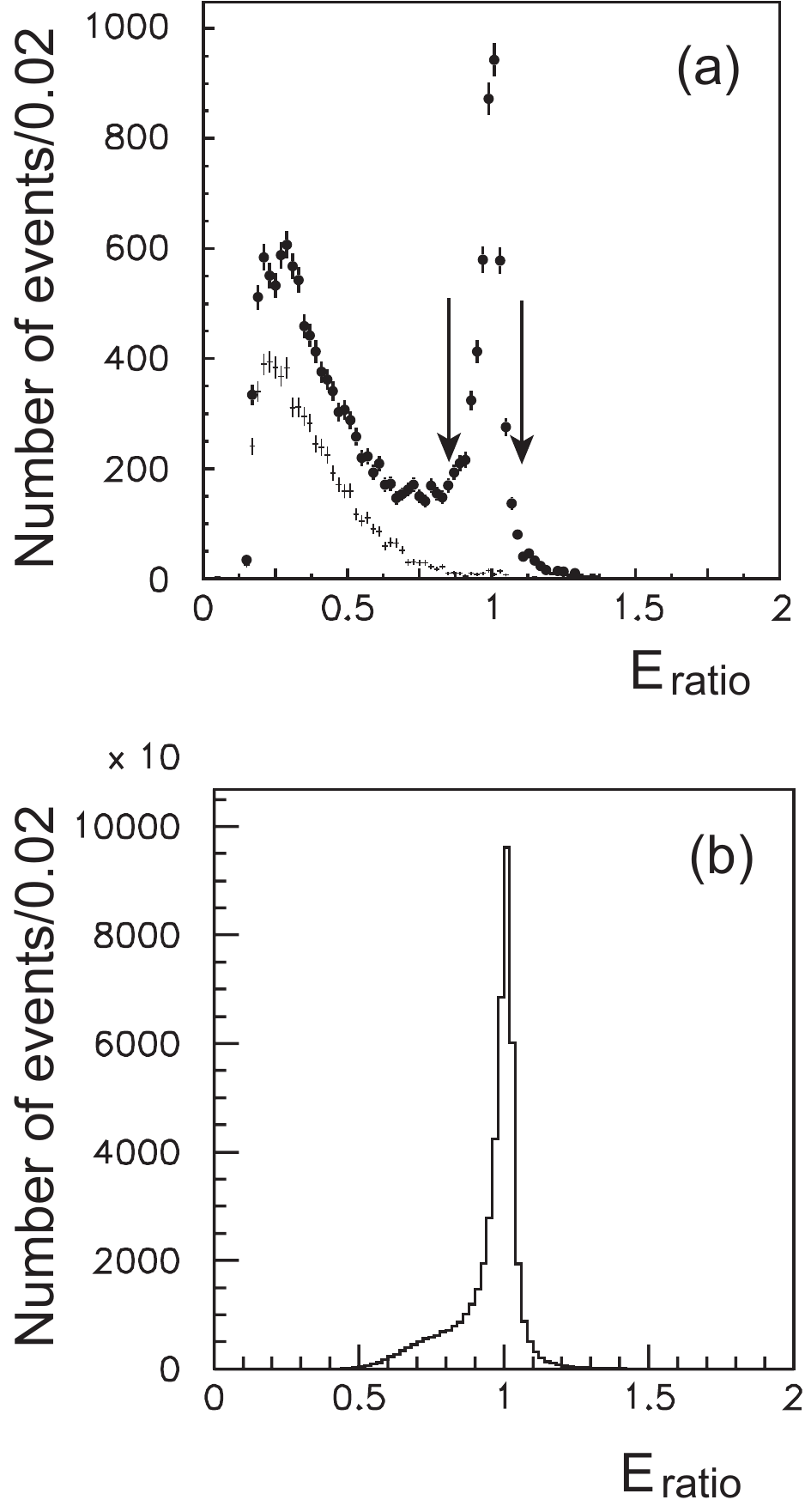}
\centering
\caption{The distributions for $E_{\rm ratio}$ for (a) the
experimental data 
and (b) 
the signal-MC sample.
The cross plots
in (a) show the distribution for the wrong-sign events, and the
arrows indicate the signal region.
Statistics of the MC figure are arbitrary.
}
\label{fig:seldist}
\end{figure}

In Fig.~\ref{fig:seldist}, we show a comparison of
$E_{\rm ratio}$ between the data and signal MC, where we observe
a sharp peak corresponding to the signal process in the data
that is consistent with the MC. 
The distribution for ``wrong-sign'' events that have the opposite 
lepton charge to Eq.~(\ref{eqn:tagchg})
is also shown; here, only a small peak is 
seen for the wrong-sign events near $E_{\rm ratio}=1$. 
This means that the backgrounds from $e^+e^-$ annihilation events,
where the charge asymmetry of the tracks is not expected,
are negligibly small.
Meanwhile, there are significant right-sign backgrounds with a small
$E_{\rm ratio}$. 
We discuss such events in Sec.~\ref{sub:other}.

\begin{figure}
\centering
\includegraphics[width=8cm]{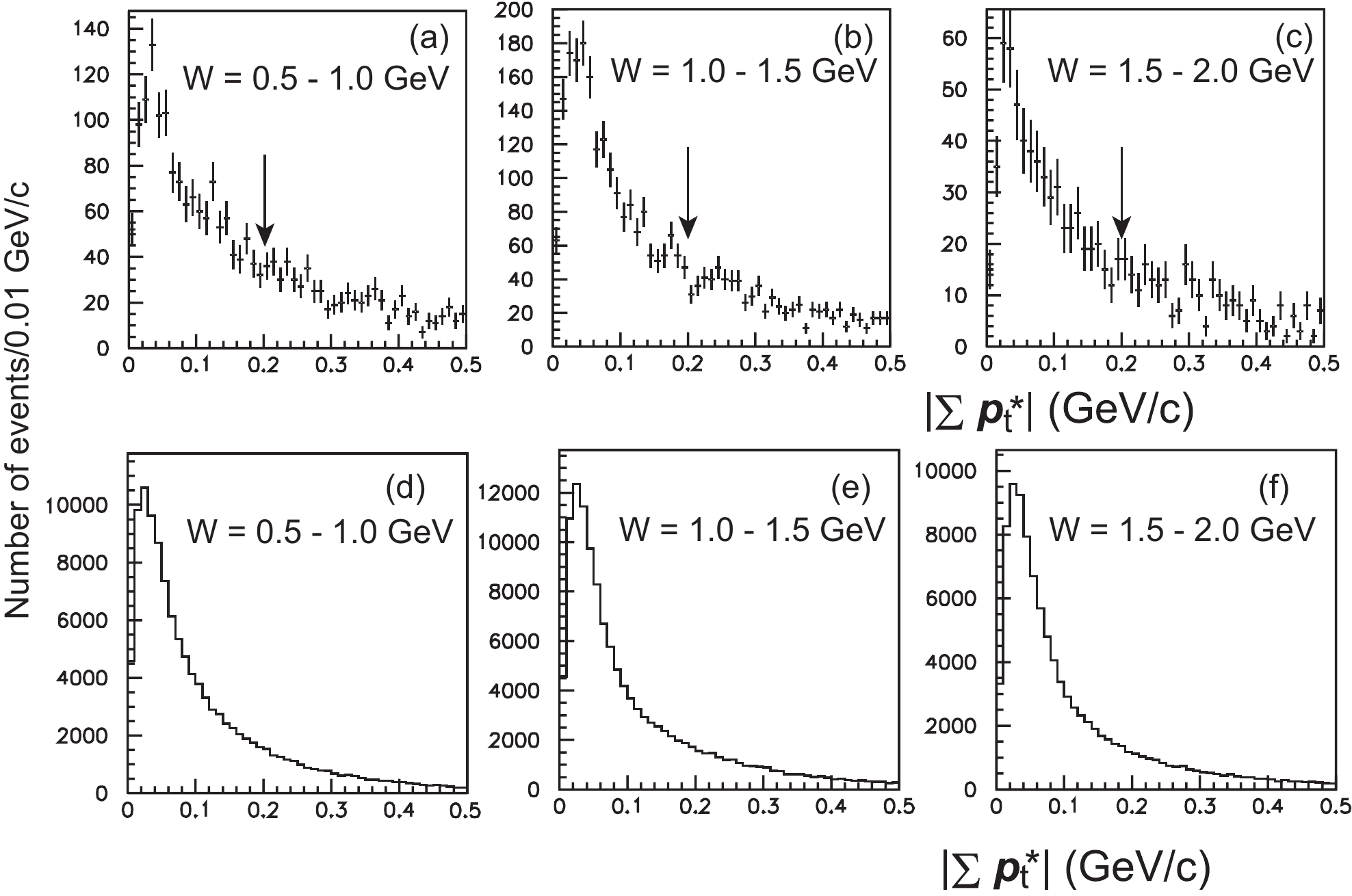}
\centering
\caption{(a,b,c) The distributions for the $p_t$ balance for
the data samples in three different $W$ ranges indicated
in each panel. 
The arrows show the selection region.
(d,e,f) The corresponding 
distributions from the signal MC.
Statistics of the MC figures are arbitrary.
}
\label{fig:ptdist}
\end{figure}

We require transverse momentum balance in the 
$e^+e^-$ c.m. frame, $|\Sigma \vec{p}_t^*| < 0.2$~GeV/$c$,
where
\begin{equation}
|\Sigma \vec{p}_t^*| =
|\vec{p}^*_{t,e} + \vec{p}^*_{t,\pi 1} + \vec{p}^*_{t,\pi 2} | .
\end{equation}
We show the distribution for $|\Sigma \vec{p}_t^*|$ 
(referred to as ``$p_t$ balance'') in a wider
range than the signal region in three different $W$ ranges in 
Fig.~\ref{fig:ptdist}, where the samples 
after the three-body kinematics condition by the 
$E_{\rm ratio}$ selection criterion applied are shown.
The experimental distributions are compared with those of the
signal-MC events. 
The signal peak near $|\Sigma \vec{p}_t^*|=0$
is a little wider in the data than in the MC. 
This is partially due to the backgrounds in the data and partially
due to non-inclusion of the finite-angle initial state radiation
(ISR) on the tag side or the non-tag side. 
However, it is expected that the transverse momentum of the 
observed system, $|\Sigma \vec{p}_t^*|$, is dominated 
by those of the colliding virtual photons, which are taken into 
account in the MC simulation because the $Q^2$ of the electron
after the ISR emission is expected to be smaller than
$Q^2$ of the colliding photon emitted by the electron. 
This is supported by our 
previous study of the $|\Sigma \vec{p}_t^*|$ distribution
for the $\gamma^* \gamma \to \pi^0$ 
process~\cite{pi0tff}.

We find that events with 
$E_{\rm ratio} < 0.7$ do not peak at $|\Sigma \vec{p}_t^*|=0$.
These features show that 
some charged or neutral hadrons or photons from 
$\pi^0$ decay escape detection in these events. 
They are considered to come from multi-hadron
production in two-photon processes or 
virtual pseudo-Compton scattering
($e^+e^- \to e (e) \gamma^*$, $\gamma^* \to {\rm hadrons}$)
according to the observed asymmetry in the correct- and 
incorrect-charge sign events,
with a leakage of these background components into the 
$E_{\rm ratio}$ signal region.

The $E_{\rm ratio}$ and $|\Sigma \vec{p}_t^*|$ distributions
are discussed in more detail for background estimation, 
in Sec.~\ref{sec:bkgest}. 

\subsection{Assignment of kinematical variables for a signal event}
\label{sub:assign}
We assign four kinematical variables $Q^2$, $W$, $|\cos \theta^*|$, and
$\varphi^*$ to each signal candidate event.  
The angles $|\cos \theta^*|$
and $\varphi^*$ are defined in the $\gamma^* \gamma$
c.m. frame, where the direction of $\gamma^*$ has  
$\cos \theta^*=1$ and the azimuthal direction 
of the recoiling electron defines $\varphi^* =0$ (Fig.~\ref{fig:coord}).
Note that only these two variables are defined in the
$\gamma^* \gamma$ c.m. frame.
In contrast, all the other variables with an asterisk appearing below
are defined  in the $e^+e^-$ c.m. frame.

The negative of the invariant mass squared, $Q^2$, of the 
virtual incident photon
is calculated using the measured four-momentum of the detected
electron ($p_e$) from
\begin{eqnarray}
Q^2_{\rm rec} &=& - (p_{\rm beam} - p_e)^2 \nonumber \\
&=& 2 E_{\rm beam}^* E_e^*(1 + q_{\rm tag} \cos \theta_e^*),
\label{eqn:q2meas}
\end{eqnarray}
where $p_{\rm beam}$ is the
nominal four-momentum of the beam particle with the same charge
as the detected electron; the right-hand side is given by
the beam energy $E_{\rm beam}^*$ and the observables of the tagged
electron in the $e^+e^-$ c.m. frame.
We do not apply a correction for
initial-state radiation (ISR) on an event-by-event basis; 
instead, this effect is taken into account in the
correction for the differential cross section, as mentioned 
in Sec.~\ref{sec:dcs}.

The c.m. energy of the incident $\gamma^* \gamma$ collision, $W$,
is the invariant mass of the final-state $\pi^0 \pi^0$ system.
The pion scattering angle $\theta^*$ is defined in the 
$\gamma^* \gamma$ c.m. frame as an angle between the virtual
photon and that of one of the produced pions. 
In case an ISR photon is emitted in the tagged-electron side, 
the direction of
the virtual photon is slightly misreconstructed and
induces an error in $\cos \theta^*$. 
However, the change due to this effect is typically 
$\pm 0.01$ in $\cos \theta^*$.
 This is smaller than the angular resolution,
typically $\sigma_{\cos \theta^*}=0.02$, and the bin width, 0.2,  
and thus we neglect the effect. 

There is a two-fold symmetry for the two sides of the plane, since
$\frac{d \sigma}{d \varphi^*} (\varphi^*) \equiv 
\frac{d \sigma}{d \varphi^*}(2\pi - \varphi^*)$, so this angle is
limited to the range $0 \leq \varphi^* \leq \pi$.

\subsection{Comparison of the experimental candidates with the signal-MC events}
\label{sub:datamc}
In this subsection, we show various distributions of the
selected signal candidates.
Backgrounds are not subtracted in the experimental data.
Data distributions are compared with the signal MC, where a uniform angular 
distribution and a representative $Q^2$ dependence are assumed.
As most $Q^2$ and angular dependence in these figures arise from 
kinematics, such comparisons are meaningful, but no perfect agreement 
between data and MC should be expected.

\begin{figure}
\centering
\includegraphics[width=7.7cm]{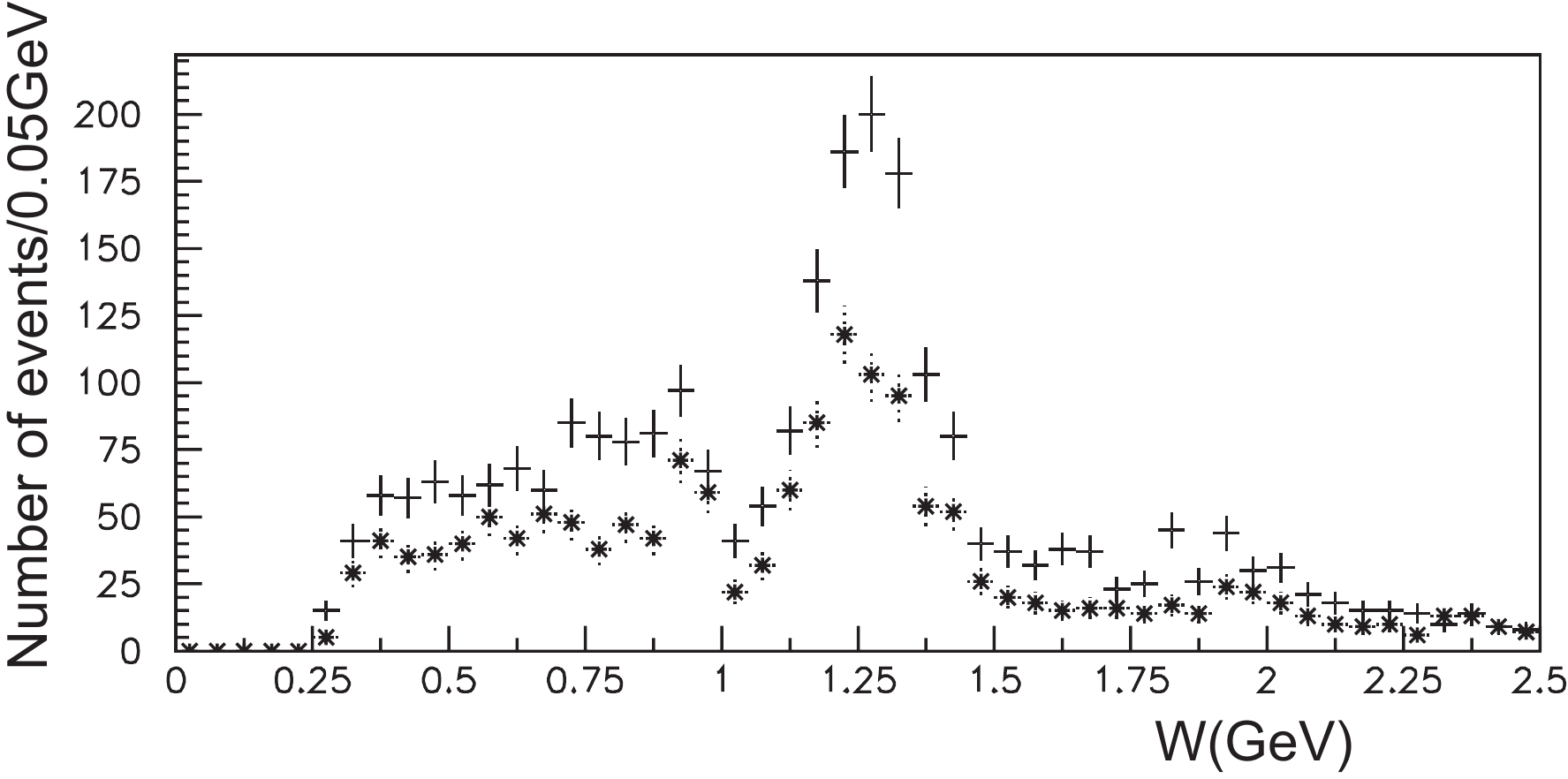}
\centering
\caption{The experimental distributions for $W$ of the
signal candidates for 3~GeV$^2$ (5~GeV$^2$) $< Q^2 < 30$~GeV$^2$ 
for the e-tag (p-tag) samples. 
Backgrounds are not subtracted. 
The cross and asterisk plots are for the e-tag and p-tag samples,
respectively.
}
\label{fig:wdist}
\end{figure}

The experimental $W$ distribution
is shown in Fig.~\ref{fig:wdist} for $W \le$ 2.5~GeV.
For comparison, the corresponding distributions from the signal MC 
are also shown in the following figures. 
In Figs.~\ref{fig:candq2} -- \ref{fig:epicost12}, all
events within $Q^2_L< Q^2 < 30$~GeV$^2$ and 0.5~GeV $<W<2.1$~GeV
are integrated, where
$Q^2_L = 3$~GeV$^2$ ($=5$~GeV$^2$) for the e-tag (p-tag) sample. 
We do not use the data below $W<0.5$~GeV because the signal 
efficiency and the signal-to-background ratio decrease steeply
below that energy. 
No large discrepancy between the data and signal MC is seen in all 
the figures.
This implies that the MC is a faithful representation of reality
and the backgrounds in the experimental
data are not very large. The difference in the pion energy 
distributions reflects the difference in its angular distribution
in the $\gamma^* \gamma$ c.m. frame between the data and signal MC,
with the uniform angular distribution for the latter.

\begin{figure}
\centering
\includegraphics[width=8cm]{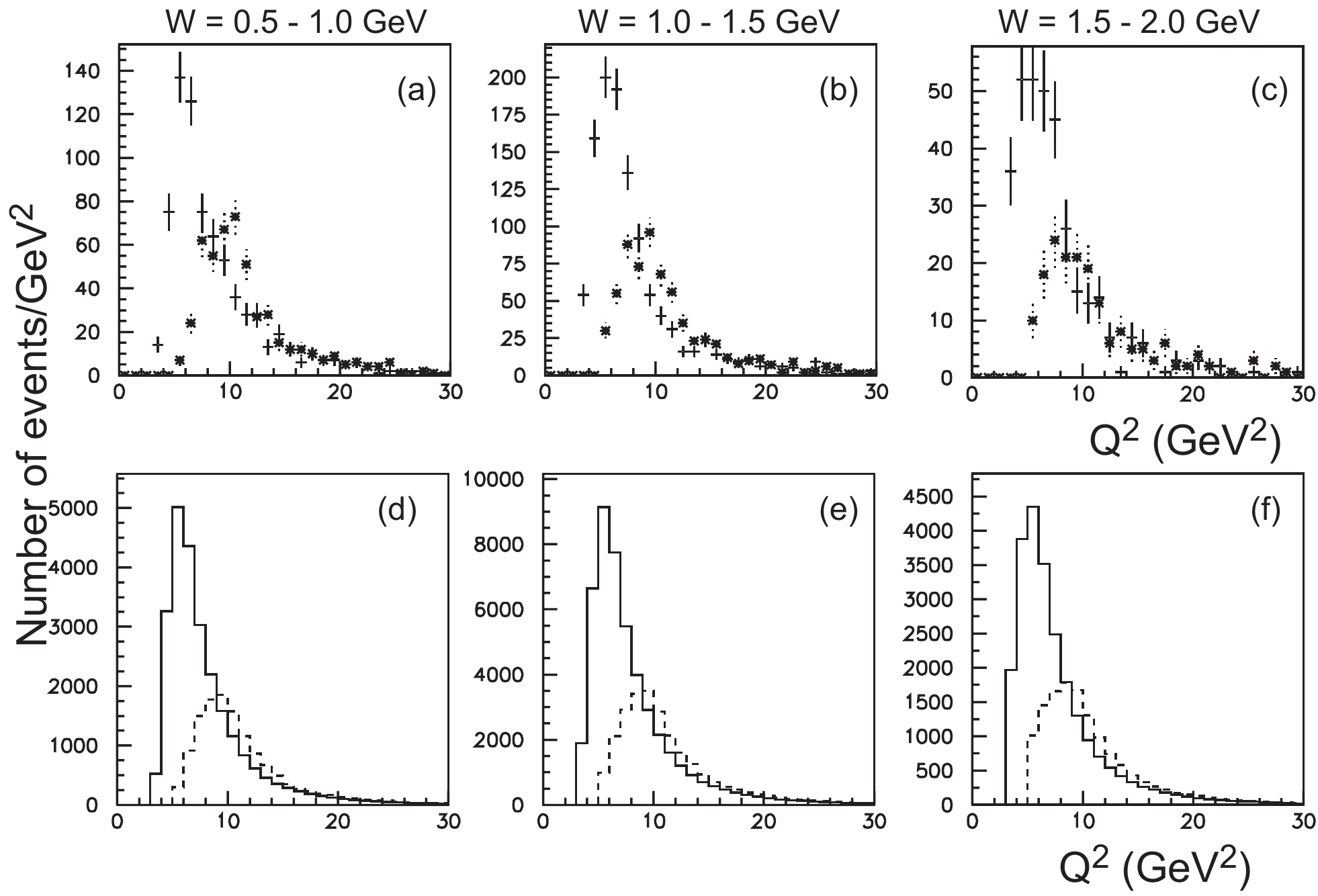}
\centering
\caption{(a,b,c) The $Q^2$ distributions for
the data samples in three different $W$ ranges indicated
above each panel. 
The cross and asterisk 
plots are for the e-tag and p-tag samples, respectively.
(d,e,f) The corresponding 
distributions from the signal MC, where 
the solid and dashed histograms
are for the e-tag and p-tag samples, respectively.
Statistics of the MC figures are arbitrary, 
but the scale is common for the e- and p-tags
so their ratio could be compared between MC and data.
}
\label{fig:candq2}
\end{figure}
 
\begin{figure}
\centering
\includegraphics[width=8cm]{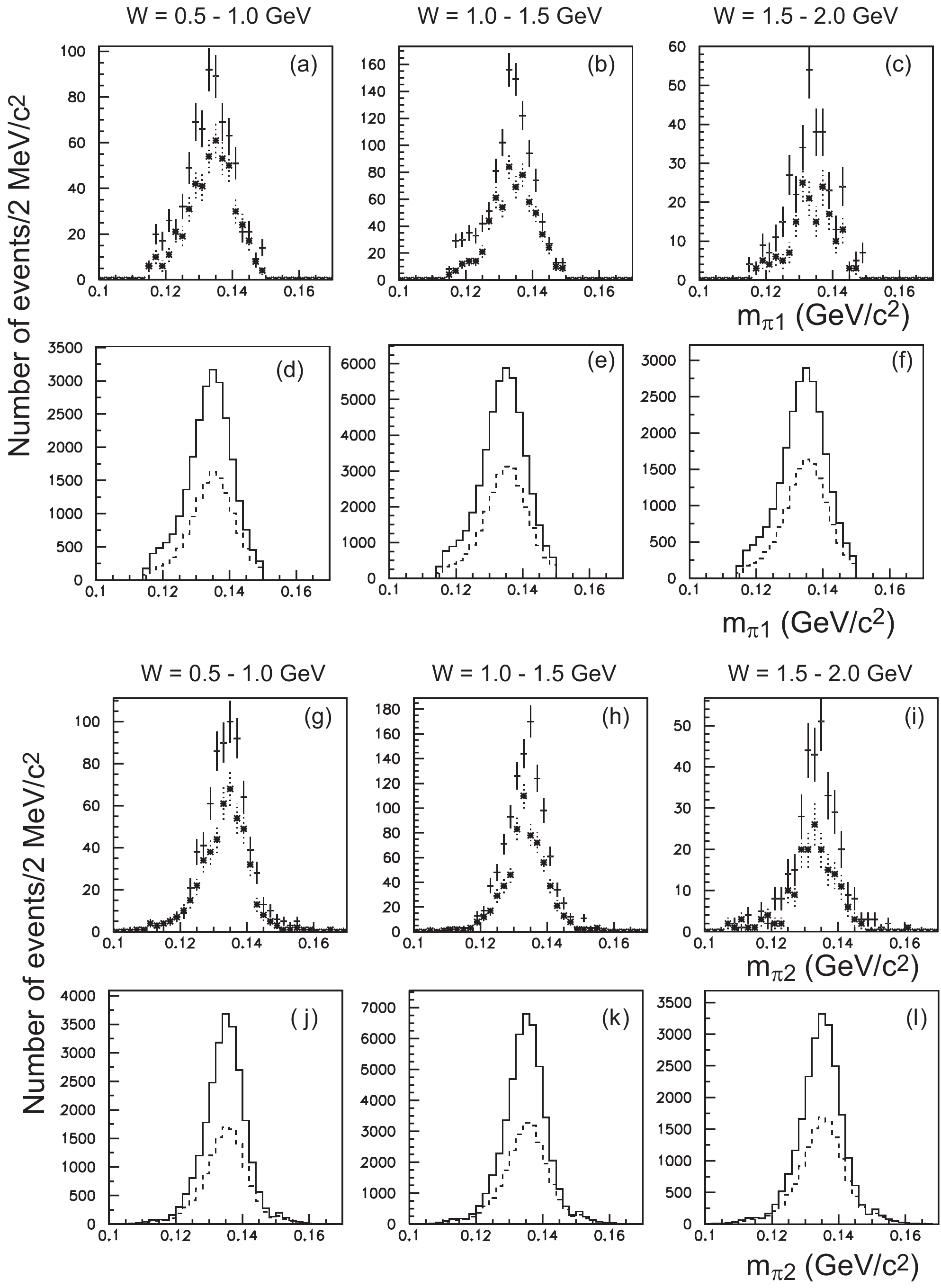}
\centering
\caption{(a,b,c,g,h,i) 
The distributions for the $\gamma \gamma$ invariant mass
to construct $\pi 1$ (a,b,c) and $\pi 2$ (g,h,i) for
the data samples in three different $W$ ranges indicated
above each panel. 
(d,e,f,j,k,l) The corresponding 
distributions from the signal MC for  $\pi 1$ (d,e,f) and $\pi 2$ (j,k,l).
The legend and comments are the same as those in Fig.~\ref{fig:candq2}. 
}
\label{fig:mpi12}
\end{figure}

\begin{figure}
\centering
\includegraphics[width=8cm]{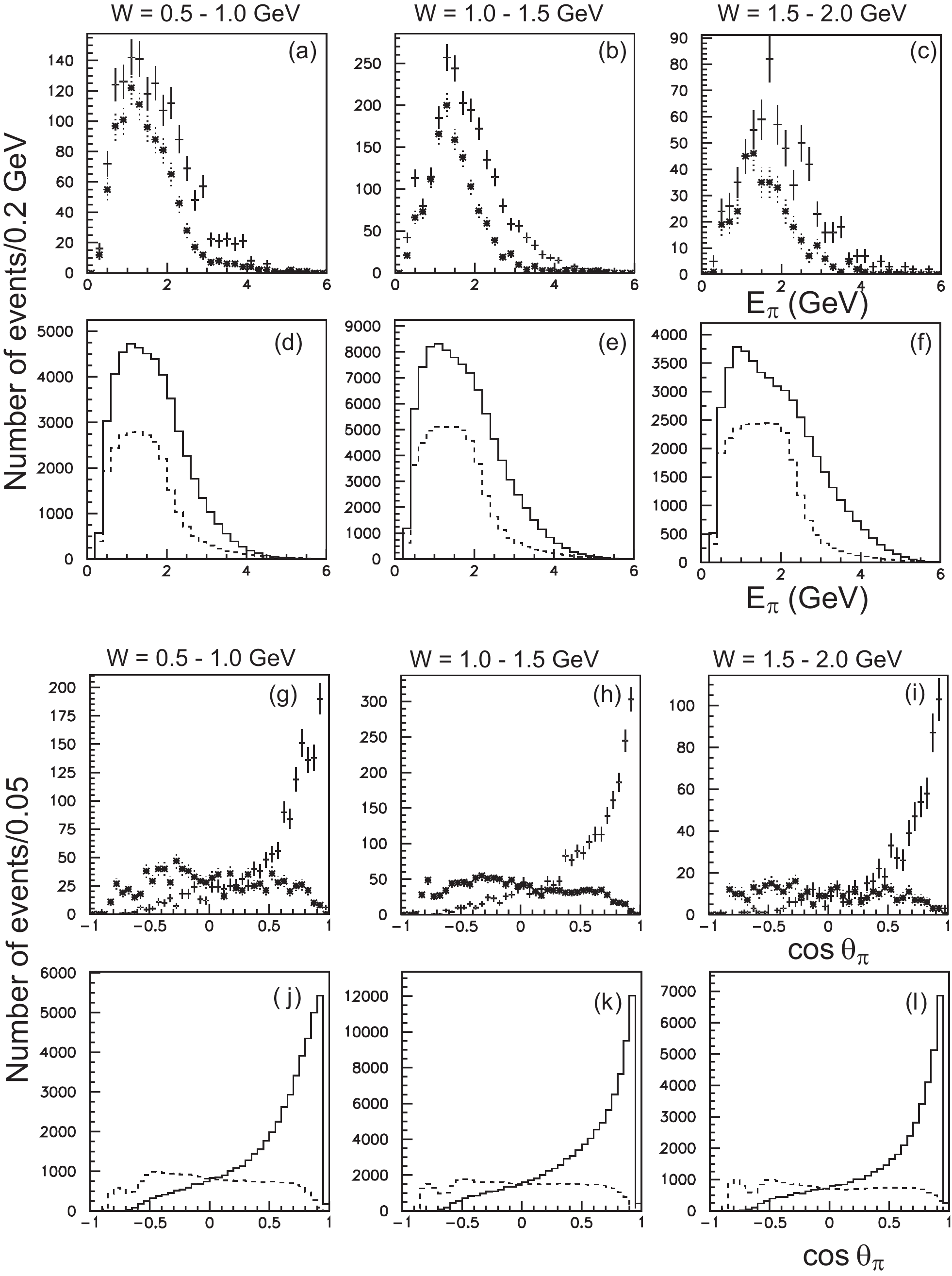}
\centering
\caption{(a,b,c) The distributions for the laboratory energy of
$\pi 1$ and $\pi 2$ for
the data samples in three different $W$ ranges indicated
above each panel.
(d,e,f) The corresponding 
distributions from the signal MC.  
(g,h,i) and (j,k,l) are the distributions for the laboratory angle 
of $\pi 1$ and $\pi 2$ for the data samples in the indicated $W$ ranges 
for the experimental samples and MC samples, respectively.
The legend and comments are the same as those in Fig.~\ref{fig:candq2}. 
}
\label{fig:epicost12}
\end{figure}

\section{Background estimation}
\label{sec:bkgest}
We consider several sources of possible background processes that could be
misidentified as the signal process. 

\subsection{Single pion production process}
\label{sub:snglpi}
Backgrounds from the single pion production, 
$e^+e^- \to e (e) \pi^0$, with one fake $\pi^0$ that is wrongly 
reconstructed, are estimated by MC simulation of the process.
From our previous measurement~\cite{pi0tff}, we know the cross section 
of this process with a sufficient accuracy for this purpose. 
From the MC study, the contamination is estimated to be less than one event 
in the whole sample of candidate events (about 3700 events) 
and is enhanced in the forward angular region in the c.m. frame, 
as shown in Fig.~\ref{fig:pi0bkgcost}(c). 
As the estimated number of events is small, we neglect this background source.
The $Q^2$ dependence of this background source is expected 
to be similar to that of the signal process.

\subsection{Radiative Bhabha process}
\label{sub:radbb}
The radiative Bhabha process with the virtual Compton scattering
topology, $e^+e^- \to e (e) \gamma$, which has a relatively large 
cross section, has been proved to give only a 
small contribution to the present measurement. 
This is verified in the $\pi 1$ and $\pi 2$ mass distributions
(Fig.~\ref{fig:mpi12}), where the pion
peaks in the experimental data with a high purity
compared to the signal-MC samples. 
This background must be less than 5\% according to the shape of 
the $m_{\pi 1,2}$ distribution. 
However, we must note that this background forms a broad enhancement near
the pion mass when the high-energy photon is converted
to an $e^+e^-$ pair in front of the ECL.
The measured $Q^2$ values for the virtual Compton scattering tend to populate
the region of higher values than in the signal process,
and the background should be 
relatively large in the high-$Q^2$ region.

The estimation of the contamination from this background process
using a background-MC sample is very difficult because of the large
MC statistics needed due to the large cross section of the process and  
the significant suppression in the event selection. 
Thus, we estimate the contamination
using distributions of the experimental signal candidates.

\begin{figure}
\centering
\includegraphics[width=8cm]{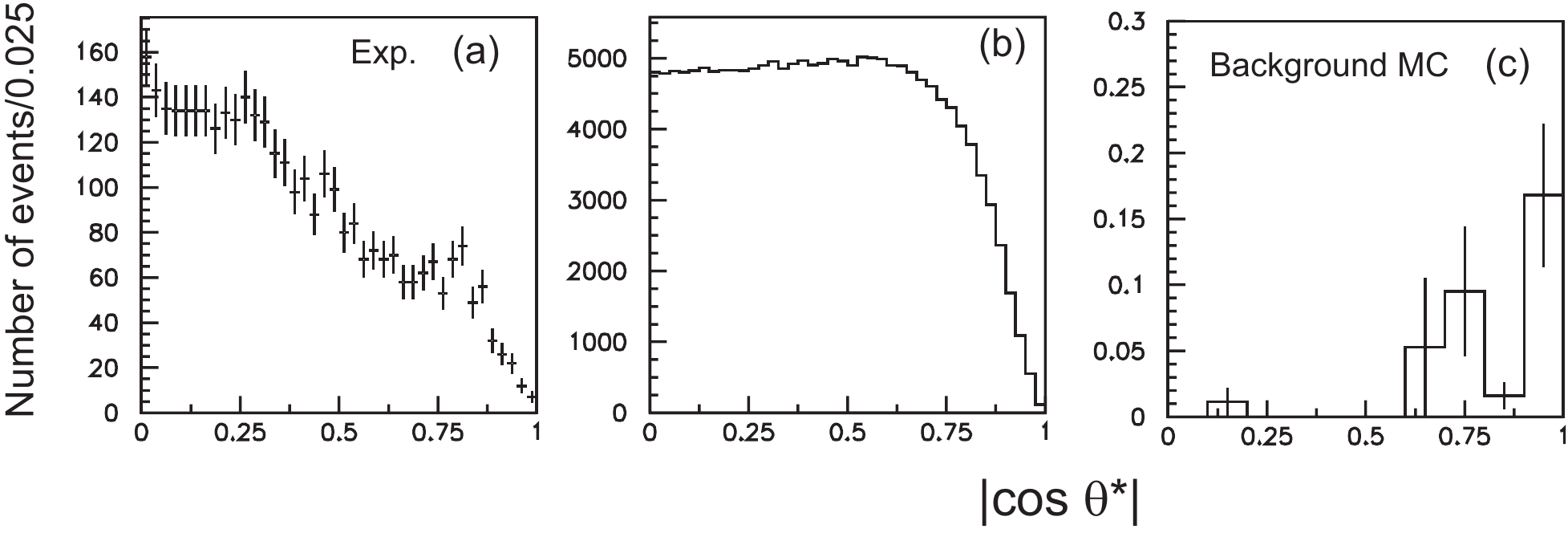}
\centering
\caption{The $|\cos \theta^*|$ distributions
for (a) the experimental signal candidates,
(b) the signal MC with the  isotropic
generation for $|\cos \theta^*|$, 
and (c) the background-MC events for the single-pion
production process, where the yield is scaled to the 
expected contamination in the set of all signal candidates.
The error bars are statistical from the MC samples
but are not proportional to square root of the number
of events because each event in the MC sample has 
a variable weight to reproduce the experimental $Q^2$ dependence. 
For (b), the normalization is arbitrary.
}
\label{fig:pi0bkgcost}
\end{figure}

\begin{figure}
\centering
\includegraphics[width=8cm]{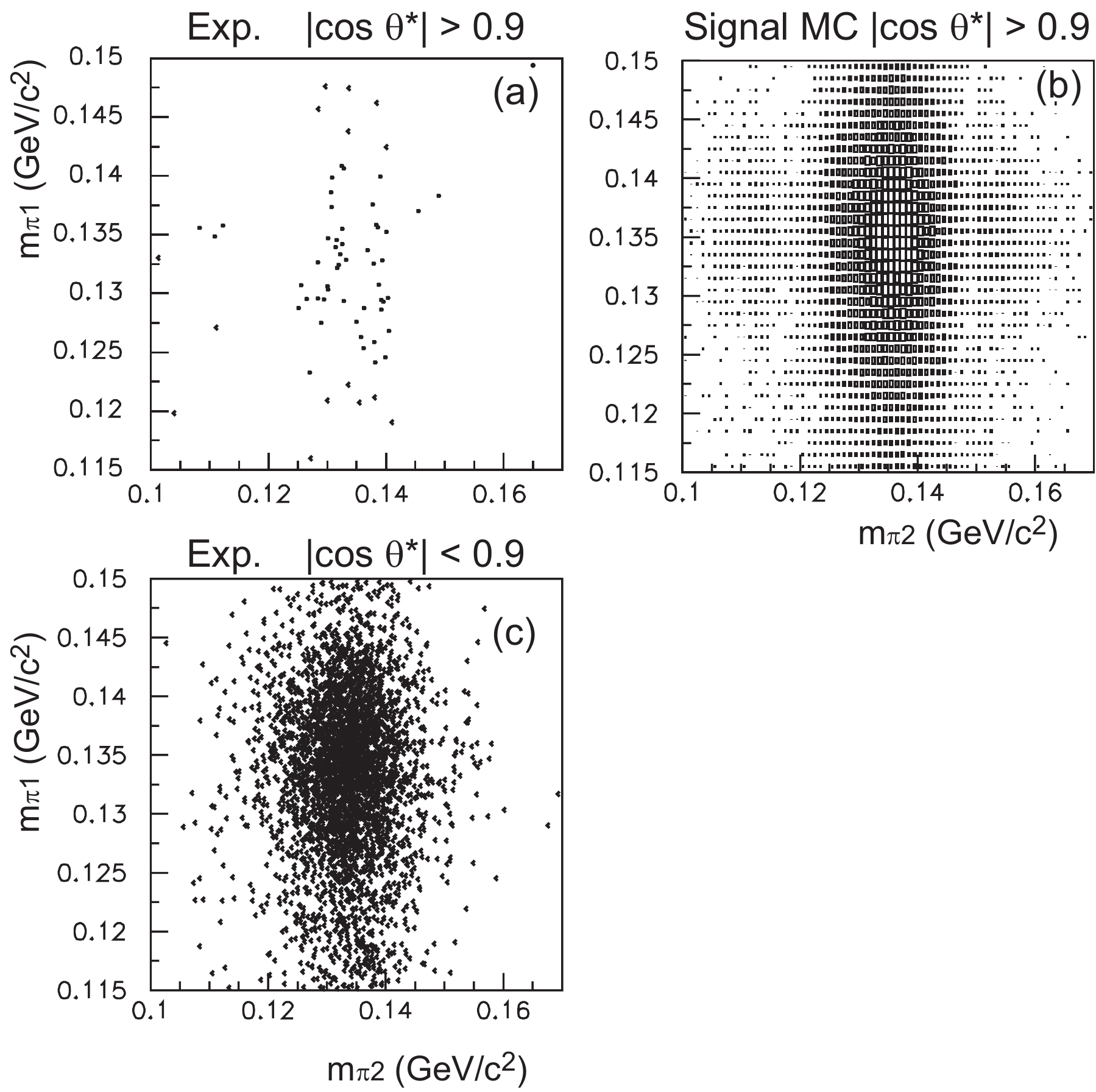}
\centering
\caption{Scatter plot for the 
$\gamma \gamma$ invariant masses for $\pi 1$
and $\pi 2$ for (a) the experimental
signal candidates for $|\cos \theta^*|>0.9$,
(b) the signal-MC samples for $|\cos \theta^*|>0.9$,
and (c) the experimental
signal candidates for $|\cos \theta^*|<0.9$.
}
\label{fig:massmass}
\end{figure}

In presence of both of the described background processes,
single pion production and the radiative Bhabha process,
at least one of the two $\pi^0$s reconstructed
is not a true $\pi^0$.
A fake $\pi^0$ is wrongly reconstructed from the
beam background or electronic noise. 
If such a background $\pi^0$ had large reconstructed energy,
the event would have been rejected by the $p_t$-balance
selection criterion.
Therefore, these fake pions remain in the signal event sample 
only if the noise has small reconstructed energy. 
In addition, such noise photons populate the lowest
energy region just above 
the energy thresholds for a photon
or a neutral pion for the selection.
Consequently, 
the c.m. scattering angle in the two-photon system tends to
be reconstructed in the forward regions 
because of the unbalanced laboratory energies between the two pions
and the Lorentz boost of the two-pion system along the beam axis,
just as expected in the single-$\pi^0$ production case.
However, as seen in Fig.~\ref{fig:pi0bkgcost}(a), 
no visible enhancement is seen even in the most
forward region $|\cos \theta^*| > 0.9$, where
the number of the signal events is small according
to the small signal acceptance (Fig.~\ref{fig:pi0bkgcost}(b))
in the angular distribution for the entire experimental signal
candidates.
Figure~\ref{fig:massmass}(a) shows the $\pi 1$ and $\pi 2$ mass distribution 
for the events with  $|\cos \theta^*|>0.9$ in comparison with the 
signal-MC expectation and the experimental data for $|\cos \theta^*|<0.9$
(Fig.~\ref{fig:massmass}(b) and (c), respectively). 
It is difficult to estimate the non-$\pi^0$ backgrounds 
quantitatively from these figures. 

\begin{figure}
\centering
\includegraphics[width=8cm]{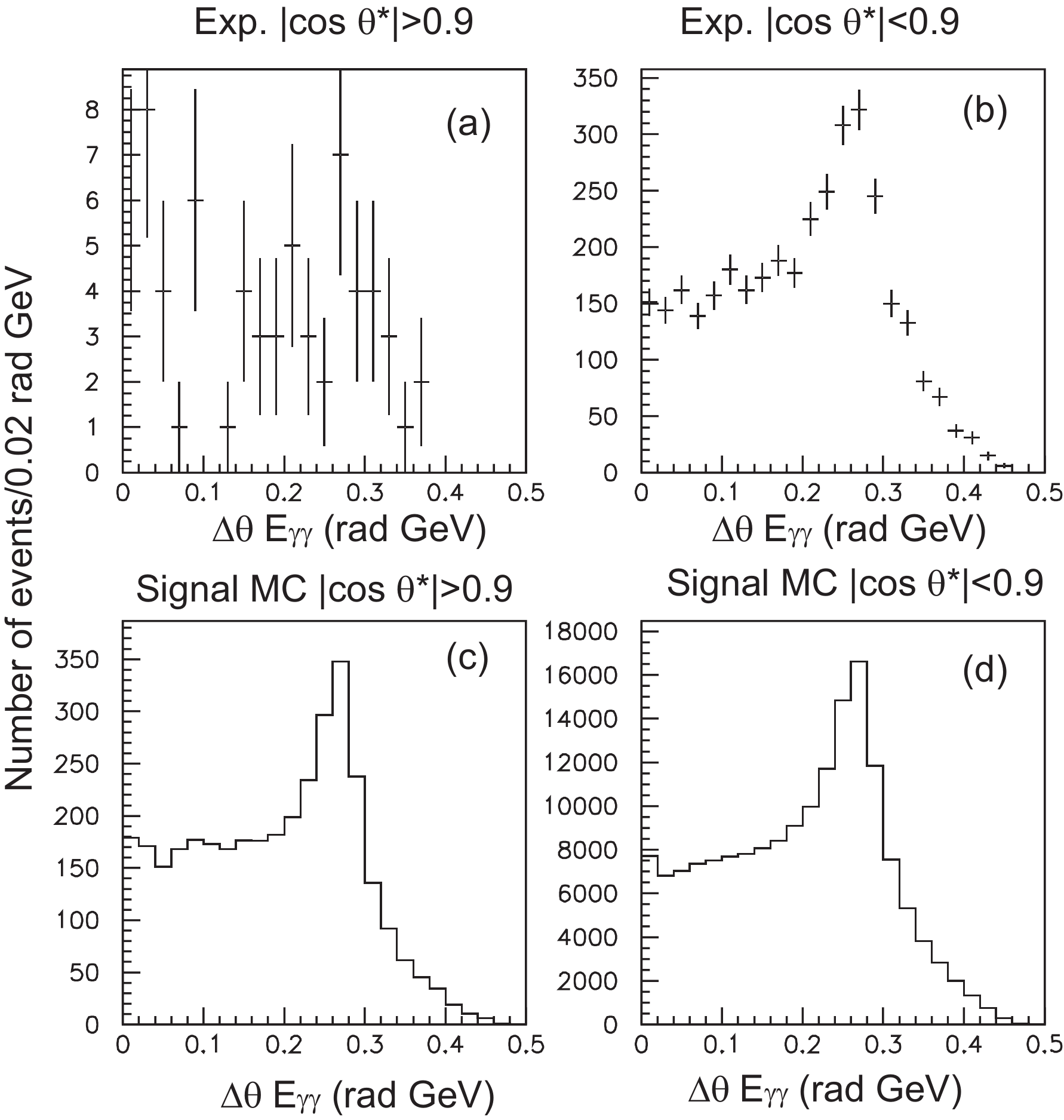}
\centering
\caption{The distributions of $\Delta \theta E_{\gamma \gamma}$
for (a) the experimental signal candidates in  $|\cos \theta^*|>0.9$
and  (b) in $|\cos \theta^*|<0.9$. Similar plots 
for (c) the signal-MC samples  in  $|\cos \theta^*|>0.9$
and  (d) $|\cos \theta^*|<0.9$.
The statistics of the MC figures are arbitrary.
}
\label{fig:poldiffe}
\end{figure}

In our previous study~\cite{pi0tff}, we found that the 
backgrounds from virtual Compton scattering 
have a high-energy neutral pion from misreconstruction of an
$e^+e^-$ conversion of the single photon, which simulates 
two photons with a high probability.
Its characteristic feature is a small polar angle difference 
$\Delta \theta$
due to the effect of the magnetic field. 
It is found experimentally that the backgrounds
concentrate strongly at  $\Delta \theta \, E_{\gamma\gamma}
<0.05$~rad$\cdot$GeV~\cite{pi0tff}.
Figure~\ref{fig:poldiffe}
shows the distribution of $\Delta \theta \, E_{\gamma\gamma}$
for $\pi 1$. 
We estimate the yield of the backgrounds 
from the virtual Compton process to be around 10 events,
which are seen as an enhancement in the region 
$\Delta \theta \, E_{\gamma\gamma}<0.05$~rad$\cdot$GeV, out of the 
67-event subsample with $|\cos \theta^*|>0.9$ and 0.5~GeV$< W < 2.1$~GeV. 
We thus estimate that the background fraction is about 15\% for
$|\cos \theta^*|>0.9$, although its uncertainty is large.
We do not find any sign of background from
this source in the region $|\cos \theta^*|<0.9$. 

\subsection{$\pi^0 \gamma$ production process}
\label{sub:pi0g}
Backgrounds from the $\pi^0 \gamma$ production, 
$e^+e^- \to e (e) \pi^0 \gamma$, are dominated by the
virtual pseudo-Compton scattering process of $\omega$-meson production,
$e^+e^- \to e(e) \omega$.
We know the cross section of this process
with sufficient 
accuracy from our previous measurement~\cite{pi0tff}
to estimate this contamination from the background-MC study,
as described below. 

\begin{figure}
\centering
\includegraphics[width=8cm]{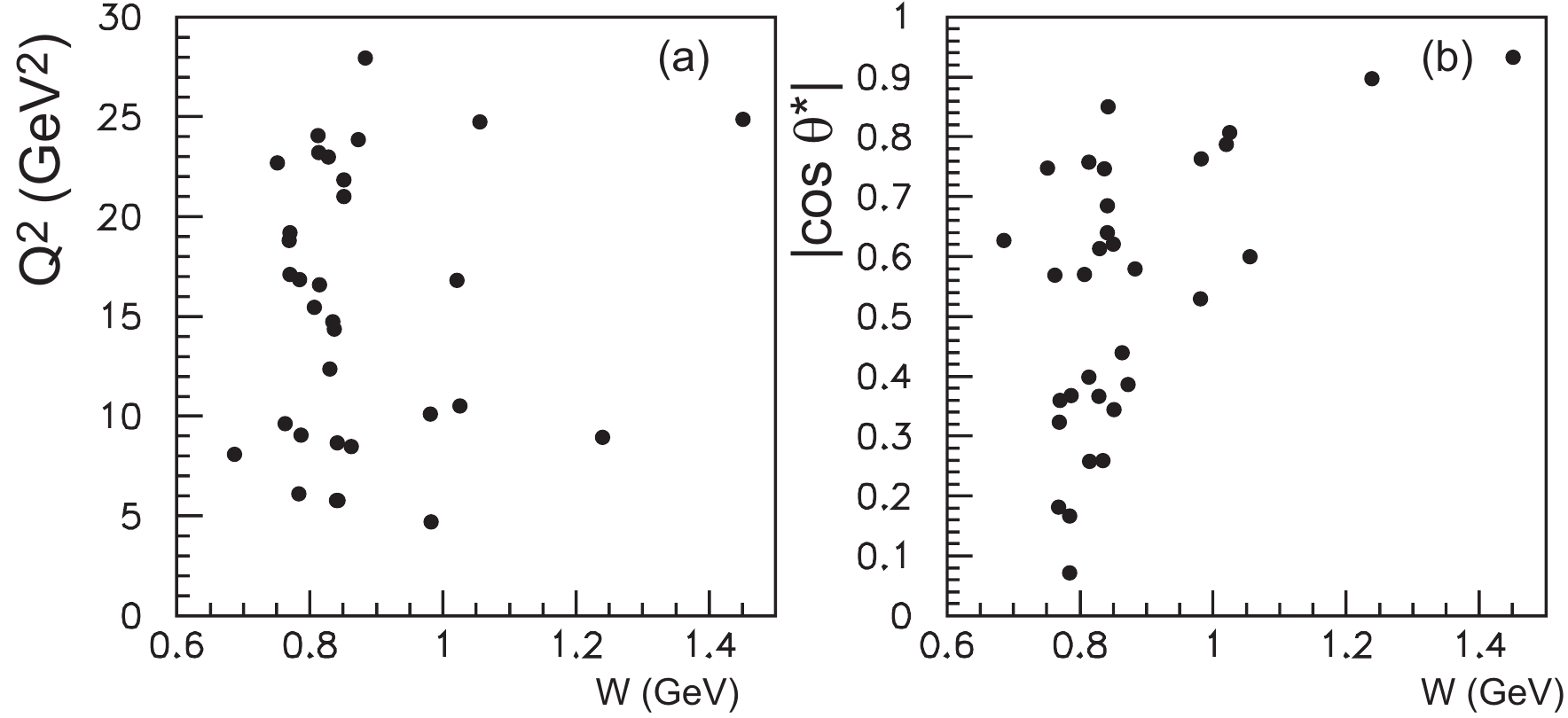}
\centering
\caption{The distribution of background events
estimated from the MC for the process 
$e^+e^- \to e (e) \omega$, $\omega \to \pi^0 \gamma$ in
the $Q^2$ versus $W$ and $|\cos \theta^*|$ versus $W$ scatter plots.
About three events in each plot correspond to a  
contamination of one event in the present signal-candidate sample.
}
\label{fig:gpi0bkg}
\end{figure}

The $W$, $Q^2$, and $|\cos \theta^*|$ dependences are based on the MC.
Figure~\ref{fig:gpi0bkg} shows the distributions of the 
background from the process $e^+e^- \to e (e) \omega$, 
$\omega \to \pi^0 \gamma$ in
the $Q^2$-$W$, and $|\cos \theta^*|$-$W$ directions. 
This background is enhanced in the $W$ region near the $\omega$ mass 
($W=0.7$ -- 0.9~GeV) in the high-$Q^2$ region 
($Q^2 > 12~\GeV^2$).
The total number of events contaminating the
signal process is estimated to be about 9 events,
of which only 6 events for $Q^2 > 
12$~GeV$^2$ are expected to have
a non-negligible effect compared to the signal yield.
The $|\cos \theta^*|$ distribution
is approximately flat so the behavior of this
background is very different from the former background
sources where both photons constituting 
one pion originated from noise.
We find no visible enhancement in the 0.7 -- 0.9~GeV region 
in the observed $W$ distribution, as expected from the small 
contamination of the background process.

\subsection{Three-$\pi^0$ production process}
\label{sub:3pi0}
The three-pion production process, 
$\gamma^* \gamma \to \pi^0 \pi^0 \pi^0$, 
is wrongly selected as the signal candidate
if one non-energetic pion escapes detection. 
As three-pion production in single-tag two-photon collisions
has not been measured to date, we estimate the contamination
by referring to the corresponding zero-tag measurement.

The $\pi^+ \pi^- \pi^0$ production in the zero-tag process has been measured 
by the L3 collaboration~\cite{l33pi}
and shows that the process is dominated
by the $a_2(1320)$ and a structure near 1.7~GeV ($a_2(1700)$ 
and potential $\pi_2(1670)$ production
decaying to $\rho^0 \pi^0$ or  $f_2(1270) \pi^0$).
However, the three-neutral-pion production is strongly
suppressed because  $\rho^0$ does not decay to $\pi^0 \pi^0$
and $a_2(1700) \to f_2(1270) \pi^0$ 
has not been observed definitively.
The L3 measurement only
provides an upper limit for $\pi_2(1670)$ production
from the two-photon process.
A measurement of a finite value for $\gamma \gamma \to
\pi_2(1670) \to \pi^0\pi^0\pi^0$ production is reported by
the Crystal Ball experiment~\cite{cballex} but it is not consistent
with the L3 upper limit.

We estimate that $\pi^0 \pi^0 \pi^0$
production is about 4\% of $\pi^0 \pi^0$ production in the zero-tag
two-photon process~\cite{pi0pi0, pi0pi02}
in the $e^+e^-$-based cross section,
according to the measurement of L3 for $\gamma \gamma
\to a_2(1700) \to f_2(1270) \pi^0$~\cite{l33pi}.
We confirm this estimate by our count of 
zero-tag  $\gamma \gamma \to \pi^0 \pi^0 \pi^0$ candidate
events, requiring three neutral pions in an event, from the Belle 
data samples.
Our estimate of the cross section is somewhat smaller than that from
the above Crystal Ball measurement~\cite{cballex}, but is
consistent with it within about a factor of two.

We assume that the cross section ratio of the two 
processes,  $\pi^0 \pi^0 \pi^0$ to $\pi^0 \pi^0$,
is the same between the zero-tag and single-tag processes.
Taking into account the selection efficiency 
for the background events and the background subtraction
using the $p_t$-balance distribution (applied in Sec.~\ref{sub:bkgsub}), 
the estimated contamination is less than 0.5\% of the signal
yield, where the $\pi^0\pi^0\pi^0$ background does not peak near 
$|\Sigma \vec{p}_t^*|=0$, like for the other non-exclusive
processes described in the following section.

The contribution from the process 
$\gamma^* \gamma \to \eta^\prime \to \pi^0\pi^0\pi^0$ is separately
estimated using the $\eta^\prime$-TFF~\cite{babar2}, the branching
fraction, and generated background-MC samples for the process.
We find that the expected contribution for $W > 0.5$~GeV is 
about 0.3~events. Thus, we conclude that the
contribution of this background is negligibly small.  

\subsection{Other non-exclusive processes}
\label{sub:other}
The other non-exclusive background processes,
$e^+e^- \to e (e) \pi^0 \pi^0 X$, where $X$ denotes multiple hadrons, 
are in general subdivided into two-photon ($C$-even) and
virtual pseudo-Compton (bremsstrahlung, $C$-odd) processes,
but they interfere with each other if the same $X$
is allowed for both processes.
The majority of such background events populate the
small $E_{\rm ratio}$ region, {\it e.g.}, less than 0.7.  
This feature is distinct from the aforementioned background processes 
that can populate the region near $E_{\rm ratio} = 1$. 
These backgrounds also do not peak near $|\Sigma \vec{p}_t^*| = 0$.

From Fig.~\ref{fig:seldist}, we expect that the low-$E_{\rm ratio}$ 
component could leak into the signal region. 
We estimate the relative ratio of the non-exclusive
backgrounds to the signal yield by counting
the number of events in the subregion
of the signal region ($0.85 < E_{\rm ratio} < 0.925 
\cap  0.1~\GeV/c < |\Sigma \vec{p}_t^*| < 0.2~\GeV/c$)
where the background component would be relatively large.
We assume that the backgrounds are distributed as
a linear function in $0 < |\Sigma \vec{p}_t^*| < 0.2~\GeV/c$
and in $0.85 < E_{\rm ratio} < 1$,  vanishing at  $|\Sigma \vec{p}_t^*| = 0$ 
and $E_{\rm ratio} = 1$, respectively, factorized for the two directions. 
The fraction of the backgrounds
falling in the above subregion is calculated to be 9/16 of 
that in the entire signal region, according to the distribution.
We also estimate the fraction of signal events coming into 
the same region using 
the signal-MC sample.
We thus determine the expected number of the
background events from this information.

In addition, we use the $W$ and $Q^2$ dependence 
of the $E_{\rm ratio}$ sideband events, which 
are extracted  from the experimental data in 
$0.7  < E_{\rm ratio} < 0.8$ as ancillary information
for the kinematical regions with low statistics.
The normalization of the background is determined
in the first method in a high-statistics region; 
we extrapolate it to different $W$ and $Q^2$ regions
with lower statistics assuming the dependence
observed in the $E_{\rm ratio}$ sideband region.

\begin{figure}
\centering
\includegraphics[width=7cm]{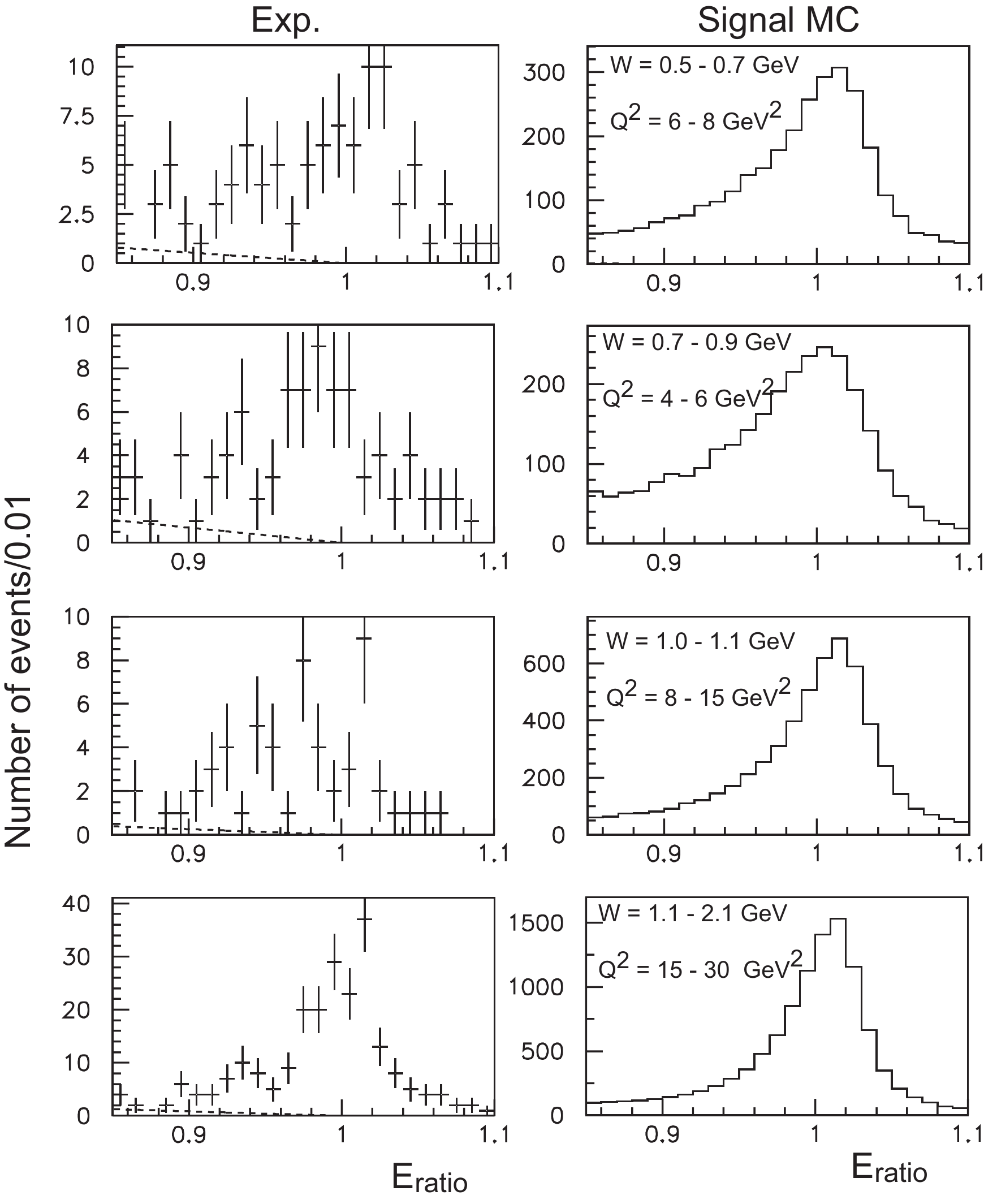}
\centering
\caption{Comparison between the experimental
(left) and signal-MC (right) distributions for 
$E_{\rm ratio}$ of events in the signal regions.
Each row corresponds to the
same $W$ and $Q^2$ regions indicated in the right panel. 
The dashed lines show the
estimated non-exclusive background, which is assumed
to distribute linearly with the horizontal variable.
Statistics of the MC figures are arbitrary.
}
\label{fig:selerdist}
\end{figure}

\begin{figure}
\centering
\includegraphics[width=7cm]{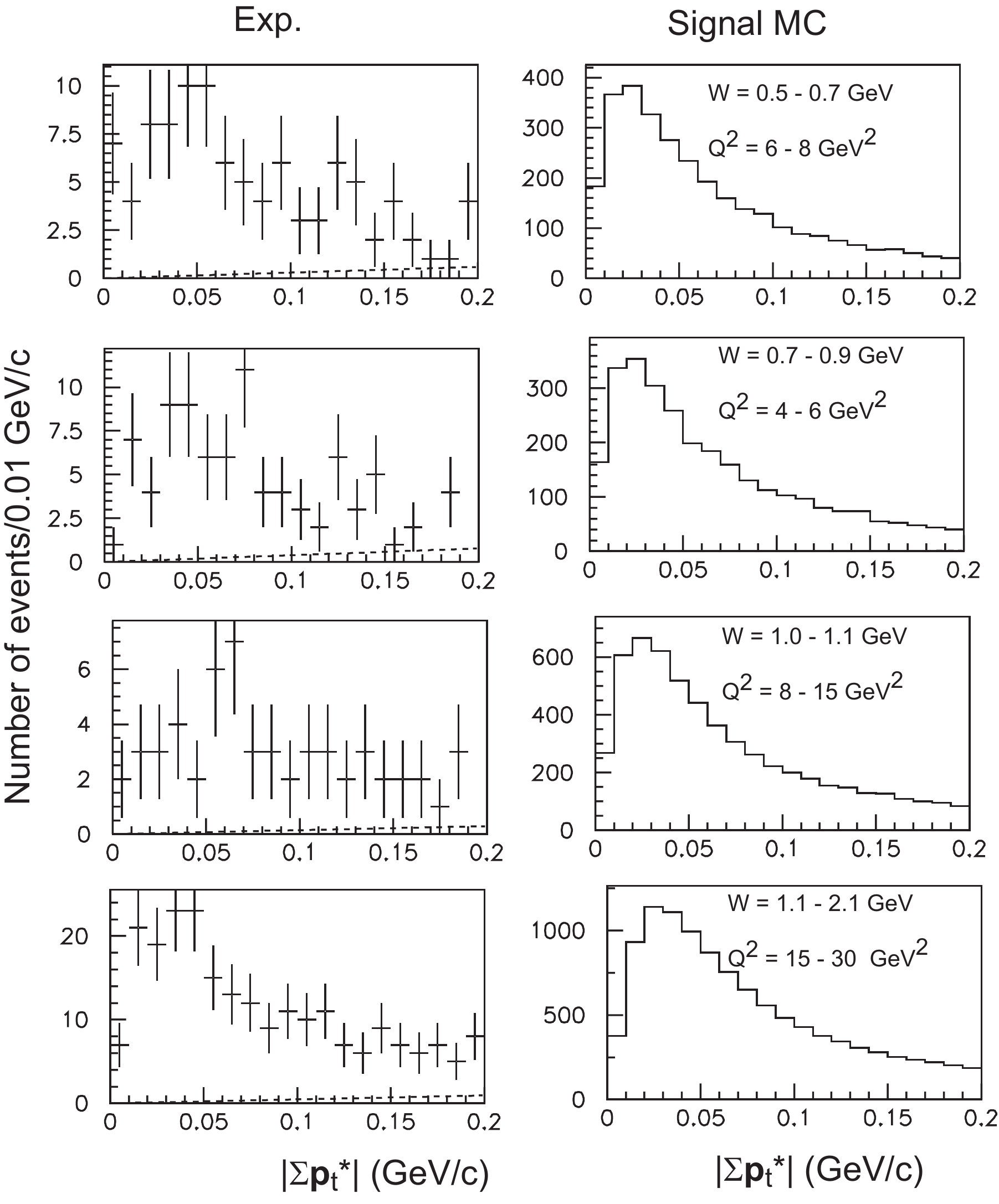}
\centering
\caption{Comparison between the experimental
(left) and signal-MC (right) distributions for 
the $p_t$ balance of events in the signal regions.
Each row corresponds to the
same $W$ and $Q^2$ regions indicated in the 
right panel. 
The dashed lines show the
estimated non-exclusive background, which is assumed
to distribute linearly with the horizontal variable.
Statistics of the MC figures are arbitrary.
}
\label{fig:selptdist2}
\end{figure}

We show the  $E_{\rm ratio}$ and $|\Sigma \vec{p}_t^*|$
distributions in the signal region of the experimental and signal-MC
samples in Figs.~\ref{fig:selerdist} and \ref{fig:selptdist2}, respectively. 
The dashed lines in the panels for the experimental data show the
estimated non-exclusive background.

The details of the background subtraction are
described in Sec.~\ref{sub:bkgsub}.

\section{Derivation of the differential cross section}
\label{sec:dcs}
We first define and evaluate the $e^+e^-$-incident-based
cross section separately for the p-tag and e-tag samples.
Then we derive the differential cross section of the process
$\gamma^* \gamma \to \pi^0 \pi^0$.

The $e^+e^-$-incident-based differential 
cross section is written as
\begin{eqnarray}
&\large( \frac{d^3 \sigma_{ee}}{dWd|\cos \theta^*| dQ^2} \large)_{x{\rm -tag}} 
= \nonumber \\
&\frac{ Y_{x{\rm -tag}}(W, |\cos \theta^*|, Q^2)}
{\varepsilon^\prime_{x{\rm -tag}}(W, |\cos \theta^*|, Q^2) 
\Delta W \Delta |\cos \theta^*| \Delta Q^2 \int {\cal L}dt {\cal B}^2}, 
\end{eqnarray}
where the yield $Y$ and the uncorrected efficiency
obtained by the signal MC $\varepsilon^\prime$ are separately evaluated 
for p-tag and e-tag, for a consistency check.
The measurement ranges of $W$, $|\cos \theta^*|$, and $Q^2$ and the 
corresponding bin widths $\Delta W$, $\Delta |\cos \theta^*|$, and 
$\Delta Q^2$ are summarized in Table~\ref{tab:bins}, where the differential 
cross section was first calculated with $\Delta W = 0.05$~GeV and 
$\Delta |\cos \theta^*| = 0.1$, and then two adjacent $W$ and 
$|\cos \theta^*|$ bins are combined (for $W$ only above $W>1.1$~GeV) with 
an arithmetic mean. 
Here, $\int {\cal L}dt$ is the
integrated luminosity of 759~fb$^{-1}$ and
${\cal B}^2 = 0.9766$ is the square
of the decay branching fraction ${\cal B}(\pi^0 \to \gamma \gamma)$.

We take into account the difference of the beam energies in evaluating 
$\varepsilon^\prime$.
Since the efficiency and the luminosity function depend on the beam 
energy, we construct the corrected efficiency using the ratio of the 
products of the efficiency, luminosity function, and the integrated luminosity
for two cases: that the products are combined for
the different beam energies in the experiment and that  
all the experiment would be done at the sole $\Upsilon(4S)$ energy 
with the same total integrated
luminosity as in the experiment.
Thus, the $e^+e^-$-based cross section measured 
for the energy of $\Upsilon(4S)$, 10.58~GeV,  is obtained.  

After confirming the consistency between the p- and e-tag measurements 
to ensure validity of the efficiency corrections (described in Sec.~\ref{sub:efplot}), 
we combine their yields and efficiencies using the formula which builds in the equality 
of the efficiency-corrected yields for both measurements, 

\begin{eqnarray}
&\frac{d^3 \sigma_{ee}}{dWd|\cos \theta^*| dQ^2}  
= \nonumber \\
&\frac{Y(W, |\cos \theta^*|, Q^2)(1-b(W, |\cos \theta^*|, Q^2))}
{\varepsilon^\prime(W, |\cos \theta^*|, Q^2) \Delta W \Delta |\cos \theta^*| 
\Delta Q^2 \int {\cal L}dt}, 
\label{eqn:cscomb}
\end{eqnarray}
where $Y = Y_{{\rm p-tag}}+ Y_{{\rm e-tag}}$,  
$\varepsilon^\prime = (\varepsilon^\prime_{{\rm p-tag}}+ 
\varepsilon^\prime_{{\rm e-tag}})/2$, 
and $b$ is the background fraction combined for p- and e-tags,
which is subtracted here. 
For the region $Q^2 = 3$ -- 5 GeV$^2$, we do not use the p-tag
data because its statistical accuracy is much worse than for the e-tag sample. 
There, as a result, the cross-section value in the e-tag measurement 
is simply doubled.

Finally, the $e^+e^-$-incident-based differential cross section
is converted to that based on $\gamma^* \gamma$-incident by dividing
by the single-tag two-photon luminosity function 
$d^2 L_{\gamma^* \gamma}/dW dQ^2$, which is a function of $W$ and $Q^2$. 
We use the relation
\begin{eqnarray}
&\frac{d \sigma_{\gamma^* \gamma}}{d|\cos \theta^*|}  = \nonumber \\
&\frac{d^3 \sigma_{ee}}{dWd|\cos \theta^*| dQ^2}  
\frac{f}{2 \frac{d^2 L_{\gamma^* \gamma}}{dW dQ^2} 
(1+\delta)(\varepsilon/\varepsilon^\prime)\varepsilon^\prime}. 
\label{eqn:csgg}
\end{eqnarray}
The factors $\delta$, $\varepsilon$, and $f$
correspond to the radiative correction, efficiency corrected 
for the $\varphi^*$ dependence of the differential cross section,
and the unfolding effect that accounts for  migrations between
the different $Q^2$ bins, respectively, which are explained in more
detail in the subsections below.

\begin{center}
\begin{table}
\caption{The measurement range and bin widths for three-dimensional
variables $(W, |\cos \theta^*|, Q^2)$.}
\begin{tabular}{c|ccc|c}
\hline \hline
Variable & Measurement & Bin width & Unit & Number \\
         & range   & & & of bins\\
\hline \hline
$W$ & 0.5 -- 1.1 & 0.05 & GeV & 12\\
    & 1.1 -- 2.1 & 0.1 &   & 10\\
\hline
$|\cos \theta^*|$ & 0.0 -- 1.0 & 0.2 & & 5\\
\hline
$Q^2$ & 3.0 -- 5.0 (e-tag only) & 1.0 & GeV$^2$ & 2\\
 & 5.0 -- 6.0 & 1.0 & & 1\\
 & 6.0 -- 12.0 & 2.0 & & 3\\
 & 12.0 -- 15.0 & 3.0 & & 1\\
 & 15.0 -- 20.0 & 5.0 & & 1\\
 & 20.0 -- 30.0 & 10.0 & & 1\\
\hline \hline
\end{tabular}
\label{tab:bins}
\end{table}
\end{center}

\subsection{Efficiency plots and consistency check of the 
p-tag and e-tag measurements}
\label{sub:efplot}
\begin{figure}
\centering
\includegraphics[width=7.7cm]{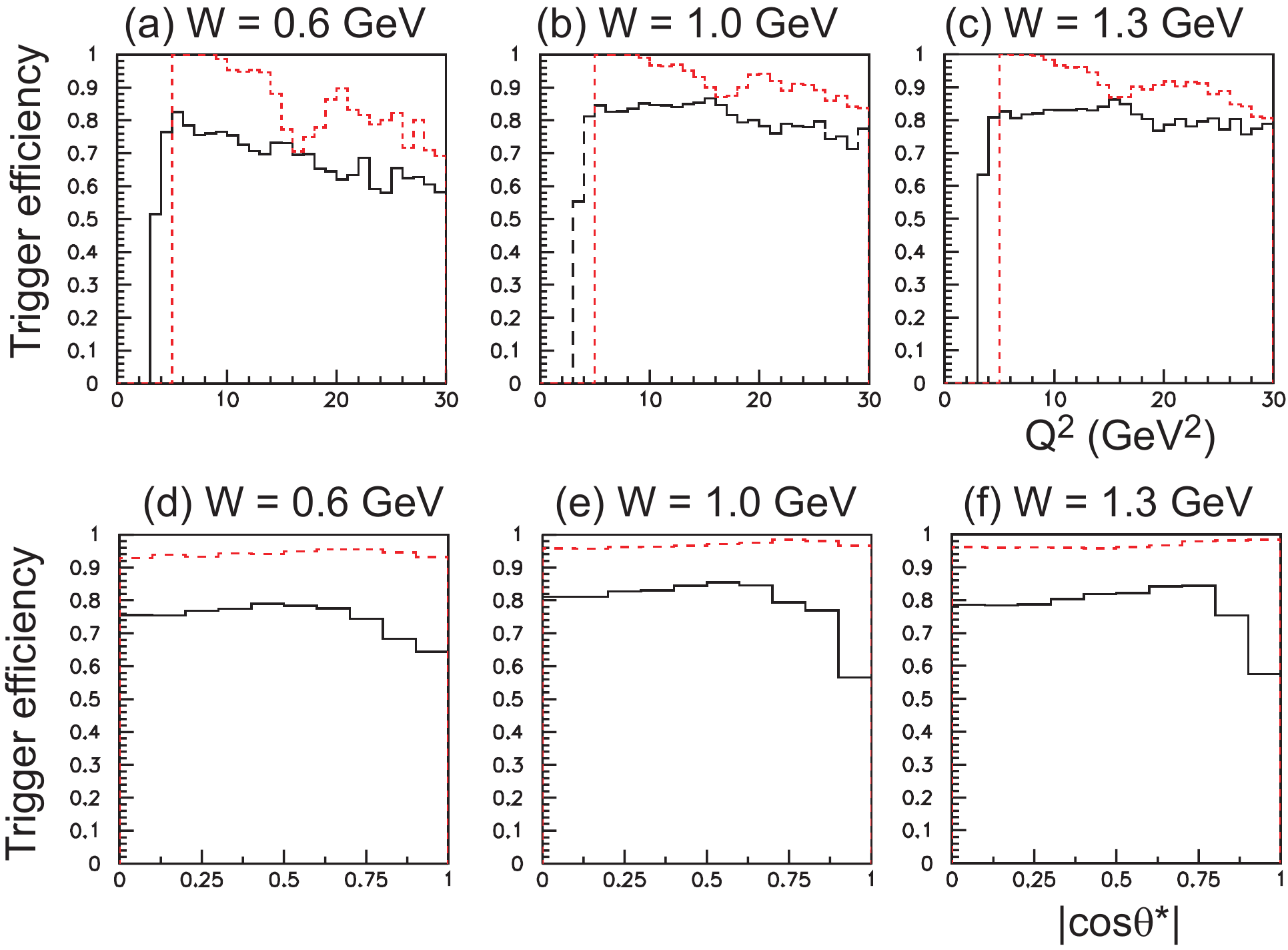}
\centering
\caption{Trigger efficiency estimated by the signal-MC samples
and trigger simulator. The solid (black) and dashed (red)
histograms are for e-tag and p-tag events, respectively.
The results are for the three $W$ points indicated above each panel:
(a,b,c) the $Q^2$ dependence for the isotropically
generated $\pi^0$ pairs in the $\gamma^* \gamma$ c.m. system;
(d,e,f) the scattering- (polar-) angle dependence of $\pi^0$
in the $\gamma^* \gamma$ c.m. system. The $Q^2$ range
3~GeV$^2$ (5~GeV$^2$) $< Q^2 < 30$~GeV$^2$ is
integrated for the e-tag (p-tag) plot.   
}
\label{fig:trgeff}
\end{figure}

Figure~\ref{fig:trgeff} shows the trigger efficiencies
obtained from the signal-MC samples and trigger simulator.
The trigger efficiency is defined for events within
the selection criteria. 
The $W$ dependence of the trigger  
efficiency is mild in the measurement region. 
The dip-bump structure seen in the $Q^2$ dependence for the p-tag 
efficiency is an artifact of the Bhabha-veto logic
in the HiE trigger.

\begin{figure*}
\centering
\includegraphics[width=15cm]{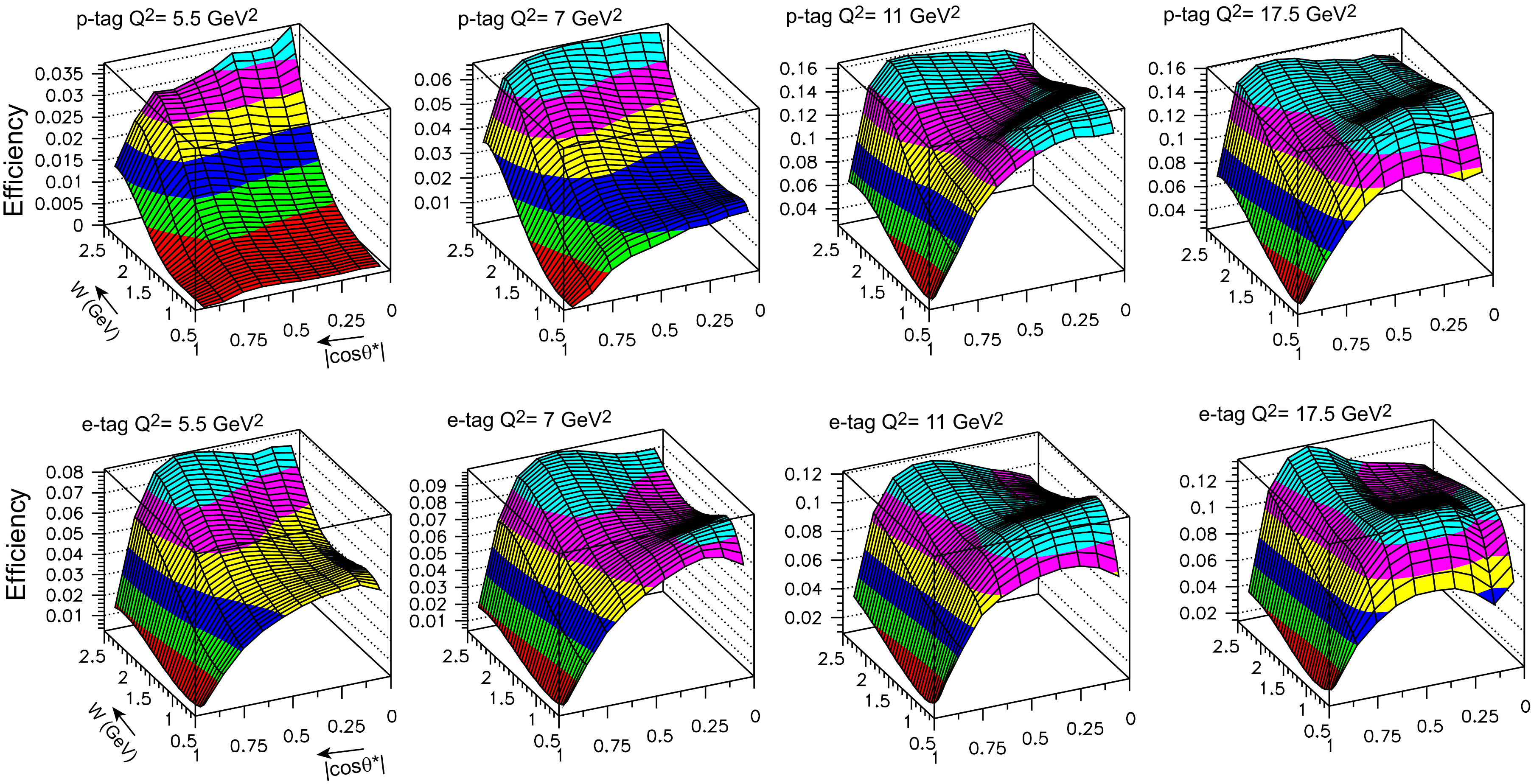}
\centering
\caption{
Efficiency estimated from the signal-MC samples passed through
all the selection and trigger conditions. 
Each panel provides 
the $W$ and $|\cos \theta^*|$ dependences of
the efficiency  for the four selected $Q^2$ bins for each of 
the p- or e-tags.
The contour levels shown by different colors are not 
common across the panels.
 }
\label{fig:effsurfs}
\end{figure*}

Figure~\ref{fig:effsurfs} shows the
efficiencies in which all the selection and trigger conditions
are taken into account.
They are provided as a function of $W$ and $|\cos \theta^*|$ for
the selected $Q^2$ bins of the p- or e-tag samples. 
These efficiencies are obtained
from the signal-MC events, which are generated assuming an
isotropic angular distribution of the pions
in the $\gamma^* \gamma$ c.m. frame. 
Efficiency corrections for the $\varphi^*$ dependence 
are not taken into account in these figures.

\begin{figure*}
\centering
\includegraphics[width=12cm]{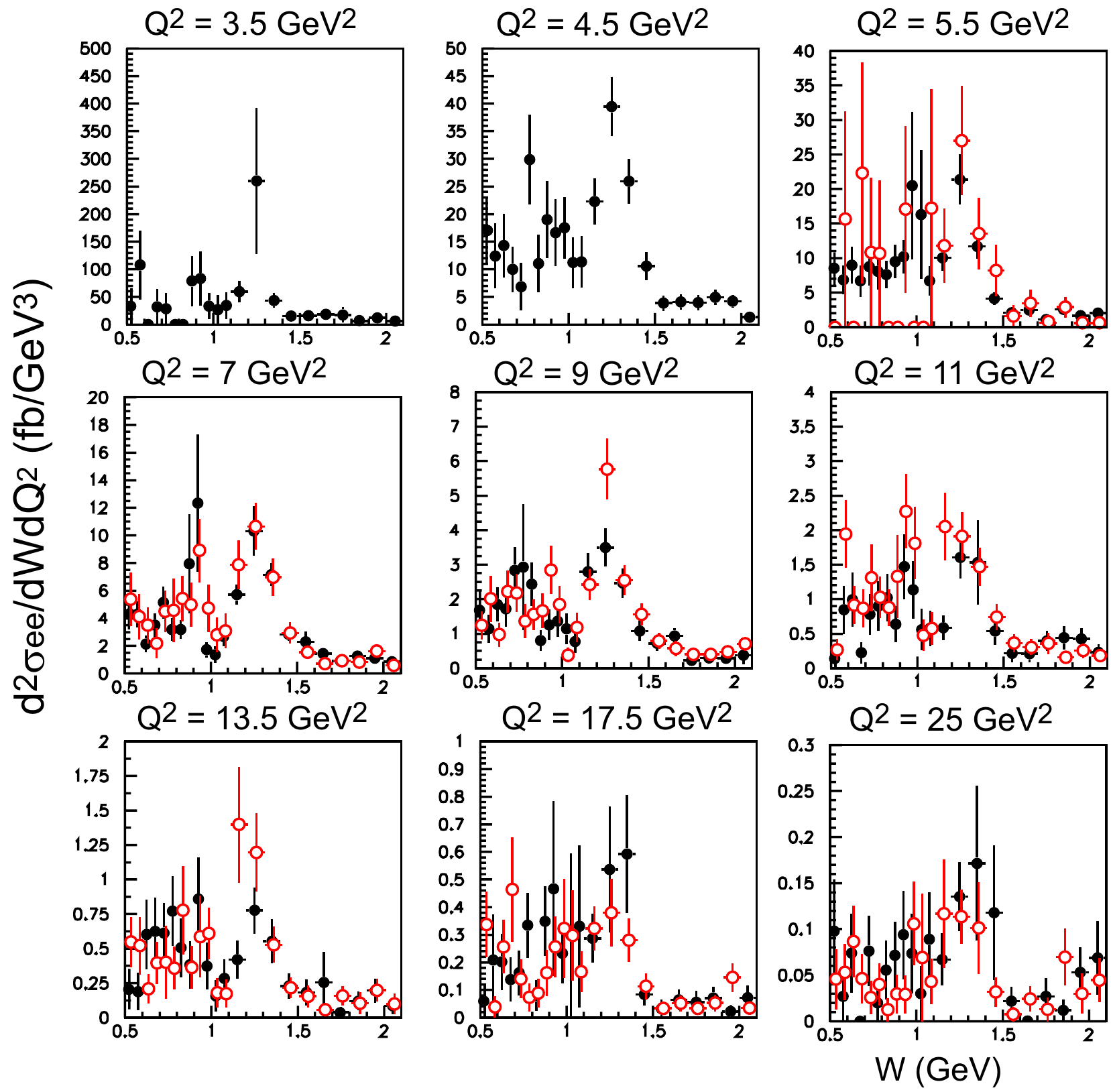}
\centering
\caption{The $W$ dependence of the $e^+e^-$-based cross section
in each $Q^2$ bin. The full $|\cos \theta^*|$ range (0 to 1) is
integrated. 
The closed circles (open circles) are for the 
e-tag (p-tag) measurements. The plotting location for each p-tag point
is shifted slightly to improve the visibility of error bars.
The efficiency corrections and background subtraction are applied.
}
\label{fig:tagtag1}
\end{figure*}

\begin{figure*}
\centering
\includegraphics[width=10cm]{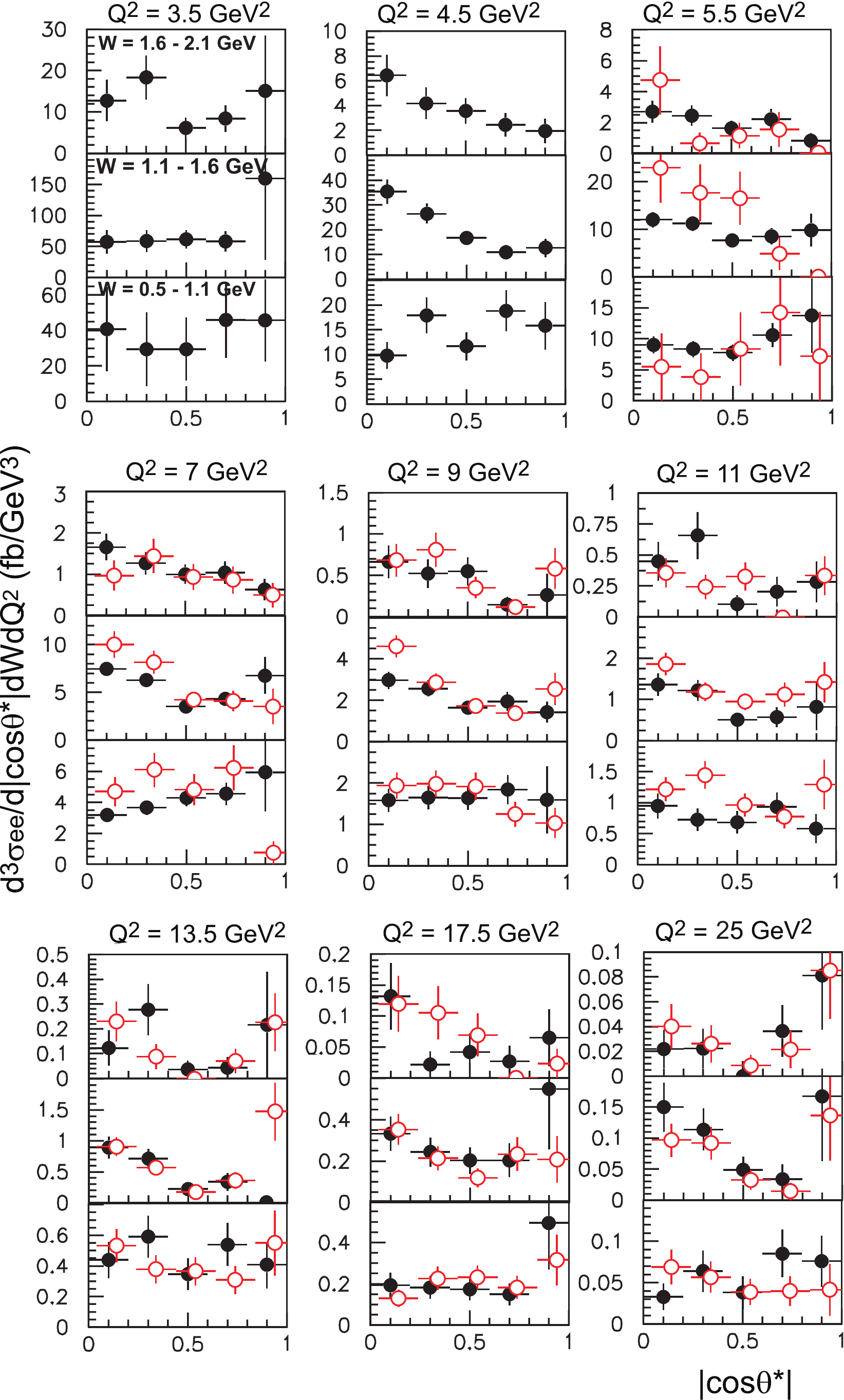}
\centering
\caption{The $|\cos \theta^*|$ dependence of the $e^+e^-$-based 
cross section in each $Q^2$ bin. 
The dependence for three
different $W$ regions is shown in above each panel, and
the $W$ regions are 0.5 -- 1.1~GeV, 1.1 -- 1.6~GeV, 1.6 -- 2.1~GeV
from bottom to top within each panel, 
as indicated in the upper left panel.
The closed circles (open circles) are for the 
e-tag (p-tag) measurements. 
The plotting location for each p-tag point
is shifted slightly to improve the visibility of the error bars.
The efficiency corrections and background subtraction are applied.
}
\label{fig:tagtag2}
\end{figure*}

Our accelerator and detector systems are asymmetric between
the positron and electron incident directions and energies, and separate
measurements of the p-tag and e-tag samples provide a good
validation check for various systematic effects of
the trigger, detector acceptance, and selection conditions.
Figures~\ref{fig:tagtag1} and \ref{fig:tagtag2} compare
the $e^+e^-$-based cross section measured separately
for the p- and e-tags.
They are expected to show the same cross section 
according to the $C$ symmetry if there is no systematic bias.
The results from the two tag conditions are consistent within 
statistical errors.
 
\subsection{Background subtraction}
\label{sub:bkgsub}
We have estimated the background yields in the signal samples
using the background-process-MC samples for the single pion production,
$e^+e^- \to e (e) \pi^0$, and the $\pi^0 \gamma$ production
dominated by $e^+e^- \to e (e) \omega$, $\omega \to \pi^0 \gamma$.
These are described in Ref.~\cite{pi0tff}, where their
yields are normalized based on the observations.

As mentioned in Sec.~\ref{sec:bkgest}, the estimated total yields 
for the sum of p- and e-tags are less than one  and nine events for the 
single pion and the $\pi^0 \gamma$ production backgrounds, respectively.  

The single-pion background is small, and we do not subtract its contribution
but rather include its effect in the systematic uncertainty due to
subtraction of the virtual Compton scattering.

For the $\pi^0 \gamma$ production, the background is estimated
to be 2\%, 6\%, and 13\% of the signal candidate yields for
the $Q^2$ bins of 12 -- 15~GeV$^2$, 15 -- 20~GeV$^2$, and
20 -- 30~GeV$^2$, respectively, independently of $W$ and 
$|\cos \theta^*|$, and is subtracted.
For the bins at $Q^2 \le 12~\GeV^2$, we neglect this background source.

The details of the estimation of the background contributions
from the virtual Compton process are discussed in Sec.~\ref{sub:radbb}. 
We estimate that the signal-candidate yield in
each $|\cos \theta^*|>0.9$ bin contains $(15 \pm 15)\%$ 
of the background from this process, and we derive the differential
cross section for the  $|\cos \theta^*|>0.8$ bin. 
We estimate  systematic uncertainties conservatively because
possible $Q^2$ dependence and the background
from the single-pion production process with similar properties
are neglected here,
and the contamination could be sensitive to the noise conditions. 
The error size is estimated from
$\Delta \theta \, E_{\gamma\gamma}$ and $|\cos \theta^*|$
distributions for several different conditions.
The uncertainties of the background estimation 
for different kinematical regions
are discussed in Sec.~\ref{sub:sysunc}.

We neglect the contribution of the $\pi^0 \pi^0 \pi^0$
production process because it is estimated
to be less than 1\% of the signal process (Sec.~\ref{sub:3pi0}). 

The contamination of the other non-exclusive
background processes is estimated using
the $E_{\rm ratio}$ and $|\Sigma \vec{p}_t^*|$ variables as described in
Sec.~\ref{sub:other}.
We estimate the fraction of the background in each of the
($W$, $Q^2$) regions to be between 3\% and 12\%. 
This background fraction has no
prominent $|\cos \theta^*|$ dependence and we neglect it.

In subtraction of the background yields, we multiply the observed yield
by the expected ratio of the background to the observed yields in each bin,
as represented by the $(1-b)$ factor in Eq.~(\ref{eqn:cscomb}).

\subsection{Efficiency corrections}
\label{sub:effcor}
The efficiency after the integration over $\varphi^*$ 
is calculated using the signal MC, assuming a flat 
$\varphi^*$ distribution. 
We correct the efficiency according to the actual non-uniformity
in $\varphi^*$ observed in the data. 
This correction is necessary
in case both the efficiency ($\varepsilon$) and 
the differential cross section ($d\sigma/d\varphi^*$) 
have a $\varphi^*$ dependence.

We partition the kinematical region of the measurement into 
three-dimensional ($W$, $|\cos \theta^*|$, $Q^2$) rectangular-prism cells 
defined by 0.5~GeV $<W<$ 1.1~GeV, 1.1~GeV $<W<$ 1.6~GeV, and  1.6~GeV $<W<$ 2.1~GeV; 
five equal-width bins in $|\cos \theta^*|$;
3~GeV$^2$ $<Q^2<$ 8~GeV$^2$, 8~GeV$^2$ $<Q^2<$ 12~GeV$^2$, and 
12~GeV$^2$ $<Q^2<$ 30~GeV$^2$. 

Due to limited statistics, we take
such a coarse binning for the $W$ and $Q^2$ directions, 
taking into account qualitative changes of  $|\cos \theta^*|$ 
dependence of the differential cross section.
We estimate the efficiency correction factor, $\varepsilon/\varepsilon^\prime$,
in each of the cells.

The signal-MC distribution after the selection 
behaves as $N_{\rm MC}(\varphi^*) \propto \varepsilon(\varphi^*)$
and the experimental distribution as 
$N_{\rm EXP}(\varphi^*) \propto \varepsilon(\varphi^*)P(\varphi^*)$,
where $P(\varphi^*) \propto d\sigma/d\varphi^*$. 
Then, the efficiency correction factor is calculated as
\begin{equation}
\frac{\varepsilon}{\varepsilon^\prime} = \frac{\pi \int_0^{\pi} 
\varepsilon(\varphi^*) P(\varphi^*) d\varphi^*}
{\int_0^{\pi} \varepsilon(\varphi^*) d\varphi^* \int_0^{\pi} P(\varphi^*) d\varphi^* }.
\end{equation}
The $\varphi^*$ dependence of the efficiency and the 
$\varphi^*$-differentiated cross section
are obtained by $\varepsilon(\varphi^*) \propto
N_{\rm MC}(\varphi^*)$ and $P(\varphi^*) \propto N_{\rm EXP}(\varphi^*)/N_{\rm MC}(\varphi^*)$,
respectively.
We then expand each function as a Fourier series, 
\begin{equation}
N_{\rm MC}(\varphi^*) = A(1+c \cos \varphi^* + d \cos 2\varphi^* + ... )
\end{equation}
and
\begin{equation}
N_{\rm EXP}(\varphi^*)/N_{\rm MC}(\varphi^*) =  B(1+a \cos \varphi^* + b \cos 2\varphi^*),
\end{equation}
where the coefficients are determined by fitting.
There are no sine terms because we expect 
symmetry between $\varphi^*$ and $-\varphi^*$
for these functions,
or we can regard them as the sum of the functions
in the positive and negative $\varphi^*$ regions,
$g(\varphi^*) = f(\varphi^*) + f(-\varphi^*)$ ($0 \leq \varphi^* \leq \pi$).     
They result in 
\begin{equation}
\frac{\varepsilon}{\varepsilon^\prime} = 1 + \frac{ac+bd}{2}.
\end{equation}

\begin{figure}
\centering
\includegraphics[width=7.7cm]{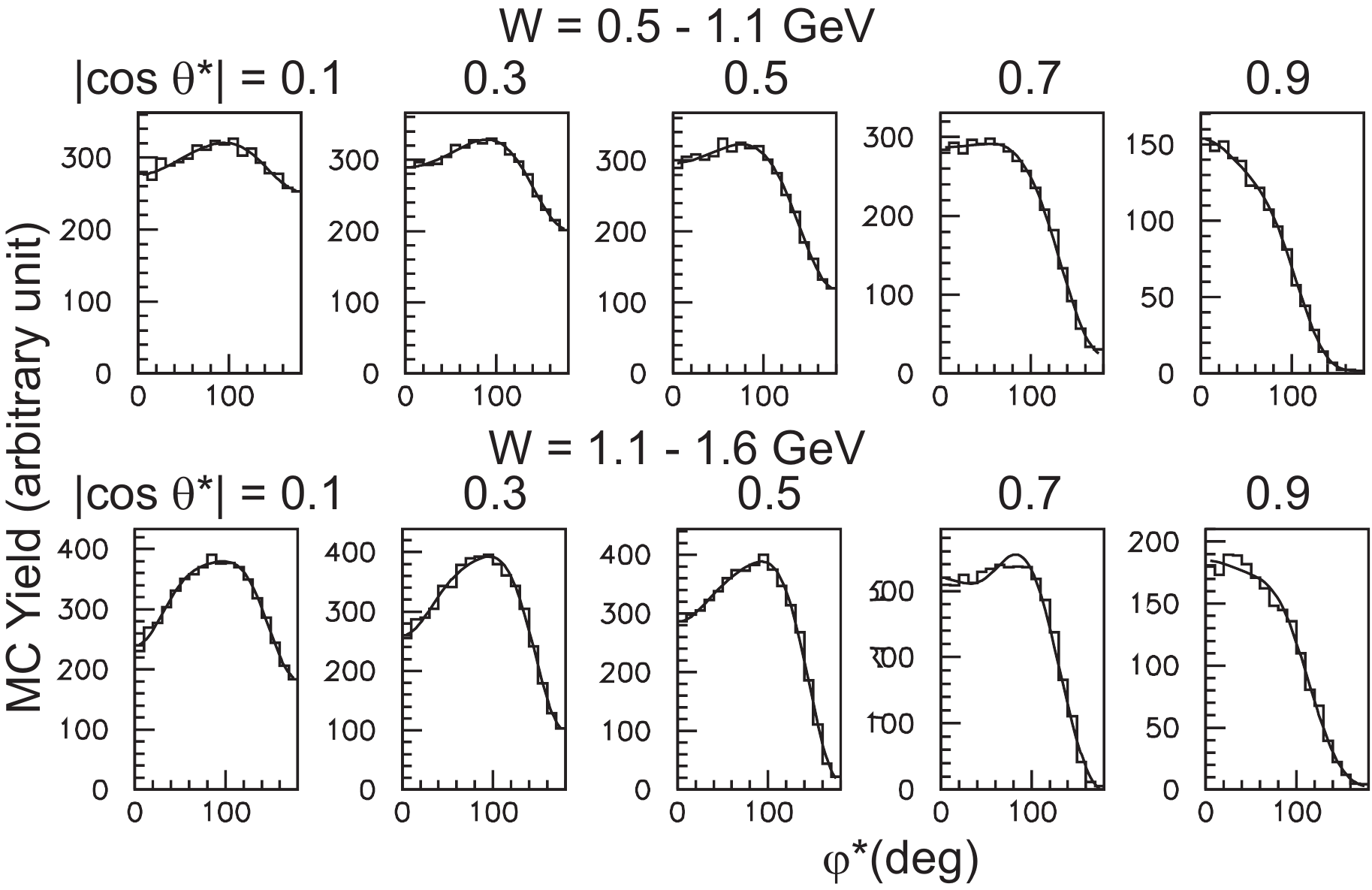}
\centering
\caption{
The $\varphi^*$ distributions for
$\varepsilon(\varphi^*) \propto N_{\rm MC}(\varphi^*)$ 
for the $W$ region 0.5 -- 1.1~GeV (upper row)
and 1.1 -- 1.6~GeV (lower row), and
8~GeV$^2<Q^2<12$~GeV$^2$. 
They are separately plotted for the five angular regions, 
whose central $|\cos \theta^*|$ value
is indicated in each panel. 
The solid curve is the 
fit to the Fourier expansion described in the text.
}
\label{fig:phidisteff}
\end{figure}

This formula is independent of the normalizations of the $N_{\rm MC}(\varphi^*)$ and
$N_{\rm EXP}(\varphi^*)$ functions. 
We approximate $N_{\rm MC}(\varphi^*)$ with
a Fourier expansion up to $\cos 4\varphi^*$; the coefficients 
of $\cos 3\varphi^*$ and $\cos 4\varphi^*$ terms do not affect
the $\varepsilon/\varepsilon^\prime$ result because $P(\varphi^*)$
is up to only $\cos 2\varphi^*$, 
but the effect of the terms
is significant in the fit to determine the coefficients $c$ and $d$. 
The $\varphi^*$ dependence of the efficiency and the fit
for two $W$ regions in the $Q^2$ range between 8~GeV$^2$ and 12~GeV$^2$
is shown in Fig.~\ref{fig:phidisteff}. 
The experimental results for
the $\varphi^*$-differentiated cross section
are discussed in Sec.~\ref{sub:phidep}.

\begin{figure}
\centering
\includegraphics[width=7cm]{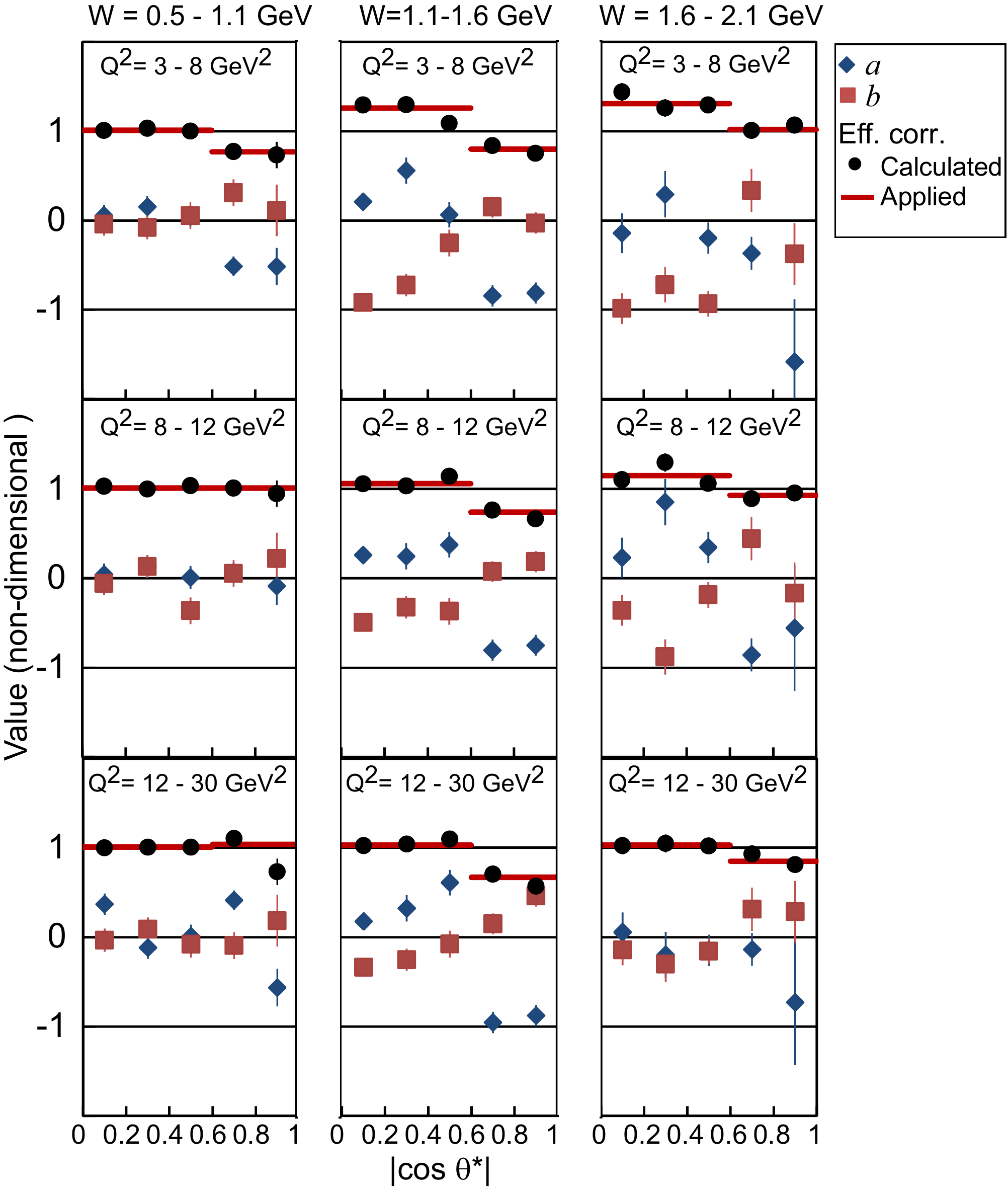}
\centering
\caption{The $|\cos \theta^*|$ dependence of
the obtained parameters of $a$, $b$, and the efficiency
correction factor $\varepsilon / \varepsilon^\prime$; $a$ and $b$ are  
the coefficients of 
$\cos \varphi^*$ and  $\cos 2\varphi^*$ in $P(\varphi^*)$
for each of the nine $(W, Q^2)$ regions (see the text). 
Thick lines indicate the efficiency 
correction factor, which is applied to obtain the $\varphi^*$-integrated
cross section.
}
\label{fig:phicor}
\end{figure}

We find that
the $\varphi^*$ dependence of the differential cross section, 
and thus the correction-factor value, 
change drastically between above and below $|\cos \theta^*| = 0.6$
and that the value for each of $|\cos \theta^*| <0.6$ and 
$|\cos \theta^*| >0.6$ is almost constant for the cells with 
the same ($W$, $Q^2$). 
Thus, we take a weighted average of the
correction factor for the three (two) bins of $|\cos \theta^*| <0.6$ 
($|\cos \theta^*| >0.6$), in order to reduce the uncertainty of
the correction.
The obtained efficiency correction factor 
$\varepsilon/\varepsilon^\prime$, as well as the fit
results for $a$ and $b$, are plotted in Fig.~\ref{fig:phicor}.
The correction factor ranges from 0.67 to 1.31 and is within 
0.93 -- 1.06 for 10 of the 18 regions. 

The trend of the $\varphi^*$ dependence is explained by
the interference term(s) with the
D$_0$ ($J=2$ and $\lambda = 0$) 
component, which changes
its sign at $1/\sqrt{3}$ $(\approx 0.577)$ 
according to the $Y_2^0$ function (see Eq.~(\ref{eqn:pwtheta})).

\subsection{Radiative correction and $Q^2$ unfolding}
\label{sub:radcor}
We apply a correction of 2\% ($\delta = 0.02$) for the cross
section as the radiative correction. 
This is the same as that evaluated in the single pion production~\cite{pi0tff}.
This correction depends very little on $W$, $|\cos\theta^*|$, and $Q^2$, 
and is treated as a constant.

We define the nominal $Q^2$ for each bin
with a finite bin width, $\bar{Q}^2$, using the formula
\begin{equation}
\frac{d\sigma_{ee}}{dQ^2}(\bar{Q}^2) = \frac{1}{\Delta Q^2}
\int_{\rm bin} \frac{d\sigma_{ee}}{dQ^2}(Q^2) dQ^2 ,
\label{eqn:q2form}
\end{equation}
where $\Delta Q^2$ is the bin width.
We assume an approximate dependence
of $d\sigma/dQ^2 \propto Q^{-7}$ for the calculation~\cite{pi0tff},
independent of $W$ and $|\cos \theta^*|$, and have omitted
the notations of $W$ and $|\cos \theta^*|$ in 
Eq.~(\ref{eqn:q2form}).
The $\bar{Q}^2$ values for the $Q^2$ bins are listed in 
Table~\ref{tab:repreq2}. 
We use the luminosity function at this $\bar{Q}^2$ point to
obtain the $\gamma^* \gamma$-based cross section for each $Q^2$ bin.
We also use the central value of the $Q^2$ bins to represent
the individual bins in tables and figures for convenience of
description.

\begin{center}
\begin{table}
\caption{The nominal $Q^2$ value ($\bar{Q}^2$) for each 
$Q^2$ bin.}
\begin{tabular}{cc|c}
\hline \hline
$Q^2$ bin (GeV$^2$) & Bin center (GeV$^2$) & $\bar{Q}^2$ (GeV$^2$) \\
\hline \hline
3 -- 4 & 3.5 & 3.45 \\
4 -- 5 & 4.5 & 4.46 \\
5 -- 6 & 5.5 & 5.47 \\
6 -- 8 & 7.0 & 6.89 \\
8 -- 10 & 9.0 & 8.92 \\
10 -- 12 & 11.0 & 10.93 \\
12 -- 15 & 13.5 & 13.37\\
15 -- 20 & 17.5 & 17.23 \\
20 -- 30 & 25.0 & 24.25 \\
\hline \hline
\end{tabular}
\label{tab:repreq2}
\end{table}
\end{center}

The $Q^2$ value measured for each event differs from the true
$Q^2$ for two reasons:
the resolution effect of the $Q^2$ determination
and the reduction of the incident electron energy due to ISR.

The signal yield is measured in bins of reconstructed $Q^2_{\rm rec}$.
For the true value of $Q^2$, we use the corrected value $Q^2_{\rm cor}$,
and we assign the number in $Q^2_{\rm cor}$ for each of the measurement bins. 
According to the ISR effect found in the signal-MC events, 
$Q^2_{\rm rec}$ is about 1\% larger than $Q^2_{\rm cor}$ on average.
We estimate the displacement of events across bins using
the signal-MC events in order to unfold the $Q^2$ distributions
folded by the displacement.

We correct the measurement by the factor
$f_i = \sum_j N_{ji}/\sum_j N_{ij}$, 
following the method applied to the $Q^2$ edge
regions in our previous analysis for $\pi^0$-TFF~\cite{pi0tff},
where $N_{ij}$ is the transfer matrix obtained from MC 
simulation with the adjusted $Q^2$-dependence close to 
the observed dependence; here $i$ ($j$) is the bin number 
for $Q^2_{\rm rec}$ ($Q^2_{\rm cor}$).

However,  the correction for this $Q^2$ migration effect is already 
partially included in the efficiency determination.
We have defined the efficiency as the selected number 
of events in a $Q^2_{\rm rec}$ bin, as in the experimental data, 
but divided by the number of generated events in the $Q^2_{\rm cor}$ bin.
This is due to the difficulty to define  $Q^2_{\rm rec}$ for 
generated events falling outside the detector acceptance.
Thus, a migration effect is also introduced in the efficiency,
but more weakly than in the experimental data,
at half the size of the latter, due to the different $Q^2$ dependencies.
By the consideration of this point, we correct
the cross section for this effect by the factor $(f_i-1)/2$
and assign a systematic
uncertainty of the same size as the correction value. 
The correction is $+5\%$ for the lowest $Q^2$ bin and within $\pm 3\%$ 
for the other $Q^2$ bins. 
We assign the systematic uncertainty 
for this correction with about half the size of the correction, 2\%.

We do not use the matrix inversion unfolding method~\cite{pi0tff} 
because the low statistics in the related multidimensional bins 
would tend to enhance the statistical errors. 
In addition, a systematic bias could appear in the edge bins of the 
measured $Q^2$ range by this method. 

The measured $\gamma^* \gamma$-based cross section 
$\sigma_{\gamma* \gamma} =
\sigma_{\rm TT} + \epsilon_0 \sigma_{\rm LT} \equiv \sigma$ 
is discussed and shown in figures in Sec.~\ref{sec:meastff}.
 
\subsection{Effect of $\sigma_{\rm LT}$ (helicity-1) component 
in the signal sample}
\label{sub:siglt}
We estimate 
$\epsilon_0$, the factor multiplying $\sigma_{\rm LT}$ in Eq.~(\ref{eqn:tots}),
in each bin.
We use the mean value of $\epsilon_0$ calculated by Eq.~(\ref{eqn:eps0}) 
for each selected event from 
the signal-MC samples
in different kinematical regions.
The value of $\epsilon_0$ has a weak dependence on 
$Q^2$, $W$, $|\cos \theta^*|$,
and $\varphi^*$. 
Since the $W$ dependence, in $W=0.5$ -- 1.5~GeV and 
the $|\cos \theta^*|$ dependence are small (within $\pm 4\%$), 
we neglect their effect.
It has some $\varphi^*$ dependence (up to $\pm 7\%$).

We have compared the $Q^2$ dependence of the experimental
events with that of the signal-MC samples and have confirmed
their consistency. 
Similar calculations for $\epsilon_1$ are also performed.

We tabulate the $Q^2$ dependence of $\epsilon_0$ in Table~\ref{tab:eps0}
for the two $W$ regions.
The expected values are $\epsilon_0 = 0.88 \pm 0.06$ and $\epsilon_1 =
1.28 \pm 0.07$ for $W = 1.1 $ -- 1.5~GeV for the $Q^2$-integrated
samples. The ranges in $\varepsilon_1$ and $\varepsilon_2$
are due to the $\varphi^*$ dependence.
These values are used for partial-wave analyses 
performed with $Q^2$-dependent and $Q^2$-integrated cross sections 
in Sec.~\ref{sub:phidep} and \ref{sub:fitres}, respectively.

\begin{center}
\begin{table}
\caption{
Value of the $\epsilon_0$ parameter depending on $Q^2$
for the two $W$ regions.
}
\begin{tabular}{c|cc}
\hline \hline
 & \multicolumn{2}{c} {$W$ region}\\
$Q^2$ bin (GeV$^2$) &  0.5 -- 1.6~GeV \ & \  1.6 -- 2.1~GeV \\
\hline \hline
3 -- 4 & 0.82 & 0.77 \\
4 -- 5 & 0.88 & 0.84 \\
5 -- 6 & 0.90 & 0.83  \\
6 -- 8 & 0.89 & 0.83 \\
8 -- 10 & 0.88 & 0.85 \\
10 -- 12 & 0.88 & 0.85  \\
12 -- 15 & 0.86 & 0.83  \\
15 -- 20 & 0.82 & 0.80 \\
20 -- 30 & 0.76 & 0.73 \\
\hline \hline
\end{tabular}
\label{tab:eps0}
\end{table}
\end{center}
 
For a quantitative study of the  $\lambda = 1$ component,
we use the azimuthal-angle differentiated cross section 
$d^2 \sigma/d |\cos \theta^*| d \varphi^*$. 
It is derived as follows.

The $\varphi^*$ dependence of $d^2 \sigma/d |\cos \theta^*| d \varphi^*$
follows $N_{\rm EXP}(\varphi^*)/N_{\rm MC}(\varphi^*)$
in each $W$ and $|\cos \theta^*|$ bin integrated in the 
$Q^2= 5$ -- 30 GeV$^2$ region.  
For this purpose, we use five  
$W$ bins, 0.7 -- 1.1~GeV, 1.1 -- 1.2~GeV, 1.2 -- 1.3~GeV, 1.3 -- 1.4~GeV,
and 1.4 -- 1.5~GeV.

The $\varphi^*$ dependence is normalized to the arithmetic mean
$\langle d \sigma/d |\cos \theta^*| \rangle$ of the differential
cross section over the corresponding ($W$, $Q^2$) bins. 
We do not use the results for
$Q^2 < 5$~GeV$^2$ because poor statistics in that region
would  diminish the accuracy of the arithmetic mean.

The normalization results in 
\begin{equation}
\int^{\pi}_{0}  \frac{d^2 \sigma}{d |\cos \theta^*| d \varphi^*} d\varphi^* = 
\langle \frac{d \sigma}{d |\cos \theta^*|} \rangle.
\end{equation}
The results for $d^2 \sigma/d |\cos \theta^*| d\varphi^*$
are shown in Sec.~\ref{sec:meastff}.

\subsection{Systematic uncertainties}
\label{sub:sysunc}
We estimate systematic uncertainties in the measurement of 
the differential cross section;
these are summarized in
Table~\ref{tab:syserr}.

\subsubsection{Uncertainties for the efficiency evaluation}
The detection efficiency is evaluated using the signal-MC events.
However, the simulation has some errors or ambiguities in the reproduction of
detector performance. 
This translates into uncertainties in the efficiency evaluation.

Tracking has a 1\% uncertainty, which is estimated from a study
of the decays 
$D^{*\pm} \to D^0 \pi^{\pm}, D^0 \to K^0_S (\to \pi^+ \pi^-) \pi^+ \pi^-$
(0.35\% per track) as well as an uncertainty of the radiation
by an electron within the CDC volume. 

The electron identification
efficiency in this measurement is very high, around 98\%, and
a 1\% systematic uncertainty is assigned for this term.
Detection of the $\gamma \gamma$ pair for reconstructing 
$\pi 1$ has a 3\% uncertainty. 
The mass resolution 
of $\pi 1$ is well reproduced in the MC simulation but the 
selection with the invariant mass
introduces an additional 3\% uncertainty, according to a comparison
of the mass distributions between the MC and data samples.
The $\pi 2$ reconstruction has a 3\% uncertainty also. 
The uncertainty of the $\pi^0$ reconstruction efficiency is estimated
by a comparison of the yields of $D^0$ to $K^\mp \pi^\pm$ and 
$K^\mp \pi^\pm \pi^0$ decays.
We combine the two uncertainties for $\pi 1$ in quadrature
(then 4.2\% for $\pi 1$) and add the uncertainty for $\pi 2$
to it linearly, because the uncertainties for the two pions 
are fully correlated:
the overall uncertainty is 7.2\% for the two pions.

A kinematical condition 
using $E_{\rm ratio}$
applied in the selection 
gives an uncertainty of 2\%. 
In addition, ambiguities in the detector edge locations and other 
geometrical-definition effects cause an uncertainty of 2\%.

The uncertainty from the trigger efficiency is estimated by changing
the energy thresholds of the Bhabha-veto in the HiE trigger (1\%)
and the cluster energy of the Clst4 trigger (2\%), as
performed in our previous analyses~\cite{pi0tff, pi0pi0, pi0pi02}. 
The estimation methodology is the same but the effects are different
among the different processes.
The total uncertainty for the trigger efficiency is 3\%.

Background tracks and photons 
overlapping with the signal events may reduce 
the efficiency; this effect is accounted in MC by embedding 
the non-triggered event pattern (random trigger) in the signal. 
We estimate this effect by
investigating the  background conditions during different
beam conditions or run periods. 
The effect on the efficiency is estimated to be 2\%. 
The largest effect for the present analysis
is estimated to come from background photons that form 
an extra $\pi^0$ with another true or background photon.

We take into account an uncertainty for the efficiency-correction 
factor arising from the $\varphi^*$ dependence. 
Half of the difference of the factors in the two neighboring $Q^2$ regions 
is assigned to the uncertainty for each ($W$,$|\cos \theta^*|$)
region for the evaluation. 
We choose the larger for the 8~GeV$^2<Q^2<12$~GeV$^2$ regions, where 
two neighboring $Q^2$ regions exist. 
The systematic error ranges from 1\% to 16\%,
according to strength of the $Q^2$ dependence of the factor.

\subsubsection{Uncertainties from the background subtraction}
We take into account the backgrounds from the single $\pi^0$ production
process and the virtual Compton process. 
We do not perform
any qualitative simulation for the latter process and use only
an estimate based on the features found in the experimental data.
We assign the subtraction size applied for 
these two background sources as the systematic uncertainty for 
$|\cos \theta^*|>0.9$.  
In addition, we assign another 1\% error, added linearly, from the same 
sources for all the angular bins.

We assign half of the subtraction size as the uncertainty of the 
background subtractions for the $\gamma \pi^0$ production and non-exclusive 
processes; they are estimated using the signal and background
MC and/or the experimental sideband events. 

We add the systematic 
uncertainty from the latter backgrounds in quadrature, but the 
others linearly,
because the first three have a common feature that photon(s) and
pion(s) are from backgrounds. 
The last one is different:
some particles escape detection.

The total uncertainty in the background subtraction 
ranges from 1\% to 23\%, depending on the ($W$, $|\cos \theta^*|$, $Q^2$) bin.
The error is relatively large for the forward angles in the
lowest $W$ or the higher $Q^2$ regions.

\subsubsection{Uncertainties from other sources}
The unfolding procedure has an uncertainty of 2\%. 
The radiative correction has an uncertainty of 3\%. 
The evaluation of
the luminosity function gives an uncertainty of 4\%, including
a model uncertainty for the form factor of the untagged side (2\%).
This model uncertainty is based on the difference
of the product of the luminosity function and efficiency
between the two models: 
one with $\sim 1/(1+Q^2_2/m_{\rho}^2)$ and the other
constant in $Q^2_2$
(effectively the same as the model with a dependence 
of $\sim 1/(Q^2_1 + Q^2_2)^d$ because $Q^2_2 \ll Q^2_1$,
where $d$ is a power parameter representing
the high-$Q_1^2$ behavior).
The integrated luminosity has a measurement uncertainty of 1.4\%.

The systematic uncertainties are added in quadrature unless
noted above. 
The total systematic uncertainty is between 11\% and 26\%,
depending on the ($W$, $|\cos \theta^*|$, $Q^2$) bin.

In the following analysis, we treat a 4\% systematic uncertainty ---
from the kinematical cut, geometrical acceptance and partially
from the trigger efficiency and unfolding for $Q^2$ --- a
bin-by-bin error that distorts the $W$ dependence. 
We also take into account the systematic uncertainty
in the efficiency correction from $\varphi^*$ dependence,
which distorts the $|\cos \theta^*|$ dependence.
The remaining part is treated as an uncertainty of
the overall normalization.

\begin{center}
\begin{table}
\caption{Sources of systematic uncertainties. The values indicated
in a range show the range sizes  in  different bins.
}
\begin{tabular}{c|c}
\hline \hline
Source & Uncertainty (\%) \\
\hline \hline
Tracking & 1 \\
Electron-ID & 1 \\
Pion-pair detection (for two $\pi^0$'s) & 7.2 \\
Kinematical selection & 2 \\
Geometrical acceptance & 2 \\
Trigger efficiency & 3 \\
Background effect for the efficiency & 2 \\
$\varphi^*$ dependence & 1 -- 16 \\
Background subtraction & 1 -- 23 \\
Unfolding for $Q^2$ & 2 \\
Radiative correction & 3 \\
Luminosity function & 4\\
Integrated luminosity & 1.4\\
\hline
Total & 11 -- 26 \\
\hline \hline
\end{tabular}
\label{tab:syserr}
\end{table}
\end{center}

\section{Measurement of transition form factors}
\label{sec:meastff}
\begin{figure}
 \centering
   {\epsfig{file=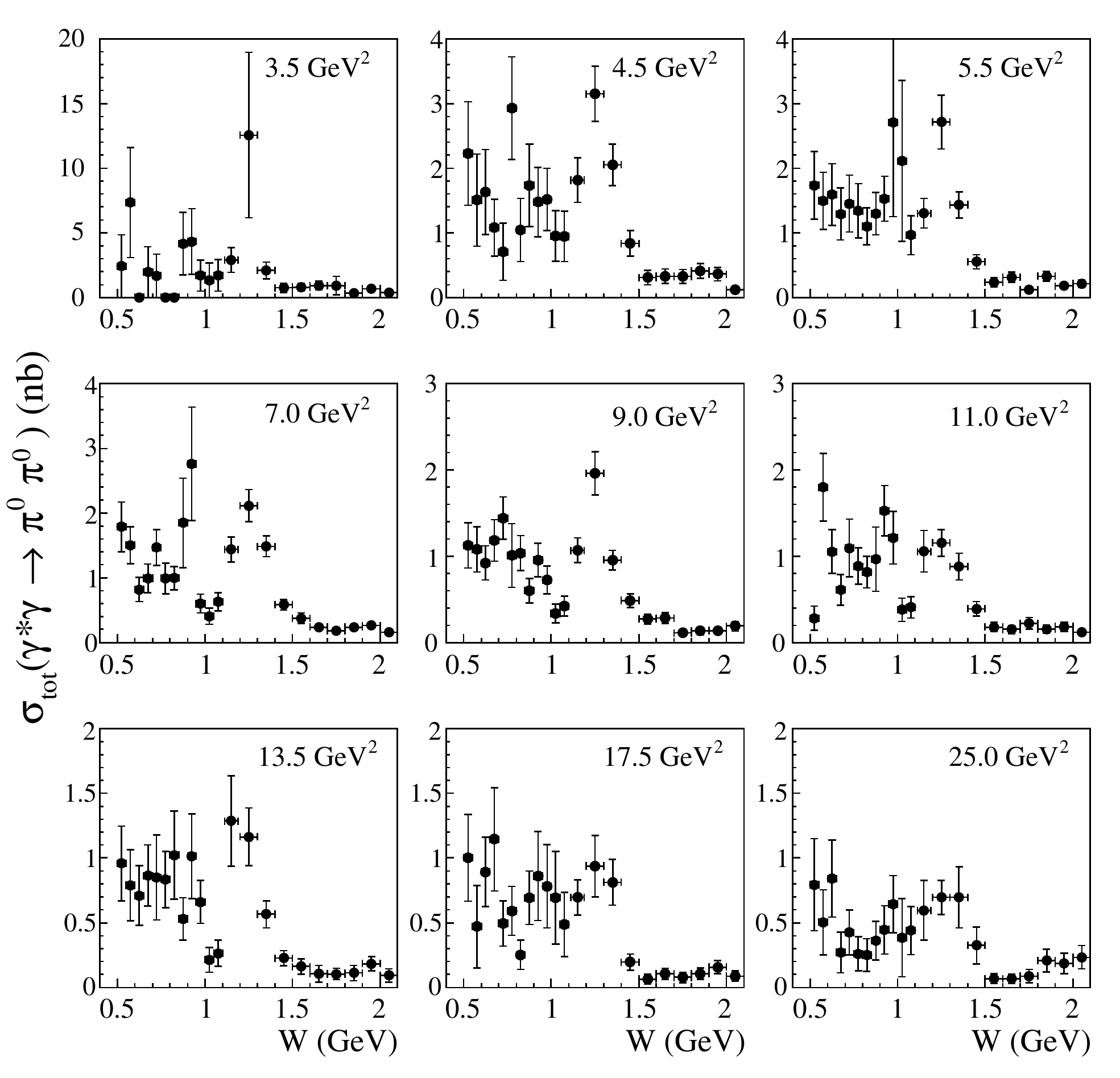,width=80mm}}
 \caption{Integrated cross section 
for $\gamma^* \gamma \to \pi^0 \pi^0$ in nine $Q^2$ bins
in GeV$^2$ indicated in each panel.}
\label{fig:tot}
\end{figure}

Figure~\ref{fig:tot} shows the integrated cross section
as a function of $W$
for the process $\gamma^* \gamma \to \pi^0 \pi^0$ in nine $Q^2$ bins.
Peaks corresponding to the $f_2(1270)$ and $f_0(980)$ are evident.
In this section, we extract the $Q^2$ dependence of the
TFF of the $f_0(980)$ and those of the helicity-0, -1, and -2 
components of the $f_2(1270)$.

\subsection{Partial-wave amplitudes}
\label{sub:pwa}
Helicity amplitudes in Eq.~(\ref{eqn:dsdcos}) are functions of 
$W$, $Q^2$, and $\theta^*$ but not of $\varphi^*$~\cite{gss} and 
can be expanded in terms of partial waves.
In this channel, only even angular-momentum partial waves contribute.
Furthermore, in the energy region $W \le 1.5~\GeV$, $J > 2$ partial 
waves (the next having $J=4$) may be neglected so that 
only S and D waves need be considered.
Then, Eqs.~(\ref{eqn:t0}) to (\ref{eqn:t2}) can be written as
\begin{eqnarray}
t_0 &=& |S Y^0_0 + D_0 Y^0_2|^2 + |D_2 Y^2_2|^2 
  + 2 \epsilon_0 |D_1 Y^1_2|^2 , \nonumber \\
t_1 &=& 2 \epsilon_1 \Re \left( (D_2^* |Y^2_2| - S^* Y^0_0 - D_0^* Y^0_2)
D_1 |Y^1_2| \right), \nonumber \\
t_2 &=& -2 \epsilon_0 \Re \left(D_2^* |Y^2_2| (S Y^0_0 + D_0 Y^0_2) \right),
\label{eqn:pwtheta}
\end{eqnarray}
where $S$ is the S-wave amplitude,
$D_0$,\; $D_1$, and $D_2$ denote the helicity-0, -1, and -2 components
of the D wave, respectively~\cite{pw},
and $Y^m_J$ are the spherical harmonics:
\begin{eqnarray}
Y^0_0 &=& \sqrt{\frac{1}{4 \pi}} \; , \nonumber \\
Y^0_2 &=& \sqrt{\frac{5}{16 \pi}}(3 \cos^2 \theta^* - 1) \; , \nonumber \\
|Y^1_2| &=& \sqrt{\frac{15}{8 \pi}} \sin\theta^* \cos \theta^* 
\; , \nonumber \\
|Y^2_2| &=& \sqrt{\frac{15}{32 \pi}} \sin^2 \theta^* 
\; .
\label{eqn:sphe}
\end{eqnarray}
We take the absolute values of the spherical
harmonics because the helicity amplitudes do not have
$\varphi^*$ dependence~\cite{gss}.
When integrated over the azimuthal angle, the differential cross section
can then be expressed as
\begin{eqnarray}
&& \frac{d \sigma (\gamma^* \gamma \to \pi^0 \pi^0)}
{4 \pi d |\cos \theta^*|} 
= \left| S \: Y^0_0 + D_0 \: Y^0_2 \right|^2   
\nonumber \\
&& + 2 \epsilon_0 \left|D_1 \: Y^1_2 \right|^2  
+ \left|D_2 \: Y^2_2 \right|^2 .
\label{eqn:dsdc}
\end{eqnarray}

The angular dependence of the cross section is governed by the 
spherical harmonics, while the $W$ and $Q^2$ dependencies are 
determined by the partial waves.

\subsection{Parameterization of amplitudes}
The $f_0(980)$ and the $f_2(1270)$
have been clearly observed in no-tag $\pi^0 \pi^0$
production~\cite{pi0pi0}.
The signal is visible in the $Q^2$-integrated spectrum
(Fig.~\ref{fig:wdist}) as well as in most of the $Q^2$ bins
in Fig.~\ref{fig:tot}.
Motivated by this, we try to extract the $Q^2$ dependence of the
TFFs: $F_{f2}(Q^2)$ of the $f_2(1270)$ together with its
helicity-0, -1, and -2 components
and $F_{f0}(Q^2)$ of the $f_0(980)$.
This is done by parameterizing $S$, $D_0$ (and/or $D_1$), and $D_2$
and by fitting the data in the
energy region $0.7~\GeV \le W \le 1.5~\GeV$.

$S$ and $D_i \; (i=0,1,2)$ are parameterized as follows
(considering the general case where the $\varphi^*$-dependent 
cross section is to be fitted):
\begin{eqnarray}
S &=& A_{f_0(980)}e^{i \phi_{f0}} + B_S e^{i \phi_{BS}}  , \nonumber \\
D_i &=& \sqrt{r_i(Q^2)} A_{f_2(1270)}e^{i \phi_{f2Di}} + 
B_{Di} e^{i \phi_{BDi}}, 
\label{eqn:param}
\end{eqnarray}
where $A_{f_0(980)}$ and $A_{f_2(1270)}$ 
are the amplitudes of the $f_0(980)$ and $f_2(1270)$, respectively;
$r_i(Q^2)$ is the fraction of the $f_2(1270)$-contribution
in the D$_i$ wave with the constraints $r_0 + r_1 + r_2 = 1$
and $r_i \ge 0$;
$B_S$ and $B_{Di}$ are non-resonant ``background'' 
amplitudes for the S and D waves;
$\phi_{BS}$, $\phi_{BDi}$, $\phi_{f0}$ 
and $\phi_{f2Di}$ are the phases of background 
amplitudes ($S$ and $D_i$), the $f_0(980)$ and $f_2(1270)$ in the D waves,
respectively.
The phases are assumed to be independent of $Q^2$.
The contributions of the $f_2'(1525)$ and $f_0(Y)$ (that are included
in no-tag $\pi^0 \pi^0$ production~\cite{pi0pi0})
are neglected.
The overall arbitrary phase is fixed by taking $\phi_{f2D2} = 0$.
When fitting the $\varphi^*$-integrated cross section,
we also set $\phi_{BS}=\phi_{BD1}=0$.  

A power behavior in $W$ is assumed for
the background amplitudes, which are 
multiplied by the threshold factor $\beta^{2 l +1}$ 
($l$ denoting the orbital angular momentum of the two-$\pi^0$ system)
with an assumed $Q^2$ dependence for all the waves:
\begin{eqnarray}
B_S &=&  \frac{\beta a_S \left( W_0/W \right)^{b_S}}
{(Q^2/m_0^2 + 1)^{c_S}}
\;, \nonumber \\
B_{D0} &=& \frac{\beta^5 a_{D0} \left( W_0/W \right)^{b_{D0}}}
{(Q^2/m_0^2 + 1)^{c_{D0}}} \;, \nonumber \\
B_{D1} &=& \frac{\beta^5 Q^2 a_{D1} \left( W_0/W \right)^{b_{D1}}}
{(Q^2/m_0^2 + 1)^{c_{D1}}} \;, \nonumber \\
B_{D2} &=& \frac{\beta^5 a_{D2} \left( W_0/W \right)^{b_{D2}}}
{(Q^2/m_0^2 + 1)^{c_{D2}}} \; .
\label{eqn:para2}
\end{eqnarray}
Here, $\beta=\sqrt{1 - 4m_{\pi^0}^2/W^2}$ is the velocity of $\pi^0$
divided by the speed of light, $m_{\pi^0}$ is the $\pi^0$ mass,
and $W_0$ and $m_0$ are assigned the values $1.1~\GeV$ and $1.0~\GeV /c^2$, 
respectively.
Note that $B_{D1}$ has an additional factor of $Q^2$ to ensure that
the amplitude becomes zero at $Q^2=0$.
We set $a_i \ge 0 \; (i=S, D_0, D_1, D_2)$ to fix
the arbitrary sign of each background amplitude
(by absorbing the sign into their corresponding phases).
We allow $b_i$ to have a negative sign because amplitudes may
still be an increasing function of $W$, 
but we limit $|b_i|<5$; large $b_i$ values give a rapid $W$ dependence, which
is considered unphysical.

We use the parameterizations of the $f_0(980)$ and  $f_2(1270)$ given in 
Refs.~\cite{mori1, mori2} that are
multiplied by their TFFs to allow a $Q^2$ dependence.
Note that ${\cal B}(f_J \rightarrow \pi^0 \pi^0)/
{\cal B}(f_J \rightarrow \pi^+ \pi^-) = 1/2$ (because the $f_J$ mesons
are isoscalar).
 For completeness, we reproduce here the parameterization of the $f_0(980)$ and
the $f_2(1270)$.
For the $f_0(980)$ meson, we need to take into account the fact that
its mass is close to the $K \bar{K}$ threshold. 
The parameterization we adopt is
\begin{equation}
A_{f_0(980)} = F_{f0} (Q^2) \sqrt{1 + \frac{Q^2}{M_{f_0}^2}}
\frac{\sqrt{8 \pi \beta_{\pi}}}{W}
\frac{g_{f_0 \gamma\gamma}g_{f_0 \pi\pi}}{16 \sqrt{3} \pi}
\frac{1}{D_{f_0}} ,
 \label{eqn:sigma}
\end{equation}
where $F_{f0} (Q^2)$ is the transition form factor of the $f_0(980)$,
$\sqrt{1 + Q^2/M_{f_0}^2}$ is the flux factor that
arises from definition of the luminosity function for a tagged
two-photon cross section,
$\beta_X = \sqrt{1-4 {M_X}^2/W^2}$ is the velocity divided by
the speed of light for 
a particle $X$ with  mass $M_X$ in the two-body final states, and
$g_{f_0XX}$ is related to the partial width
of the $f_0(980)$ meson via
$\Gamma_{XX} (f_0) = \beta_X g_{f_0 XX}^2/(16 \pi M_{f_0})$.
The factor $D_{f_0}$ is given by the following expression~\cite{denom}:
\begin{eqnarray}
 D_{f_0}(W) &=& M_{f_0}^2 - W^2
    + \Re{\Pi_{\pi}^{f_0}}\lr{M_{f_0}}-\Pi_{\pi}^{f_0}\lr{W}
\nonumber\\
&&    + \Re{\Pi_K^{f_0}}\lr{M_{f_0}} - \Pi_K^{f_0}\lr{W} ,
\end{eqnarray}
with
\begin{equation}
 \Pi_X^{f_0}(W) =  \frac{\beta_X {g^2_{f_0XX}}}{16\pi}
 \left[i + \frac{1}{\pi}
 \ln\frac{1-\beta_X}
 {1+\beta_X}\right] ,
\end{equation}
where $X = \pi$ or $K$.
The factor $\beta_K$ is real in the region $W \geq 2M_K$
and becomes imaginary for $W < 2M_K$.
The parameter values are summarized in Table~\ref{tab:f0par}.
\begin{center}
\begin{table}[H]
\caption{Parameters of the $f_0\lr{980}$ used in the fit.
When two errors are provided, the first is statistical,
and the second systematic.
}
\label{tab:f0par}
\begin{tabular}{ccc} \hline \hline
Parameter & Value & Reference \\ \hline \hline
 Mass ($\MeV/c^2$) & $980 \pm 20$ & \cite{pdg2012}\\
$g_{f_0(980) \pi\pi} (\GeV)$ & $1.82 \pm 0.03 
                     ~_{-0.17}^{+0.24}$ & \cite{pi0pi0}\\
$\Gamma_{\gamma\gamma} (\keV)$ & $0.29~^{+0.07}_{-0.06}$ 
& \cite{pdg2012} \\
$g^2_{f_0 K \bar{K}}/g^2_{f_0 \pi\pi}$ 
& $4.21 \pm 0.25 \pm 0.21$ 
& \cite{bes} \\
\hline \hline
\end{tabular}
\end{table}
\end{center}

We use a parameterization of the $f_2(1270)$
in Ref.~\cite{pi0pi0} multiplied by the TFF, $F_{f2}(Q^2)$.
The relativistic Breit-Wigner resonance amplitude
$A_R(W)$ for a spin-$J$ resonance $R$ of mass $m_R$ is given by
\begin{eqnarray}
A_R^J(W) &=& F_R(Q^2) \sqrt{1 + \frac{Q^2}{M_{R}^2}} 
\sqrt{\frac{8 \pi (2J+1) m_R}{W}} 
\nonumber \\
&& \times \frac{\sqrt{ \Gamma_{\rm tot}(W)
\Gamma_{\gamma \gamma}(W) \B(\pi^0 \pi^0)}}
{m_R^2 - W^2 - i m_R \Gamma_{\rm tot}(W)} \; ,
\label{eqn:arj}
\end{eqnarray}
where $F_R(Q^2)$ is the TFF of the resonance $R$ and
$\sqrt{1 + Q^2/M_{R}^2}$ is the flux factor mentioned above.
Hereinafter, we consider the case $J=2$.
The energy-dependent total width $\Gamma_{\rm tot}(W)$ is given by
\begin{equation}
\Gamma_{\rm tot}(W) = \sum_X \Gamma_{X \bar{X}} (W) 
+ \Gamma_{\rm other}(W) \; ,
\label{eqn:gamma}
\end{equation}
where $X$ is $\pi$, $K$, $\eta$, $\gamma$, etc.
For $J=2$,
the partial width $\Gamma_{X \bar{X}}(W)$ is parameterized as~\cite{blat}
\begin{eqnarray}
\Gamma_{X \bar{X}}(W) &=& \Gamma_R \B(R \rightarrow X \bar{X}) 
\left( \frac{q_X(W^2)}{q_X(m_R^2)} \right)^5 \nonumber \\
&& \times
\frac{D_2\left( q_X(W^2) r_R \right)}{D_2 \left( q_X(m_R^2) r_R 
\right)} \;,
\label{eqn:gamx}
\end{eqnarray}
where $\Gamma_R$ is the total width at the resonance mass,
\begin{eqnarray}
q_X(W^2) &=& \sqrt{\frac{W^2}{4} - m_X^2}, \nonumber \\
D_2(x) &=& \frac{1}{9 + 3 x^2 +x^4} \; ,
\label{eqn:qx}
\end{eqnarray}
and $r_R$ is an effective interaction radius that varies from 
1~$(\GeV/c)^{-1}$ to 7~$(\GeV/c)^{-1}$ in different hadronic 
reactions~\cite{grayer}.
For the three-body and other multi-body decay modes,
\begin{equation}
\Gamma_{\rm other} (W) = \Gamma_R \B(R \rightarrow {\rm other})
\frac{W^2}{m_R^2}
\label{eqn:gam3}
\end{equation}
is used instead of Eq.~(\ref{eqn:gamx}).
All parameters of the $f_2(1270)$
are fixed at the PDG values~\cite{pdg2012}
except for $r_R$, which is fixed at the value determined in 
Ref.~\cite{mori2}, as summarized in Table~\ref{tab:f2par}.

The normalizations of TFFs are such that 
$F_{f0}(0)=1.00$ $\pm$ $0.11$ and $F_{f2}(0)=1.00 \pm 0.06$;
the errors reflect the uncertainties of the two-photon decay
widths (at $Q^2 = 0$) of the $f_0(980)$ and $f_2(1270)$~\cite{pdg2012}.
The TFF of the $f_2(1270)$ for the helicity-$i$ component is given by
$\sqrt{r_i(Q^2)}F_{f2}(Q^2)$, according to Eq.~(\ref{eqn:param}).

\begin{center}
\begin{table}
\caption{Parameters of the $f_2\lr{1270}$.}
\label{tab:f2par}
\begin{tabular}{ccc} \hline \hline
Parameter & $f_2\lr{1270}$ & Reference \\ 
\hline \hline
 Mass ($\MeV/c^2$ ) & $1275.1 \pm 1.2$ 
& \cite{pdg2012}\\
$\Gamma_{\rm tot}$ (MeV) & $185.1 ^{+2.9}_{-2.4}$ 
& \cite{pdg2012}\\
$\B(\pi \pi)$ (\%) & $84.8^{+2.4}_{-1.2}$ 
& \cite{pdg2012} \\
$\B(K \bar{K})$ (\%) & $4.6 \pm 0.4$ 
&\cite{pdg2012} \\
$\B(\eta \eta)$ (\%) & $ 0.40 \pm 0.08$
& \cite{pdg2012} \\
$\B(\gamma \gamma)$ ($10^{-6}$) & $16.4 \pm 1.9$ 
& \cite{pdg2012}\\
$r_R$ ($(\GeV/c)^{-1}$) & $3.62 \pm 0.03$ 
& \cite{mori2} \\
\hline \hline
\end{tabular}
\end{table}
\end{center}

Here, we note that there is a limitation in the partial-wave 
analysis based on Eq.~(\ref{eqn:dsdc}) ({\it i.e.}, through the 
$\varphi^*$-integrated cross section).
That is, we can extract information on partial waves for
three out of the four 
(S, D$_0$, D$_1$, and D$_2$) waves only.
This can be understood from the fact that Eq. (\ref{eqn:dsdc}) 
can be written as a function of 
$z \equiv |\cos \theta^*|$ as $a + b z^2 + c z^4$, {\it i.e.},
only three coefficients $a, b, c$ are independent, where
$a, b, c$ are given by combinations of the
S, D$_0$, D$_1$, and D$_2$ waves.
This conclusion holds whether or not
the interference term $\Re(S^* D_0)$ exists.

To partially overcome this limitation, we first fit the $\varphi^*$-dependent
(but $Q^2$-integrated) differential cross section 
in Sec.~\ref{sub:phidep}, to obtain information 
on the fractions of the $f_2(1270)$ in the 
D$_0$, D$_1$, and D$_2$ waves.
Then in Sec.~\ref{sub:fitres}, this information is used in the fit of the
$\varphi^*$-integrated cross section.

\subsection{Analysis of the $\varphi^*$-dependent cross section}
\label{sub:phidep}
As described above, the $\varphi^*$-integrated cross section
does not give information on all the partial waves 
(S, D$_0$, D$_1$, and D$_2$).
Here, we analyze the $\varphi^*$-dependent cross section to partially
overcome this problem.
Because of limited statistics, the $\varphi^*$-dependent cross section is
integrated over $Q^2$; it is divided into nine $\varphi^*$ bins
with a bin width of $20^\circ$,
five equal-width $|\cos \theta^*|$ bins,
and five $W$ bins of 0.7 -- 1.1~GeV, 1.1 -- 1.2~GeV, 1.2 -- 1.3~GeV, 
1.3 -- 1.4~GeV, and 1.4 -- 1.5~GeV.
The average value of $Q^2 $($Q^2_{\rm av}$) is 9.6~GeV$^2$.
The cross section is fitted using the parameterization described above
by ignoring the contribution of the $f_0(980)$ and $Q^2$ dependence.
The values of $\epsilon_0$ and $\epsilon_1$ are evaluated at
$Q^2_{\rm av}$.

Parameters describing the assumed amplitudes 
are obtained by minimizing $\chi^2$.
To search for the true minimum and to identify possible multiple 
solutions, 1000 sets of randomly generated initial parameters 
are employed for fits performed using MINUIT~\cite{minuit}.
Fitted values are accepted as satisfactory solutions
when their $\chi^2$ values are within $\chi^2_{\rm min} + 10$ 
(corresponding to $3.2 \sigma$), where $\chi^2_{\rm min}$ is the
$\chi^2$ value at the true minimum. 

There are too many parameters to be fitted simultaneously, particularly from the non-resonant 
(background) amplitude that interferes with the resonant one in each wave. 
Thus we study the sensitivity of parameters especially in the D waves
by investigating several sets of assumptions in the parameters.

Two categories of fits are made: $B_{D2} \ne 0$ and $B_{D2} = 0$,
because we consider that the interference between the $f_2(1270)$ and 
a possible non-resonant background is important in the D$_2$ wave.
In each category, we try cases where both $r_0$ and $r_1$ are
non-zero or one of them is zero in Eq.~(\ref{eqn:param}).
The number of solutions found is one or two
as listed in Table~\ref{tab:phifit}.
Here, solutions in which any of powers ($b$'s) in Eq.~(\ref{eqn:para2}) 
exceed 5 are discarded as unphysical provided that the corresponding 
$a$ parameter is not consistent with zero.

\begin{figure}
 \centering
  {\epsfig{file=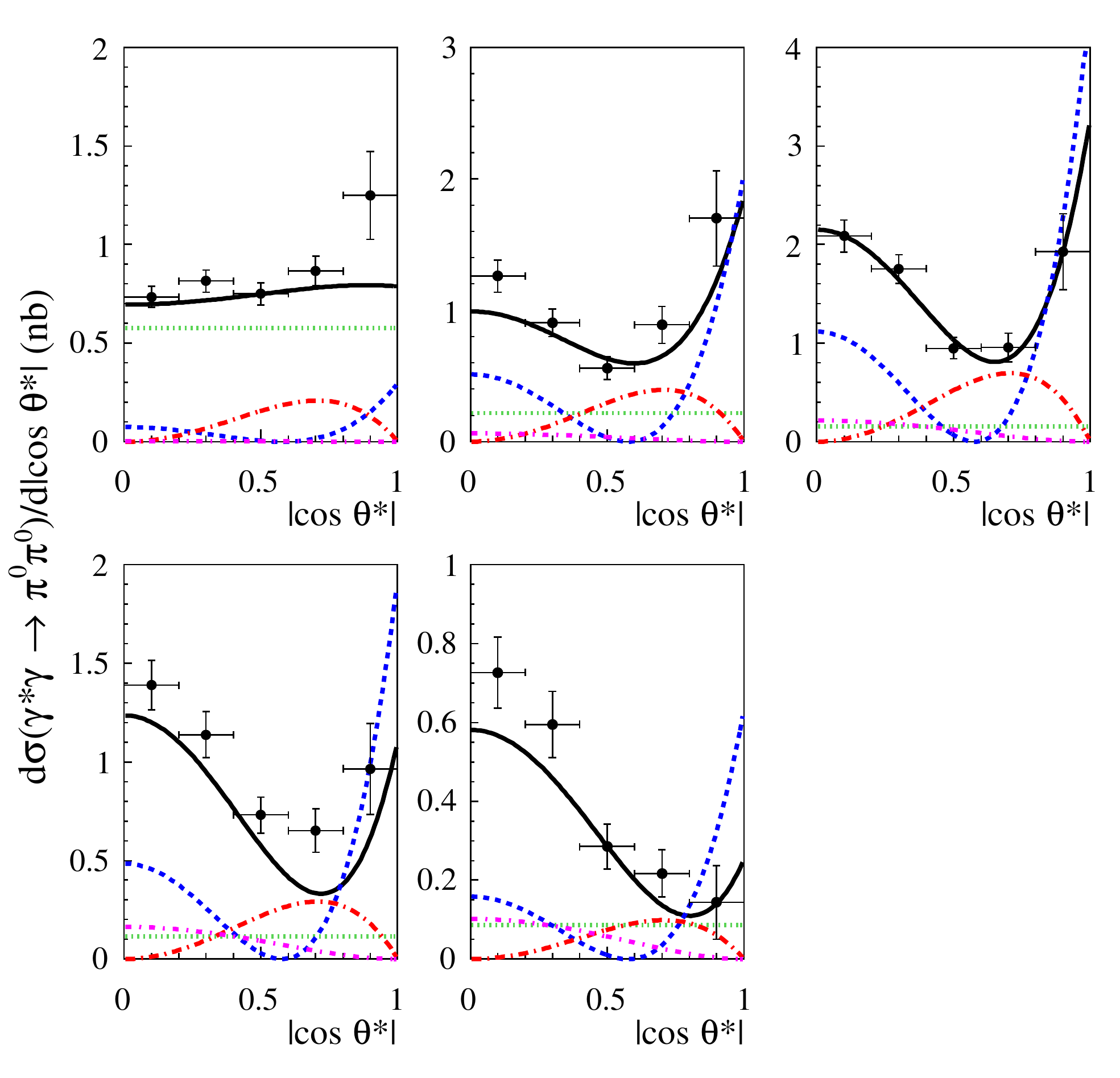,width=77mm}}
 \caption{$|\cos \theta^*|$ dependence in
five $W$ bins: 0.9, 1.15, 1.25, 1.35, and 1.45~GeV from left to right
and top to bottom and fitted results
of the $\varphi^*$ nominal fit.
Legend of lines are: solid lines show the total (black), 
dotted $|S|^2$ (green), dashed $|D_0|^2$ (blue),
long dashed $|D_1|^2$ (red),
and dash-dotted $|D_2|^2$ (purple).
}
 \label{fig:dcos_w1}
\end{figure}

\begin{figure}
 \centering
  {\epsfig{file=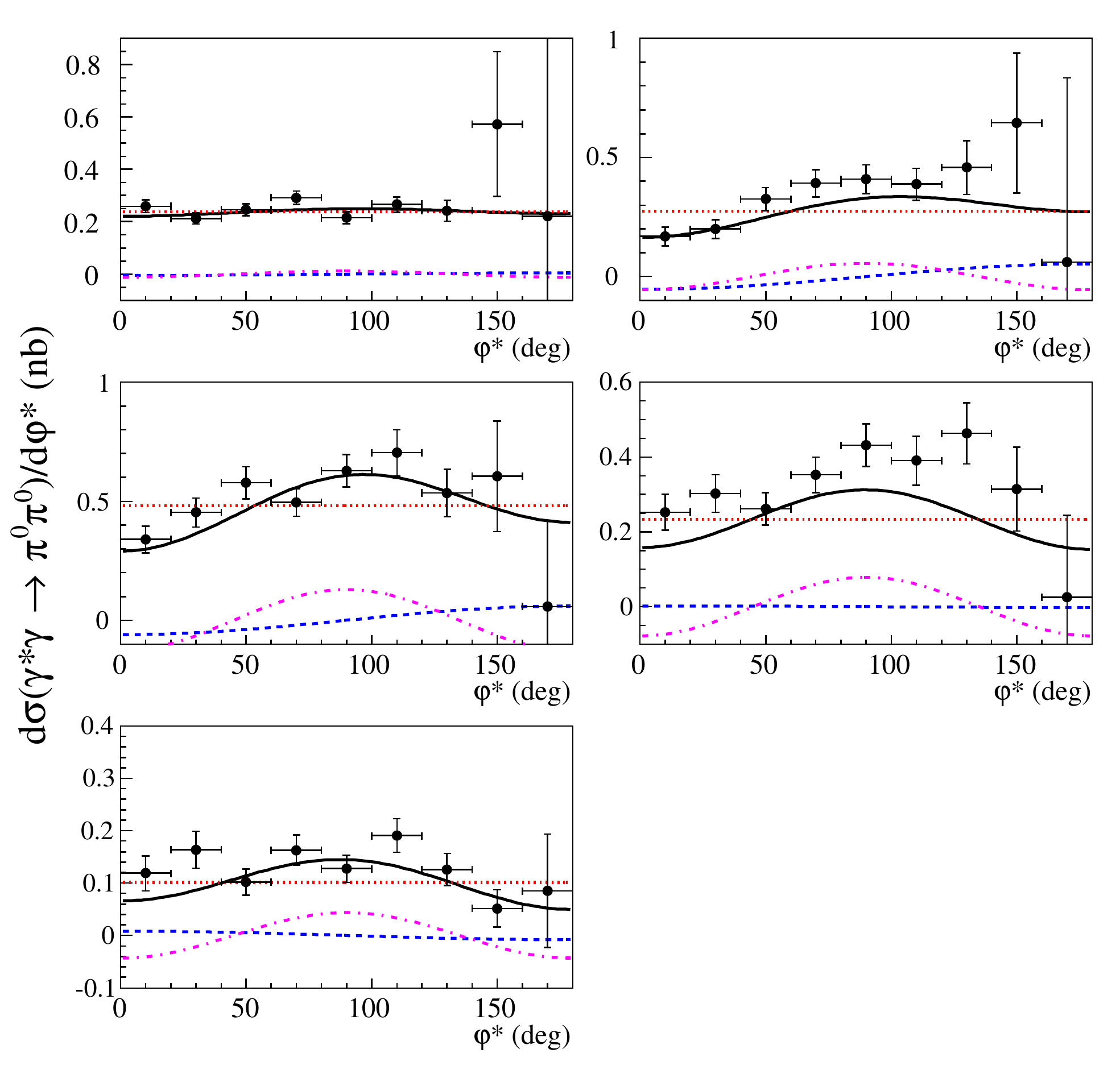,width=77mm}}
 \caption{$\varphi^*$ dependence in
five $W$ bins: 0.9, 1.15, 1.25, 1.35, and 1.45~GeV from left to right
and top to bottom and fitted results
of $\varphi^*$ nominal fit.
Legend of lines: solid lines for the total (black),
dotted $t_0$ (red), dashed $t_1 \cos \varphi^*$ (blue),
and dash-dotted $t_2 \cos 2 \varphi^*$ (purple).
}
\label{fig:dphi_w1}
\end{figure}

\begin{figure}
 \centering
  {\epsfig{file=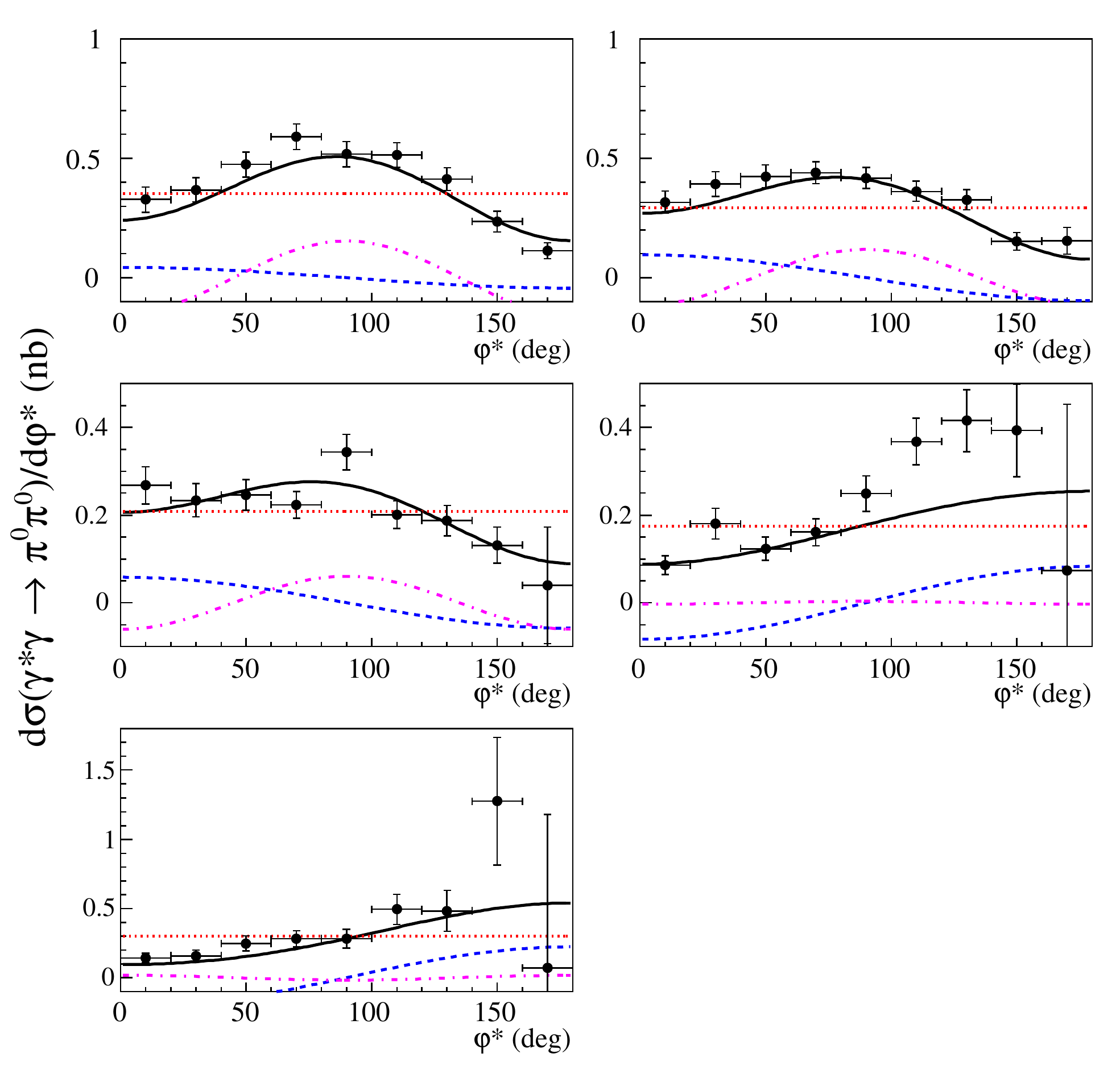,width=77mm}}
 \caption{$\varphi^*$ dependence in
five $|\cos \theta^*|$ bins: 0.1, 0.3, 0.5, 0.7, and 0.9
from left to right and top to bottom and fitted results
of $\varphi^*$ nominal fit.
The lines are defined in Fig.~\ref{fig:dphi_w1}.
}
 \label{fig:fig32v}
\end{figure}

\begin{figure}
 \centering
  {\epsfig{file=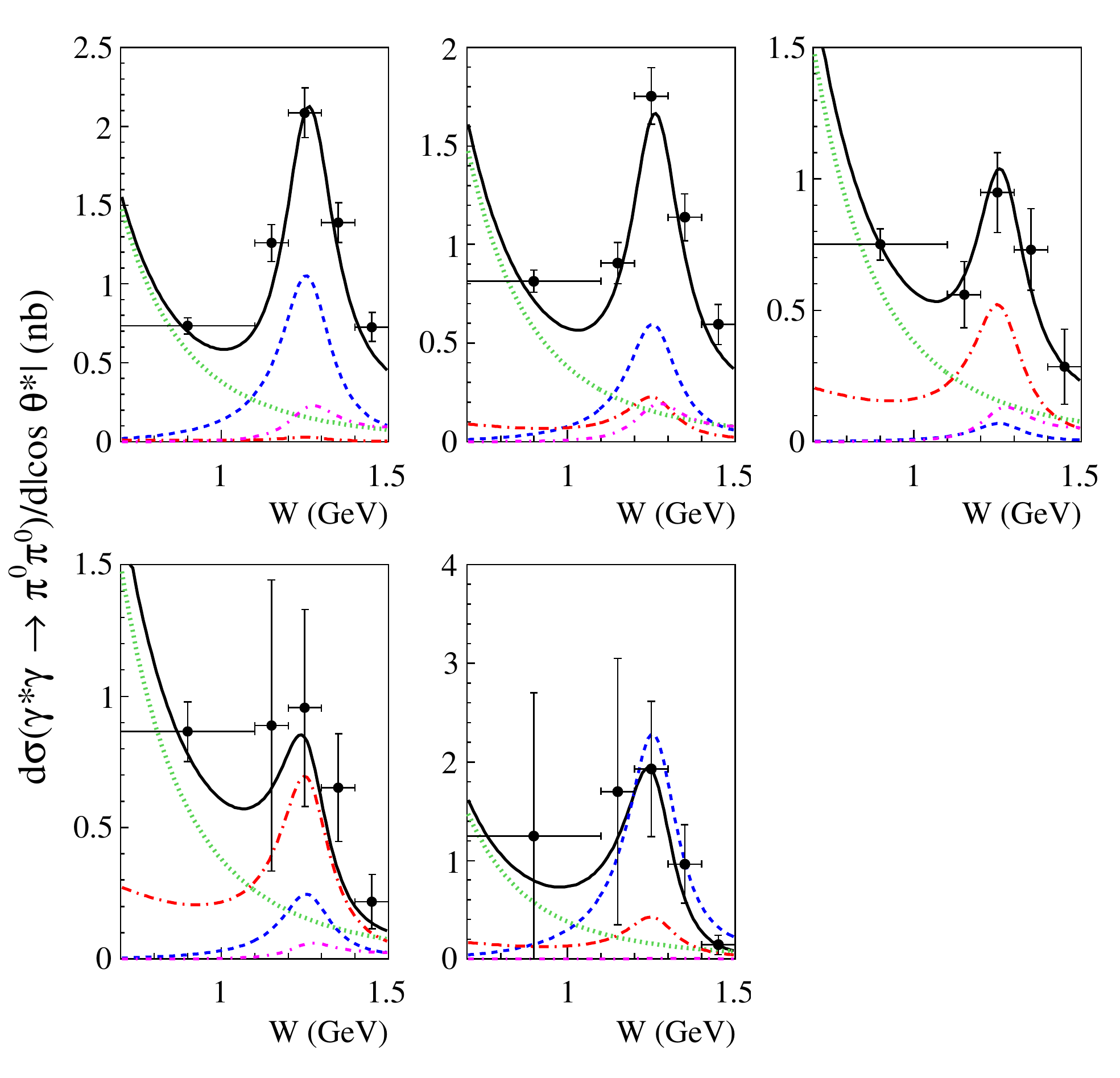,width=77mm}}
 \caption{$W$ dependence in
five $|\cos \theta^*|$ bins: 0.1, 0.3, 0.5, 0.7, and 0.9
from left to right and top to bottom and fitted results
of $\varphi^*$ nominal fit.
The lines are defined in Fig.~\ref{fig:dcos_w1}.
}
\label{fig:wcos_w1}
\end{figure}

\begin{figure}
 \centering
  {\epsfig{file=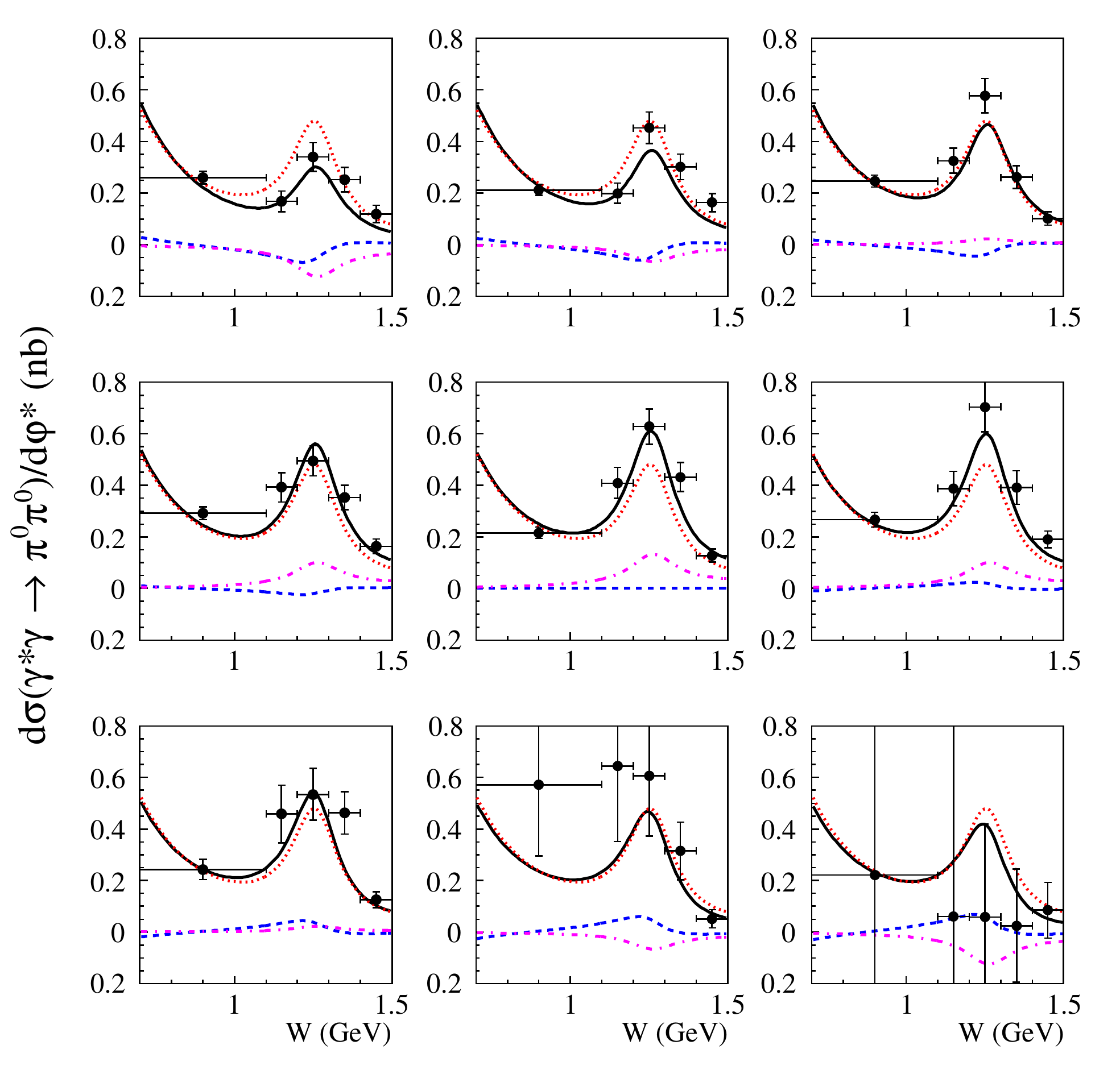,width=77mm}}
 \caption{$W$ dependence in
nine $\varphi^*$ bins: $10^\circ$ to $170^\circ$ in $20^\circ$ step
from left to right
and top to bottom and fitted results of $\varphi^*$ nominal fit.
The lines are defined in Fig.~\ref{fig:dphi_w1}.
}
 \label{fig:wphi_w1}
\end{figure}

Two solutions are obtained for $B_{D2} \ne 0$ where neither
$r_0$ nor $r_1$ is zero.
The favored solution that has a smaller $\chi^2$
(referred to as a ``$\varphi^*$ nominal fit'') has 
a small $B_{D2}$, which naturally gives almost identical values with 
respect to the fit with $B_{D2}=0$.
The fitted values of $F_{f2}(Q^2_{\rm av})$ are rather similar for 
all the fits.
The fitted fraction of the $f_2(1270)$ in the 
D$_0$ wave ($r_0$)
is large; the assumption of $r_0=0$ is disfavored as it gives 
a much worse $\chi^2$.
If, in addition, $r_1$ is fitted, the obtained
value is $r_1 = 0.15^{+0.05}_{-0.03}$ for the $\varphi^*$ nominal fit,
whose value is used in the ``$r_1$ fit'' described in 
Sec.~\ref{sub:fitres}.
In the second solution, $B_{D2}$ interferes destructively with the 
$f_2(1270)$ giving a smaller value of 
$r_1 = 0.11 \pm 0.03$.
Figures~\ref{fig:dcos_w1} -- \ref{fig:wphi_w1} show the fitted results 
of the $\varphi^*$ nominal fit that are projected onto the variable plotted
and integrated over the other variables.

\begin{center}
\begin{table*}
\caption{Fitted parameters of the $\varphi^*$-dependent cross section.
Here, $Q^2_{\rm av}=9.6~\GeV^2$.}
\label{tab:phifit}
\begin{tabular}{l|cccc|cccc} \hline \hline
Parameter & \multicolumn{4}{c|}
{$B_{D2} \ne 0$} 
& \multicolumn{4}{c}{$B_{D2} = 0$} \\ 
condition & \multicolumn{2}{c}{$r_0 \ne 0 \bigcap r_1 \ne 0$}
 & $r_0 = 0$ & $r_1 = 0$
& $r_0 \ne 0 \bigcap r_1 \ne 0$ & \multicolumn{2}{c}{$r_0 = 0$} & $r_1 = 0$ \\
\hline
No. sol.& \multicolumn{2}{c}{2} & 1 & 1 & 1 & \multicolumn{2}{c}{2} & 1 \\
& Sol.1 & Sol.2 & & & & Sol.1 & Sol.2 & \\
$\chi^2/ndf$ & 236.6/208 & 243.1/208 & 357.4/210 & 289.6/210 & 241.3/211 & 357.5/213 & 366.0/213 & 308.0/213 \\
\hline
$F_{f2}(Q^2_{\rm av})$($\times 10^{-2}$) & $1.67 \pm 0.15$ & $1.76 \pm 0.11$ & $1.68^{+0.05}_{-0.06}$ & $1.46 \pm 0.09$ & $1.70 \pm 0.08$ & $1.68^{+0.05}_{-0.06}$ & $1.65 \pm 0.06$ & $1.41 \pm 0.06$ \\
\hline
$a_{D2}~(\sqrt{\rm nb})$ & $0.03 \pm 0.03$ & $0.17^{+0.04}_{-0.05}$ & $0\pm 0.003$ & $0.23^{+0.02}_{-0.03}$ & \multicolumn{4}{c}{0 (fixed)} \\
$b_{D2}$ & $-3.8^{+2.6}_{-8.8}$ & $4.0^{+1.9}_{-1.0}$ & $-4.6 \pm 7.9$ & $2.3^{+0.6}_{-0.5}$ & \multicolumn{4}{c}{--} \\
$\phi_{BD2}~(^\circ)$ & $89 \pm 123$ & $114^{+17}_{-16}$ & $198 \pm 1$ & $101 \pm 13$ & \multicolumn{4}{c}{--} \\
\hline
$r_0$ & $0.76 \pm 0.06$ & $0.80^{+0.04}_{-0.05}$ & 0 (fixed) & $0.81^{+0.06}_{-0.07}$ & $0.70 \pm 0.04$ & \multicolumn{2}{c}{0 (fixed)} & $0.68^{+0.06}_{-0.07}$ \\
$a_{D0}~(\sqrt{\rm nb})$ & $0.13 \pm 0.03$ & $-0.13^{+0.04}_{-0.03}$ & $0.06 \pm 0.02$ & $0.15 \pm 0.03$ & $-0.13 \pm 0.02$ & $0.06 \pm 0.02$ & $0.23^{+0.02}_{-0.03}$ & $0.16 \pm 0.02$ \\
$b_{D0}$ & $-1.0^{+0.9}_{-1.5}$ & $-0.8^{+0.9}_{-1.4}$ & $-1.5^{+1.3}_{-1.7}$ & $-0.6^{+0.8}_{-1.1}$ & $-0.6^{+0.8}_{-1.0}$ & $-1.5^{+1.3}_{-1.7}$ & $1.5^{+0.6}_{-0.4}$ & $-0.3 \pm 0.6$ \\
$\phi_{BD0}~(^\circ)$ & $186^{+22}_{-22}$ & $350 \pm 21$ & $300^{+18}_{-20}$ & $161 \pm 15$ & $171 \pm 12$ & $300^{+18}_{-20}$ & $271 \pm 9$ & $151^{+10}_{-9}$ \\
$\phi_{f2D0}~(^\circ)$ & $161^{+18}_{-18}$ & $160^{+19}_{-16}$ & -- & $144^{+14}_{-13}$ & $20 \pm 12$ & \multicolumn{2}{c}{--} & $184 \pm 9$ \\
\hline
$r_1$ & $0.15^{+0.05}_{-0.03}$ & $0.11 \pm 0.03$ & $0.03^{+0.03}_{-0.02}$ & 0 (fixed) & $0.16 \pm 0.03$ & $0.03^{+0.03}_{-0.02}$ & $0.07 \pm 0.04$ & 0 (fixed) \\
$a_{D1}~(\sqrt{\rm nb})$ & $-0.11 \pm 0.03$ & $0.15^{+0.02}_{-0.03}$ & $-0.09 \pm 0.03$ & $0.21 \pm 0.01$ & $-0.10 \pm 0.02$ & $0.09 \pm 0.03$ & $0.15 \pm 0.02$ & $0.17 \pm 0.01$ \\
$b_{D1}$ & $1.8^{+1.7}_{-1.7}$ & $3.4^{+0.9}_{-0.7}$ & $2.9^{+1.9}_{-1.3}$ & $1.9^{+0.3}_{-0.2}$ & $2.1^{+1.2}_{-1.1}$ & $2.9^{+1.9}_{-1.3}$ & $4.3^{+0.9}_{-0.9}$ & $1.3^{+0.3}_{-0.3}$ \\
$\phi_{BD1}~(^\circ)$ & $2 \pm 5$ & $225^{+21}_{-20}$ & $166 \pm 10$ & $209 \pm 17$ & $140 \pm 12$ & $345 \pm 10$ & $356 \pm 8$ & $180 \pm 9$ \\
$\phi_{f2D1}~(^\circ)$ & $148 \pm 19$ & $155^{+20}_{-19}$ & $278 \pm 29$ & -- & $163 \pm 13$ & $28^{+23}_{-36}$ & $69^{+11}_{-21}$ & -- \\
\hline
$a_S~(\sqrt{\rm nb})$ & $-0.30 \pm 0.03$ & $0.16 \pm 0.02$ & $0.29 \pm 0.02$ & $0.16 \pm 0.02$ & $0.30 \pm 0.01$ & $0.29 \pm 0.02$ & $0.18 \pm 0.02$ & $0.32 \pm 0.01$ \\
$b_S$ & $2.0 \pm 0.6$ & $-0.6^{+0.4}_{-0.5}$ & $2.2^{+0.3}_{-0.3}$ & $-0.1^{+0.4}_{-0.4}$ & $1.9^{+0.2}_{-0.2}$ & $2.2^{+0.3}_{-0.3}$ & $0.6^{+0.4}_{-0.4}$ & $1.4^{+0.2}_{-0.2}$ \\
$\phi_{BS}~(^\circ)$ & $262 \pm 19$ & $73^{+15}_{-12}$ & $76^{+6}_{-6}$ & $60 \pm 8$ & $95^{+12}_{-11}$ & $76^{+6}_{-6}$ & $74 \pm 7$ & $80^{+8}_{-7}$ \\
\hline \hline
\end{tabular}
\end{table*}
\end{center}
%

\subsection{Fitted results for the $\varphi^*$-integrated cross section}
\label{sub:fitres}
We fit the $\varphi^*$-integrated differential cross section, Eq.~(\ref{eqn:dsdcos}), 
in the $W$ region $0.7~\GeV \le W \le 1.5~\GeV$ for all $Q^2$
with Eq.~(\ref{eqn:param}) and Eq.~(\ref{eqn:para2}).
In the fit, the usual $\chi^2$ is replaced by its corresponding quantity
using the Poisson likelihood $\lambda$
defined as~\cite{BC}
\begin{equation}
\chi_{\rm P}^2 \equiv \ln \lambda 
= 2 \sum_i \left[ p_i - n_i + n_i \ln \left( \frac{n_i}{p_i} 
\right) \right] ,
\label{eqn:plike}
\end{equation}
where, $n_i$ and $p_i$ are the numbers of events observed
and predicted in the $i$-th bin and the sum is over the bins
within the fitted $W$ range.
We fit the predicted number of events in each bin that is related 
to the differential cross section. 
The latter is converted to the number of events by multiplying 
by a known conversion factor given from 
Eqs.~(\ref{eqn:cscomb}) and (\ref{eqn:csgg}).  
Here, we include zero-event 
bins in calculating $\chi_{\rm P}^2$ in Eq.~(\ref{eqn:plike}).
In fitting with Eq.~(\ref{eqn:plike}), systematic uncertainties in the cross section
are not taken into account. 
Their effects are considered separately in the study of 
systematic uncertainties in Sec.~\ref{sub:systff}.

The TFFs for the $f_0(980)$ and $f_2(1270)$ and the fractions 
$r_i(Q^2)$ are floated in each $Q^2$ bin,
{\it i.e.}, $F_{f0}(Q^2)$, $F_{f2}(Q^2)$, and $r_i(Q^2)$ 
are obtained in each $Q^2$ bin.

Because of the limitation mentioned above, 
we cannot determine 
$D_0$ and $D_1$ simultaneously together with $S$ and $D_2$. 

Here, we prioritize the determination of $D_0$ and $D_2$ over $D_1$
by setting $B_{D1}=0$ and
\begin{equation}
r_1(Q^2) = r_1(Q^2_{\rm av}) \left( \frac{Q^2}{Q^2_{\rm av}} \right)^d,
\label{eqn;r1q2}
\end{equation}
where $Q^2_{\rm av} = 9.6~\GeV^2$ is the average $Q^2$ value
for the $Q^2$-integrated cross section and $r_1(Q^2_{\rm av}) = 0.15$
determined in the $\varphi^*$ nominal fit, and $d$ is a free parameter.
The effect of setting $B_{D1}=0$ is considered in systematic studies.
In this ``$r_1$ fit,'' we can obtain information 
on the helicity-1 TFF simultaneously with those of helicity-0 and -2
in spite of the limitation.

The issue with the $f_2(1270)$ is determination of its fractions in $D_0$, $D_1$, 
and $D_2$ in each $Q^2$ bin.
Thus, in addition to the $r_1$ fit, we perform fits assuming either 
$D_0 \ne 0$ with $D_1=0$ (denoted as the $D_0$ fit) or 
$D_1 \ne 0$ with $D_0=0$ (denoted as the $D_1$ fit).
In each category, we put either $B_{D2} \ne 0$ or $B_{D2} = 0$.
In the $D_1$ fit, the evaluation of $\epsilon_0$ is necessary.
It turns out that it has little dependence on $W$ and $\cos \theta^*$
and we use the values listed in Table~\ref{tab:eps0}
for $Q^2$ bins from 3.5 to 25.0 $\GeV^2$.  
We also use the average $Q^2$ value given in Table~\ref{tab:repreq2}
in each $Q^2$ bin in the fitting.

The $r_1$ fit with $B_{D2} \ne 0$ (also with $B_{D2} = 0$)
gives a unique solution
as summarized in Table~\ref{tab:d0r1fit}.
In the $D_0$ fit, two solutions each are obtained for both
 $B_{D2} \ne 0$ and $B_{D2} =0$
as also summarized in Table~\ref{tab:d0r1fit}.
In the $D_1$ fit with $B_{D2} \ne 0$, no solutions are obtained, 
{\it i.e.}, all solutions have one or more powers beyond the limit.
Also, in the $D_1$ fit with $B_{D2} = 0$, two solutions were found
with $\chi_{\rm P, min}^2/ndf = 719.7/504$, which is much worse compared
to that of the $D_0$ fit.

\begin{figure*}
 \centering
  {\epsfig{file=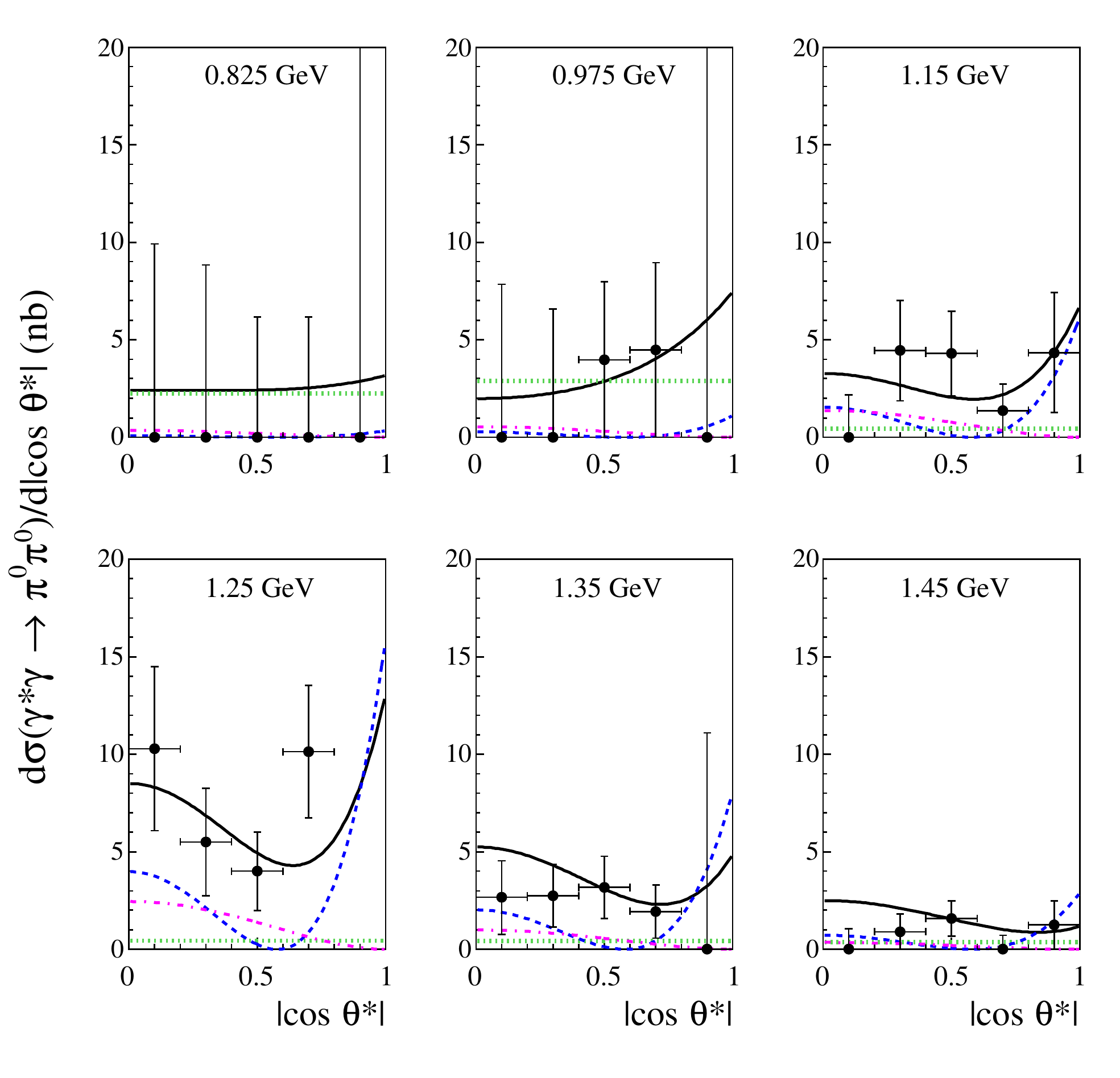,width=77mm}}
  \hspace{6mm}
  {\epsfig{file=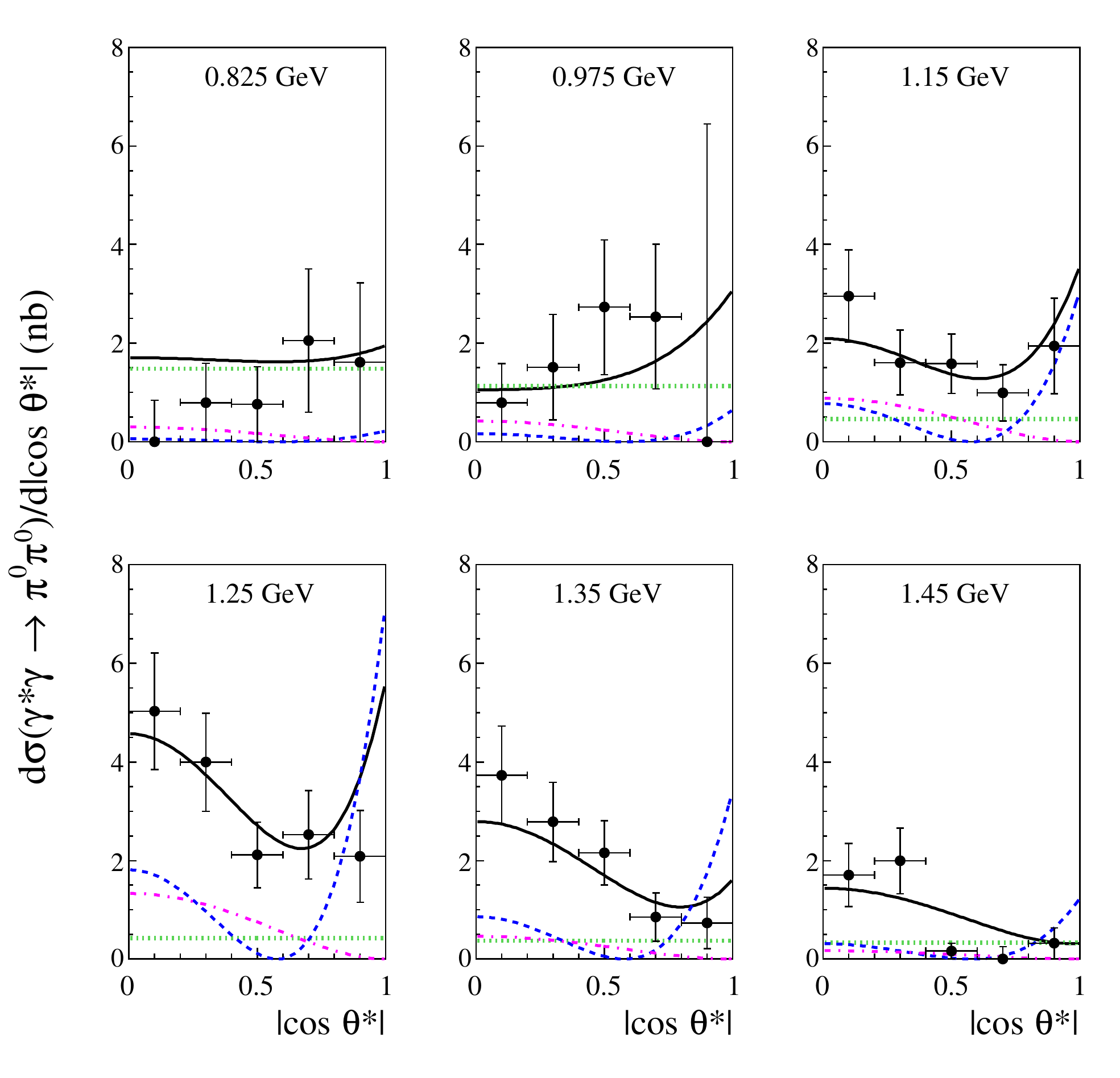,width=77mm}}
 \caption{Angular dependence of the cross section in the indicated $W$ bin
and results of the $r_1$ fit
at $Q^2=3.5~\GeV^2$ (left) and $Q^2=4.5~\GeV^2$ (right).
Legend of lines: solid lines for the total (black), 
dotted $|S|^2$ (green), 
dashed $|D_0|^2$ (blue), and dash-dotted $|D_2|^2$ (purple). 
}
 \label{fig:ds1r1}
\end{figure*}

\begin{figure*}

 \centering
  {\epsfig{file=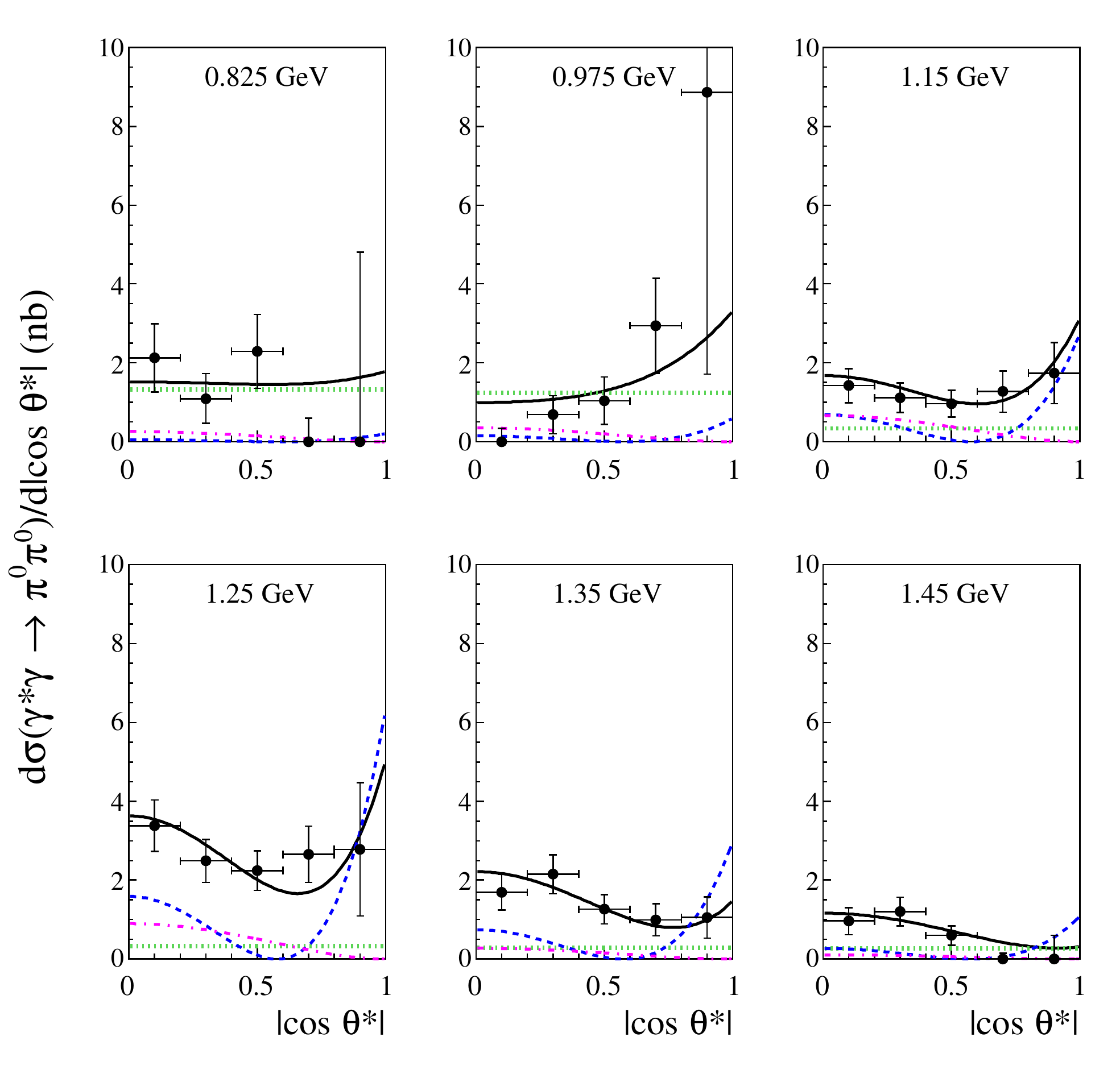,width=77mm}}
  \hspace{6mm}
  {\epsfig{file=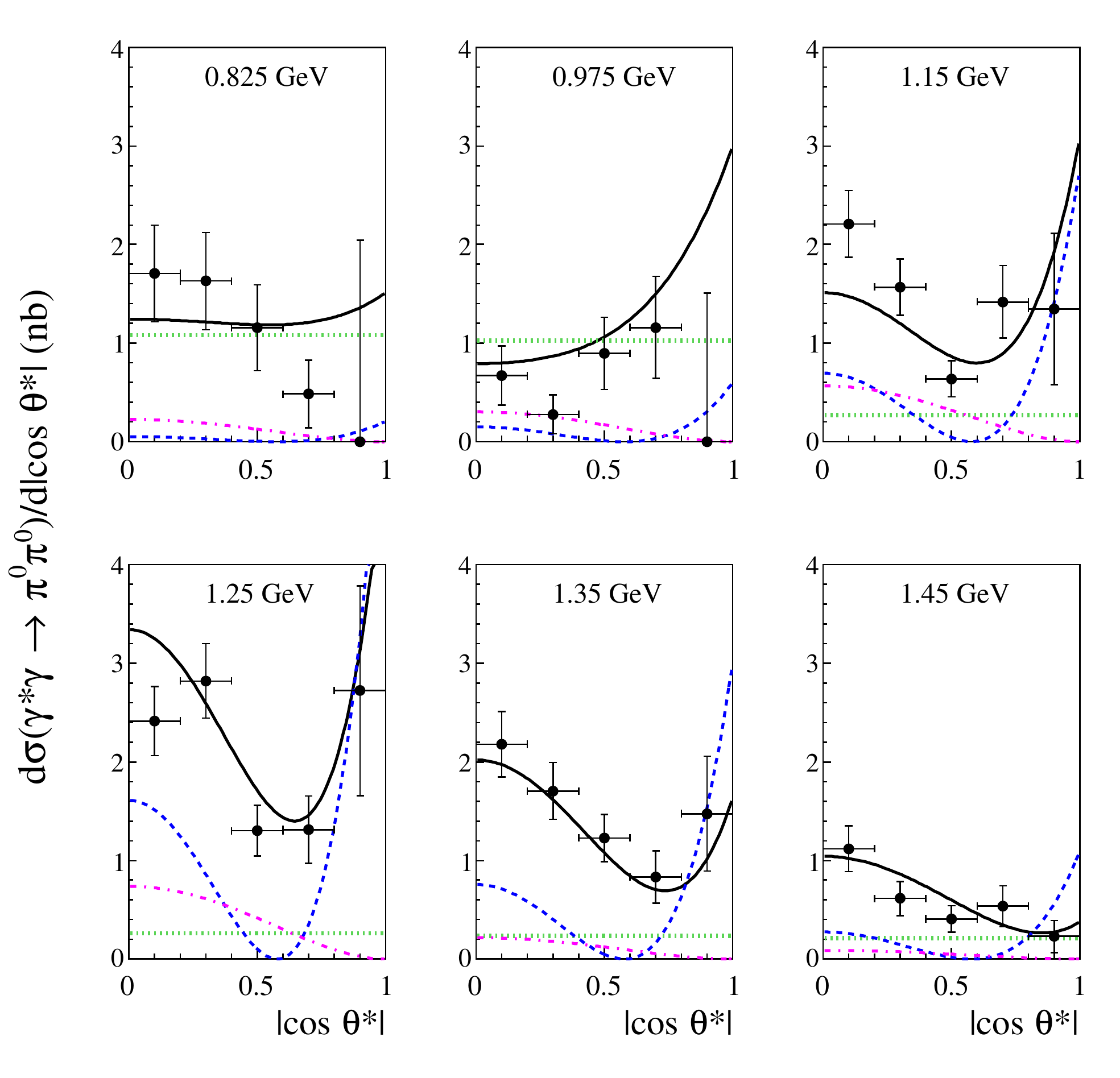,width=77mm}}
 \caption{Angular dependence of the cross section in the indicated $W$ bin
and results of the $r_1$ fit
at $Q^2=5.5~\GeV^2$ (left) and $Q^2=7.0~\GeV^2$ (right).
The lines are defined in Fig.~\ref{fig:ds1r1}.
}
 \label{fig:ds3r1}
\end{figure*}

\begin{figure*}
 \centering
  {\epsfig{file=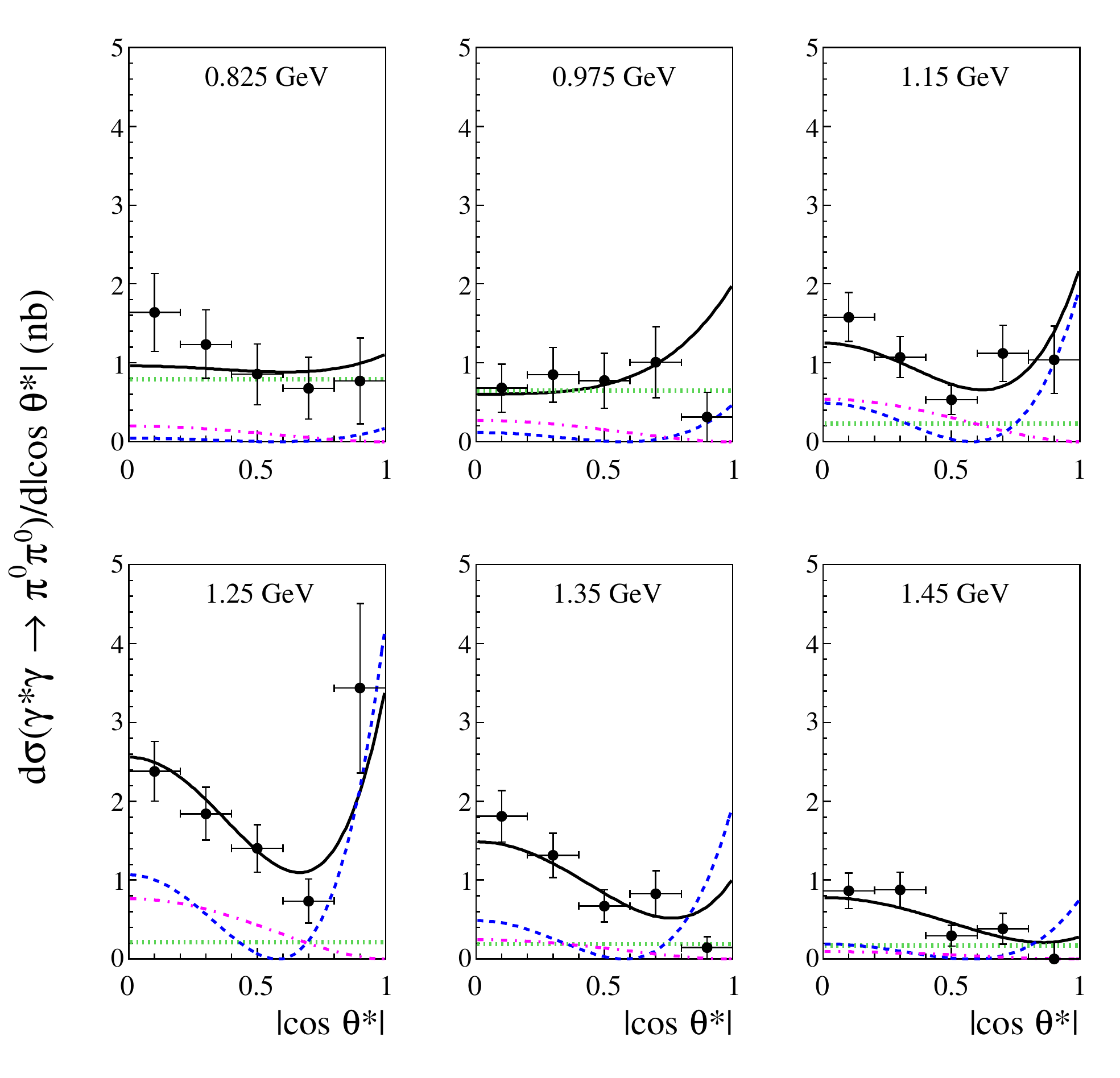,width=77mm}}
  \hspace{6mm}
  {\epsfig{file=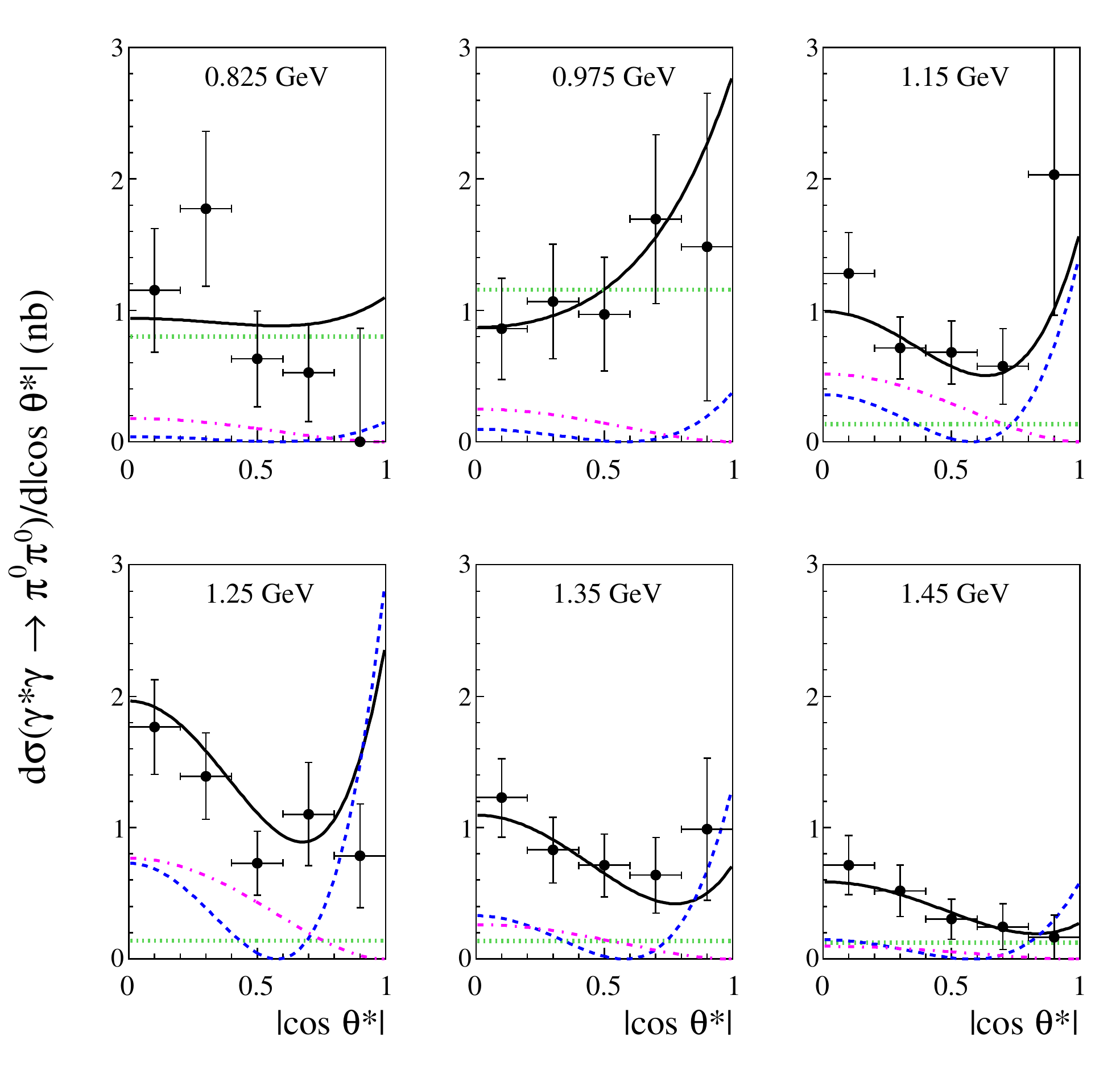,width=77mm}}
 \caption{Angular dependence of the cross section in the indicated $W$ bin
and results of the $r_1$ fit
at $Q^2=9.0~\GeV^2$ (left) and $Q^2=11.0~\GeV^2$ (right).
The lines are defined in Fig.~\ref{fig:ds1r1}.
}
 \label{fig:ds5r1}
\end{figure*}

\begin{figure*}
 \centering
  {\epsfig{file=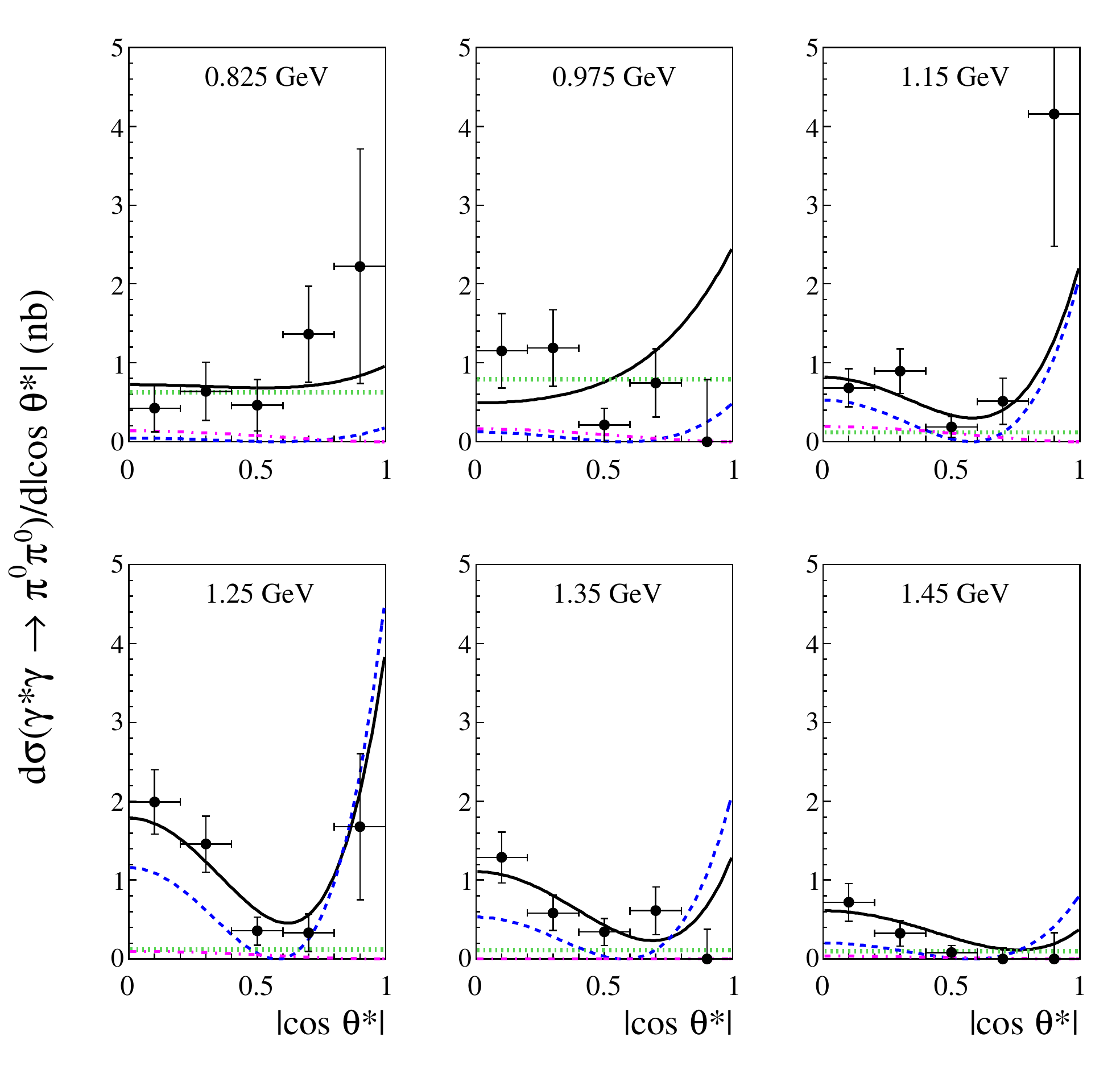,width=77mm}}
  \hspace{6mm}
  {\epsfig{file=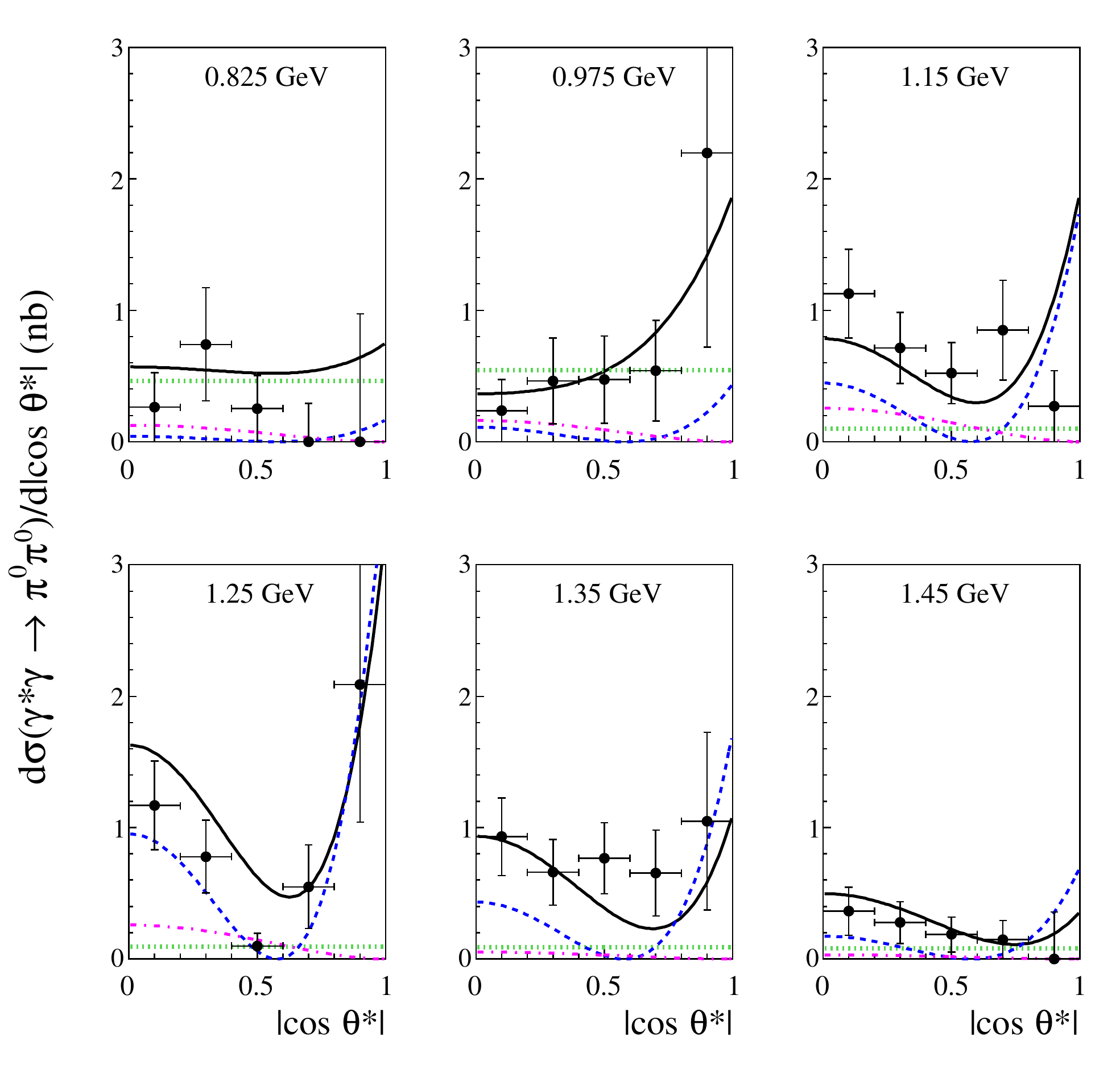,width=77mm}}
 \caption{Angular dependence of the cross section in the indicated $W$ bin
and results of the $r_1$ fit
at $Q^2=13.5~\GeV^2$ (left) and $Q^2=17.5~\GeV^2$ (right).
The lines are defined in Fig.~\ref{fig:ds1r1}.
}
 \label{fig:ds7r1}
\end{figure*}

\begin{figure}
 \centering
  {\epsfig{file=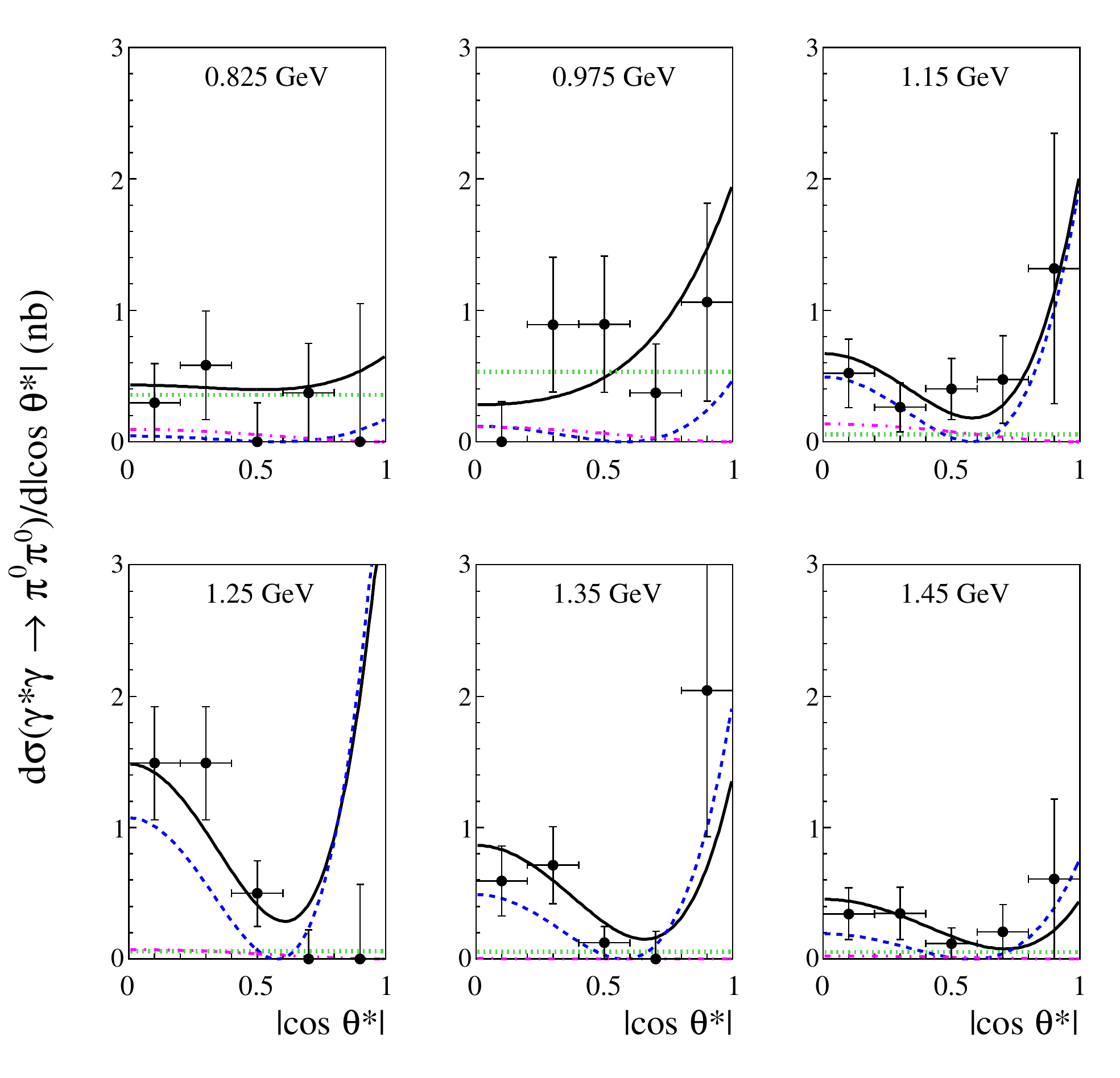,width=80mm}}
 \caption{Angular dependence of the cross section in the indicated $W$ bin
and results of the $r_1$ fit at $Q^2=25~\GeV^2$.
The lines are defined in Fig.~\ref{fig:ds1r1}.
}
 \label{fig:ds9r1}
\end{figure}

\begin{figure}
 \centering
  {\epsfig{file=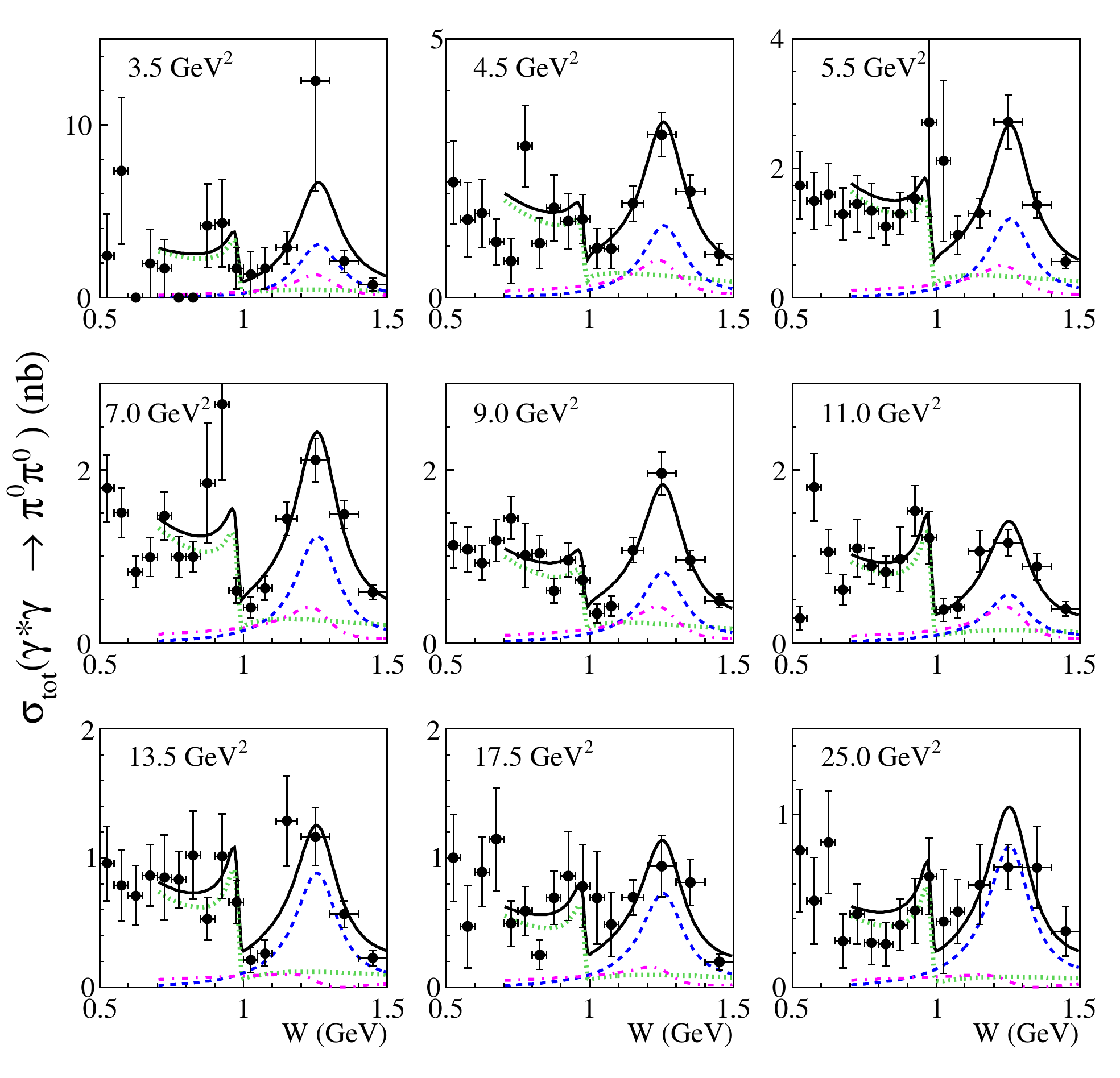,width=80mm}}
 \caption{Integrated cross section 
in the indicated $Q^2$ bin and results of the $r_1$ fit.
The lines are defined in Fig.~\ref{fig:ds1r1}.
}
 \label{fig:totr1}
\end{figure}

As a result, 
it is apparent that there is a significant helicity-0 component of 
the $f_2(1270)$ in two-photon production when one of the photons 
is highly virtual.
The results obtained in the $\varphi^*$-dependent fit that
$B_{D2}$ is non-zero and $r_1$ is non-zero are also strongly supported by the $\varphi^*$-integrated fit. 
The results of the $r_1$ fit with $B_{D2} \ne 0$ 
(denoted as the ``$r_1$ nominal fit'')
give a unique solution with the minimum $\chi_{\rm P}^2$ and  
are shown in Figs.~\ref{fig:ds1r1} -- \ref{fig:totr1}.
Figures~\ref{fig:ds1r1} -- \ref{fig:ds9r1} 
show the angular dependence in selected $W$ bins in each $Q^2$ bin 
while Fig.~\ref{fig:totr1} 
shows the integrated cross section in each $Q^2$ bin.

To test the contribution of the $f_0(980)$,
we perform a fit in which the $f_0(980)$-TFF is set to zero.
This results in $\chi_{\rm P}^2/ndf=650.2/511$ to be compared to $572.4/501$
for the $r_1$ nominal fit.
The significance of the $f_0(980)$ contribution is $7.1 \sigma$.
This strongly supports the signature of the $f_0(980)$ and
the validity of its TFF measurement. 

\begin{center}
\begin{table*}
\caption{
Fitted parameters of the $\varphi^*$-integrated cross section,
where TFF parameters, $F_{f2}(Q^2)$, $r_0(Q^2)$, and $F_{f0}(Q^2)$, are 
obtained at each $Q^2$ bin (in GeV$^2$).
}
\label{tab:d0r1fit}
\begin{tabular}{l|cc|cccc} \hline \hline
Parameter & \multicolumn{2}{c|}{$r_1$ fit} 
& \multicolumn{4}{c}{$D_0$ fit} \\ \cline{2-7}
 & $B_{D2} \ne 0$ & $B_{D2} = 0$ &
\multicolumn{2}{c}{$B_{D2} \ne 0$} & \multicolumn{2}{c}{$B_{D2} = 0$} \\
& &  & Sol.1 & Sol.2 & Sol.1 & Sol.2 \\
\hline
$\chi_{\rm P}^2/ndf$  & 572.4/501 & 689.9/505 & 589.2/502 & 591.0/502 & 621.2/506 & 622.1/506 \\
\hline
$F_{f2}(0.0);(\times 10^{-2})$ 
& \multicolumn{6}{c}{ $100 \pm 6$ (def.)} \\
$F_{f2}(3.5);(\times 10^{-2})$ & $11.5 \pm 1.2$ & $12.6^{+1.3}_{-1.2}$ & $10.9^{+1.6}_{-1.5}$ & $11.6 \pm 1.5$ & $13.6^{+1.5}_{-1.4}$ & $13.3^{+1.3}_{-1.3}$ \\
$F_{f2}(4.5);(\times 10^{-2})$ & $7.2^{+0.7}_{-0.8}$ & $7.0 \pm 0.4$ & $7.0^{+1.2}_{-0.9}$ & $7.6 \pm 1.0$ & $7.5 \pm 0.5$ & $7.6 \pm 0.5$ \\
$F_{f2}(5.5);(\times 10^{-2})$ & $5.9^{+0.5}_{-0.6}$ & $5.8 \pm 0.4$ & $5.5^{+1.0}_{-0.8}$ & $6.0^{+0.9}_{-0.8}$ & $6.6 \pm 0.4$ & $6.5 \pm 0.4$ \\
$F_{f2}(7.0);(\times 10^{-2})$ & $5.1^{+0.3}_{-0.4}$ & $4.8 \pm 0.3$ & $4.8^{+0.8}_{-0.7}$ & $5.2^{+0.7}_{-1.4}$ & $5.8 \pm 0.3$ & $5.7 \pm 0.3$ \\
$F_{f2}(9.0);(\times 10^{-2})$ & $4.0^{+0.3}_{-0.4}$ & $3.8 \pm 0.2$ & $3.8^{+0.7}_{-0.6}$ & $4.1^{+0.6}_{-1.2}$ & $4.4 \pm 0.2$ & $4.3 \pm 0.2$ \\
$F_{f2}(11.0);(\times 10^{-2})$ & $3.3^{+0.3}_{-0.4}$ & $3.0 \pm 0.2$ & $3.0^{+0.6}_{-0.5}$ & $3.3^{+0.6}_{-1.1}$ & $3.6 \pm 0.3$ & $3.4 \pm 0.3$ \\
$F_{f2}(13.5);(\times 10^{-2})$ & $2.5 \pm 0.3$ & $2.5^{+0.3}_{-0.2}$ & $2.7^{+0.5}_{-0.4}$ & $2.9 \pm 0.5$ & $3.4 \pm 0.3$ & $3.3 \pm 0.3$ \\
$F_{f2}(17.5);(\times 10^{-2})$ & $2.3 \pm 0.3$ & $2.3 \pm 0.3$ & $2.2 \pm 0.4$ & $2.4^{+0.4}_{-0.5}$ & $2.9 \pm 0.2$ & $2.8 \pm 0.3$ \\
$F_{f2}(25.0);(\times 10^{-2})$ & $1.8 \pm 0.2$ & $1.4 \pm 0.2$ & $1.8 \pm 0.3$ & $2.0 \pm 0.4$ & $2.5 \pm 0.2$ & $1.6^{+1.2}_{-0.2}$ \\
$\phi_{f2}$($^{\circ}$) & $64^{+14}_{-32}$ & -- & $5^{+53}_{-46}$ & $32^{+30}_{-89}$ & \multicolumn{2}{c}{--} \\
\hline
$r_0(3.5);(\%)$ & $45.1^{+21.8}_{-22.1}$ & $51.0^{+14.9}_{-16.5}$ & $37.0^{+18.4}_{-18.1}$ & $33.5^{+16.7}_{-14.9}$ & $56.6^{+12.0}_{-14.2}$ & $52.9^{+12.1}_{-13.8}$ \\
$r_0(4.5);(\%)$ & $37.6^{+26.3}_{-15.1}$ & $2.0^{+4.9}_{-2.0}$ & $17.8^{+11.7}_{-8.5}$ & $15.8^{+10.0}_{-7.1}$ & $11.7^{+7.8}_{-6.5}$ & $10.3^{+7.3}_{-5.9}$ \\
$r_0(5.5);(\%)$ & $41.7^{+15.2}_{-14.0}$ & $13.8^{+8.9}_{-7.5}$ & $28.0^{+13.7}_{-10.7}$ & $25.6^{+12.3}_{-9.1}$ & $31.2^{+8.7}_{-9.1}$ & $28.8^{+8.6}_{-8.7}$ \\
$r_0(7.0);(\%)$ & $46.3^{+12.9}_{-12.1}$ & $6.1^{+6.3}_{-4.6}$ & $24.6^{+11.8}_{-9.0}$ & $22.5^{+10.6}_{-7.4}$ & $31.4^{+7.7}_{-8.2}$ & $28.5^{+7.9}_{-8.2}$ \\
$r_0(9.0);(\%)$ & $35.8^{+14.0}_{-11.8}$ & $4.7^{+5.4}_{-3.7}$ & $21.4^{+11.9}_{-8.8}$ & $20.3^{+10.9}_{-7.7}$ & $21.8^{+7.5}_{-7.2}$ & $18.9^{+7.4}_{-6.8}$ \\
$r_0(11.0);(\%)$ & $26.2^{+17.1}_{-14.2}$ & $7.5^{+8.1}_{-5.6}$ & $21.1^{+15.2}_{-10.9}$ & $20.7^{+13.6}_{-9.4}$ & $28.3^{+9.9}_{-10.4}$ & $22.5^{+11.1}_{-10.7}$ \\
$r_0(13.5);(\%)$ & $71.3 \pm 19.3$ & $10.9^{+10.6}_{-8.2}$ & $52.5^{+20.0}_{-18.1}$ & $40.2^{+18.1}_{-14.3}$ & $47.9^{+12.1}_{-13.2}$ & $43.0^{+12.8}_{-13.8}$ \\
$r_0(17.5);(\%)$ & $53.8^{+21.8}_{-21.0}$ & $26.0^{+13.5}_{-13.2}$ & $44.5^{+21.8}_{-18.1}$ & $37.7^{+19.9}_{-14.7}$ & $55.6^{+9.9}_{-11.1}$ & $52.6^{+11.0}_{-13.2}$ \\
$r_0(25.0);(\%)$ & $75.3 \pm 24.5$ & $0.1 \pm 4.3$ & $59.9^{+21.6}_{-23.0}$ & $49.7^{+21.4}_{-18.6}$ & $62.4^{+10.6}_{-12.9}$ & $3.7 \pm 7.7$ \\
$\phi_{f2D0}$($^{\circ}$) & $59^{+10}_{-11}$ & $296^{+10}_{-11}$ & $359 \pm 10$ & $348^{+28}_{-7}$ & $300^{+6}_{-7}$ & $304^{+5}_{-6}$ \\
\hline
$d$ in $r_1(Q^2)$ & $-0.2 \pm 0.3$ & $0.00 \pm 0.04$ & -- & -- & -- & -- \\
\hline
$F_{f0}(0.0);(\times 10^{-2})$ 
& \multicolumn{6}{c}{ $100 \pm 11$ (def.)} \\
$F_{f0}(3.5);(\times 10^{-2})$ & $18.6^{+8.1}_{-8.4}$ & $11.8^{+7.0}_{-7.1}$ & $16.2^{+8.4}_{-11.8}$ & $15.7^{+8.3}_{-10.8}$ & $10.9^{+7.2}_{-7.8}$ & $11.5^{+7.2}_{-7.8}$ \\
$F_{f0}(4.5);(\times 10^{-2})$ & $7.6 \pm 4.9$ & $3.8^{+2.7}_{-3.0}$ & $8.7^{+4.0}_{-6.2}$ & $8.3^{+3.9}_{-5.6}$ & $3.0 \pm 3.2$ & $3.4 \pm 3.2$ \\
$F_{f0}(5.5);(\times 10^{-2})$ & $8.8 \pm 2.2$ & $3.0^{+1.9}_{-2.1}$ & $7.8^{+2.4}_{-2.6}$ & $7.8^{+2.3}_{-2.4}$ & $2.9^{+2.1}_{-2.4}$ & $3.2^{+2.1}_{-2.4}$ \\
$F_{f0}(7.0);(\times 10^{-2})$ & $7.4 \pm 1.5$ & $3.0 \pm 1.1$ & $7.2 \pm 1.5$ & $7.2^{+1.4}_{-1.4}$ & $3.6 \pm 1.1$ & $3.7 \pm 1.1$ \\
$F_{f0}(9.0);(\times 10^{-2})$ & $4.7 \pm 1.5$ & $2.1 \pm 0.9$ & $5.2^{+1.4}_{-1.5}$ & $5.1^{+1.4}_{-1.5}$ & $2.8 \pm 1.0$ & $2.7 \pm 1.0$ \\
$F_{f0}(11.0);(\times 10^{-2})$ & $7.4^{+1.2}_{-1.2}$ & $4.0 \pm 0.9$ & $7.4 \pm 1.1$ & $7.3 \pm 1.1$ & $4.8 \pm 0.9$ & $4.6 \pm 0.9$ \\
$F_{f0}(13.5);(\times 10^{-2})$ & $5.4^{+1.2}_{-1.4}$ & $2.3 \pm 0.8$ & $5.5 \pm 1.1$ & $5.5^{+1.1}_{-1.2}$ & $3.2 \pm 0.8$ & $3.0 \pm 0.8$ \\
$F_{f0}(17.5);(\times 10^{-2})$ & $3.9 \pm 1.0$ & $0.7 \pm 0.8$ & $3.3^{+1.0}_{-1.2}$ & $3.4^{+1.0}_{-1.1}$ & $1.8 \pm 0.9$ & $1.3 \pm 0.9$ \\
$F_{f0}(25.0);(\times 10^{-2})$ & $3.5^{+0.8}_{-0.9}$ & $1.0 \pm 0.8$ & $3.1^{+0.9}_{-1.0}$ & $3.3^{+0.8}_{-0.9}$ & $1.9 \pm 0.8$ & $1.4^{+0.8}_{-0.9}$ \\
$\phi_{f0}$($^{\circ}$) & $23\pm 10$ & $342^{+14}_{-16}$ & $30^{+9}_{-10}$ & $28^{+9}_{-10}$ & $356^{+11}_{-12}$ & $357^{+11}_{-12}$ \\
\hline
$a_{S}$($\sqrt{\rm nb}$) & $2.1^{+0.5}_{-0.4}$ & $1.8 \pm 0.3$ & $2.4^{+0.5}_{-0.4}$ & $2.4^{+0.4}_{-0.3}$ & $2.1 \pm 0.3$ & $2.0 \pm 0.3$ \\
$b_{S}$ & $1.1 \pm 0.2$ & $1.8 \pm 0.2$ & $0.8 \pm 0.2$ & $0.8^{+0.5}_{-0.1}$ & $1.5 \pm 0.1$ & $1.5 \pm 0.1$ \\
$c_{S}$ & $0.6 \pm 0.1$ & $0.5 \pm 0.1$ & $0.6 \pm 0.1$ & $0.5 \pm 0.1$ & $0.5 \pm 0.1$ & $0.5 \pm 0.1$ \\
\hline
$a_{D0}$($\sqrt{\rm nb}$) & $0.2 \pm 0.1$ & $0.5^{+0.1}_{0.0}$ & $0.2^{+0.4}_{-0.1}$ & $0.2^{+0.8}_{-0.1}$ & $0.7^{+0.6}_{-0.4}$ & $0.4 \pm 0.1$ \\
$b_{D0}$ & $-1.3^{+1.3}_{-1.1}$ & $-0.7^{+0.4}_{-0.4}$ & $-0.3 \pm 1.1$ & $1.7^{+1.6}_{-4.3}$ & $-2.2^{+1.2}_{-1.6}$ & $-1.6^{+0.7}_{-0.8}$ \\
$c_{D0}$ & $0.0 \pm 1.0$ & $0.0 \pm 0.1$ & $0.0^{+0.4}_{-0.4}$ & $0.0 \pm 0.8$ & $0.4 \pm 0.3$ & $0.0 \pm 0.1$ \\
$\phi_{D0}$($^{\circ})$ & $93^{+9}_{-10}$ & $271^{+4}_{-4}$ & $57^{+16}_{-19}$ & $95^{+8}_{-77}$ & $265^{+8}_{-10}$ & $272 \pm 5$ \\
\hline
$a_{D2}$($\sqrt{\rm nb}$) & $0.8^{+0.6}_{-0.4}$ & 0 (fixed)& $0.5^{+0.3}_{-0.2}$ & $0.4^{+0.3}_{-0.2}$ & 0 (fixed)& 0 (fixed)\\
$b_{D2}$ & $0.3^{+1.2}_{-1.2}$ & 0 (fixed)& $2.1^{+0.5}_{-0.9}$ & $-0.9^{+3.8}_{-1.4}$ & 0 (fixed)& 0 (fixed)\\
$c_{D2}$ & $0.3 \pm 0.3$ & 0 (fixed)& $0.2 \pm 0.2$ & $0.2 \pm 0.1$ & 0 (fixed)& 0 (fixed)\\
 \hline\hline
\end{tabular}
\end{table*}
\end{center}


\subsection{Estimation of systematic uncertainties of TFFs}
\label{sub:systff}
In this subsection, we estimate the systematic uncertainties for the TFFs of the
$f_0(980)$ and $f_2(1270)$.
They can be divided in two: the overall normalization uncertainty that
affects all $Q^2$ bins simultaneously and the individual uncertainties
that vary in each $Q^2$ bin.
The former arises from the uncertainty of 
$\Gamma_{\gamma \gamma}$ and amounts to $\pm$ 6\% for the $f_2(1270)$
and $\pm$ 11\% for the $f_0(980)$, respectively.

Individual uncertainties are estimated in each $Q^2$ bin
for the $f_2(1270)$-TFF, $r_0$, and $f_0(980)$-TFF.
The individual uncertainties considered include those related
to the differential cross section: normalization and distortion.
The normalization uncertainty $n$ in each bin of $|\cos \theta^*|$, 
$W$, and $Q^2$ varies between 0.104 and 0.253;
the systematic uncertainty from this source is estimated by
multiplying the cross section by the factor $1 \pm n$.
The systematic uncertainties from distortion
are estimated in each bin by multiplying the cross section: in $W$ 
by the factor 
$1 \pm 0.10(W-1.1~\GeV)$ 
($\pm 4$\% distortion in the full range) and
in $|\cos \theta^*|$ by the factor 
$1 \mp [(a+b)|\cos \theta^*| - a$], where $a$ ($b$) has values
$0.01-0.09$ ($0.02-0.16$) depending on $W$ and $Q^2$
($^{+a}_{-b}$ or $^{-a}_{+b}$ distortion in the full range).
The values of $n,a,b$ are estimated in Sec.~\ref{sub:sysunc}.
The other individual uncertainties considered are from the fitted 
$W$ range, from the parameterization and from assumed constants.
The fitted $W$ ranges are changed to $0.65-1.4$~GeV or to $0.75-1.6$~GeV.
The parameterization of the background $B_i$ is changed to 
$W/(1.1~\pm~0.2)~\GeV$ or to $Q^2/(1.0~\pm~0.5)~\GeV^2$.
We also set $B_1 \ne 0$ to study its effect.
The properties of the $f_0(980)$ and $f_2(1270)$ are changed 
using the world-average values~\cite{pdg2012}.
Also $r_R$ in the $f_2(1270)$ parameterization is varied by its error.
We have also taken into account the uncertainty of 
$r_1(Q^2_{\rm av}) $ obtained in the $\varphi^*$-dependent analysis, 
which is evaluated by setting the distortion of ${\pm 4\%}$ in $\varphi^*$.
The estimated systematic uncertainty of $r_1(Q^2_{\rm av})$ is much smaller than
the statistical uncertainties obtained in the fit.
Thus, we set $r_1(Q^2_{\rm av})= 0.15 \pm 0.05$.
A shift of the $f_0(980)$ mass (($980 \pm 20)~\MeV/c^2$) gives rise to a
large effect due to its proximity to the $K \bar{K}$ threshold.
All the uncertainties are summed in quadrature in each $Q^2$ bin.
For the $f_2(1270)$-TFF, systematic uncertainties for the 
helicity-0, -1, and -2 components 
are calculated from the values of $f_2(1270)$-TFF, $r_0$, and
$d$ in $r_1$ (Eq.~(\ref{eqn;r1q2})).
The results are summarized in Table~\ref{tab:tfff0f2}.

\begin{center}
\begin{table*}
\caption{
Resulting transition form factors for the $f_0(980)$
and $f_2(1270)$ ($\times 10^{-2}$).
The first and second errors are statistical and systematic, respectively.
}
\label{tab:tfff0f2}
\begin{tabular}{c|c|cccc} \hline \hline
 & $f_0(980)$-TFF & \multicolumn{4}{c}{$f_2(1270)$-TFF} \\
 $\bar{Q}^2$ & & Hel. = 0 & Hel. = 1 & Hel. = 2 & Total \\ \hline \hline
0 & 100 (def.) & $18.9^{+0.6}_{-0.7}$$^{+21.6}_{-9.8}$ & -- 
& $98.2 \pm 0.1^{+1.4}_{-6.5}$ & 100 (def.) \\ \hline 
3.45 & $18.7^{+8.1}_{-8.4}$$^{+5.6}_{-2.0}$ & $7.7^{+2.0}_{-2.1}$$^{+2.4}_{-0.8}$ & $5.0 \pm 0.9$$^{+0.9}_{-0.5}$ & $7.2 \pm 2.2$$^{+0.8}_{-2.6}$ & $11.5 \pm 1.2$$ \pm 1.0$ \\
4.46 & $7.6^{+4.9}_{-4.9}$$^{+4.2}_{-0.6}$ & $4.4^{+1.6}_{-1.0}$$^{+0.6}_{-1.6}$ & $3.1^{+0.4}_{-0.5}$$^{+0.5}_{-0.9}$ & $4.9^{+0.9}_{-1.5}$$^{+1.1}_{-1.6}$ & $7.2^{+0.7}_{-0.8}$$^{+0.9}_{-2.2}$ \\
5.47 & $8.8 \pm 2.2$$^{+3.3}_{-1.0}$ & $3.8 \pm 0.8$$^{+1.1}_{-0.9}$ & $2.5 \pm 0.3$$^{+0.4}_{-0.5}$ & $3.9^{+0.7}_{-0.8}$$^{+0.6}_{-1.3}$ & $5.9^{+0.5}_{-0.6}$$^{+0.7}_{-1.2}$ \\
6.89 & $7.4 \pm 1.5$$^{+2.3}_{-0.8}$ & $3.5^{+0.6}_{-0.5}$$^{+0.9}_{-0.7}$ & $2.1 \pm 0.2$$^{+0.3}_{-0.4}$ & $3.2 \pm 0.6$$^{+0.5}_{-1.1}$ & $5.1^{+0.3}_{-0.4}$$^{+0.6}_{-0.9}$ \\
8.92 & $4.5 \pm 1.5$$^{+1.2}_{-0.5}$ & $2.4 \pm 0.5$$ \pm 0.7$ & $1.6^{+0.1}_{-0.2}$$^{+0.2}_{-0.4}$ & $2.8^{+0.4}_{-0.5}$$^{+0.4}_{-1.0}$ & $4.0^{+0.3}_{-0.4}$$^{+0.5}_{-1.1}$ \\
10.93 & $7.4 \pm 1.2$$^{+1.8}_{-0.7}$ & $1.7^{+0.6}_{-0.5}$$^{+1.4}_{-0.5}$ & $1.3 \pm 0.1$$^{+0.1}_{-0.3}$ & $2.5^{+0.4}_{-0.5}$$^{+0.3}_{-1.1}$ & $3.3^{+0.3}_{-0.4}$$^{+0.4}_{-0.8}$ \\
13.37 & $5.4^{+1.2}_{-1.4}$$^{+0.9}_{-0.6}$ & $2.1 \pm 0.4$$^{+0.4}_{-0.2}$ & $0.9 \pm 0.1$$^{+0.2}_{-0.1}$ & $0.9 \pm 0.7$$^{+0.3}_{-0.4}$ & $2.5 \pm 0.3$$^{+0.5}_{-0.2}$ \\
17.23 & $3.9 \pm 1.0$$^{+1.0}_{-0.5}$ & $1.7 \pm 0.4$$^{+0.6}_{-0.2}$ & $0.8 \pm 0.1$$ \pm 0.1$ & $1.3 \pm 0.5$$^{+0.3}_{-0.8}$ & $2.3 \pm 0.3$$^{+0.3}_{-0.2}$ \\
24.25 & $3.5^{+0.8}_{-0.9}$$^{+0.7}_{-0.5}$ & $1.6 \pm 0.3$$^{+0.3}_{-0.2}$ & $0.6 \pm 0.1$$ \pm 0.1$ & $0.6 \pm 0.7$$^{+0.4}_{-0.5}$ & $1.8 \pm 0.2$$^{+0.4}_{-0.1}$ \\
\hline \hline
\end{tabular}
\end{table*}
\end{center}

\subsection{$Q^2$ dependence of TFFs for the $f_2(1270)$ and
$f_0(980)$}
\begin{figure}
 \centering
  {\epsfig{file=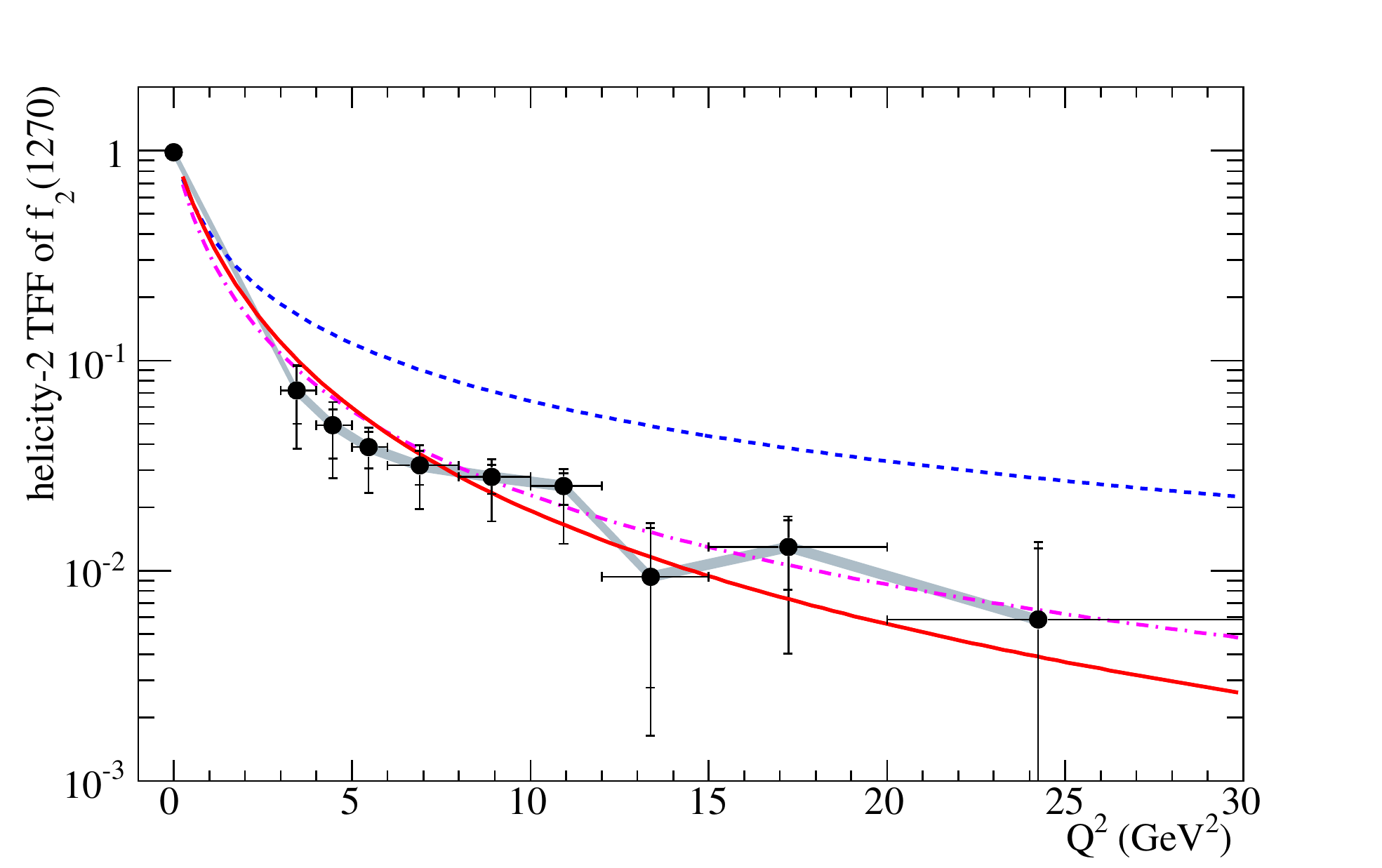,width=80mm}}
 \caption{
The measured helicity-2 TFF of the $f_2(1270)$
as a function of $Q^2$.
Short (long) vertical bars indicate statistical (statistical and systematic
combined) errors.
The shaded area corresponds to the overall systematic uncertainty
arising from that of $\Gamma_{\gamma \gamma}$.
Correlations of uncertainties between neighboring bins exist
and are included in the long vertical bars. 
The solid line shows the predicted $Q^2$ dependence in
Table~\ref{tab:pred} by Ref.~\cite{schuler}
and those by Ref.~\cite{ppv}: Eq.~(\ref{eqn:pseudo}) (dashed line)
and Eq.~(\ref{eqn:axial}) (dot-dashed line).
The theoretical curves for the helicity-2 TFF are normalized to 1 at $Q^2$ = 0.
}
 \label{fig:f22q2}
\end{figure}

The obtained $Q^2$ dependence of TFF for the $f_2(1270)$
for the $r_1$ nominal fit is shown in Fig.~{\ref{fig:f22q2}
for the helicity-2 component.}
Also shown are the predicted $Q^2$ dependence by Ref.~\cite{schuler}
and those from Ref.~\cite{ppv} (Eqs.~(\ref{eqn:pseudo}) 
and (\ref{eqn:axial})).
The theoretical prediction of Ref.~\cite{schuler}
and of Eq.~(\ref{eqn:axial}) in Ref.~\cite{ppv}  
agree well with data.

\begin{figure}
 \centering
  {\epsfig{file=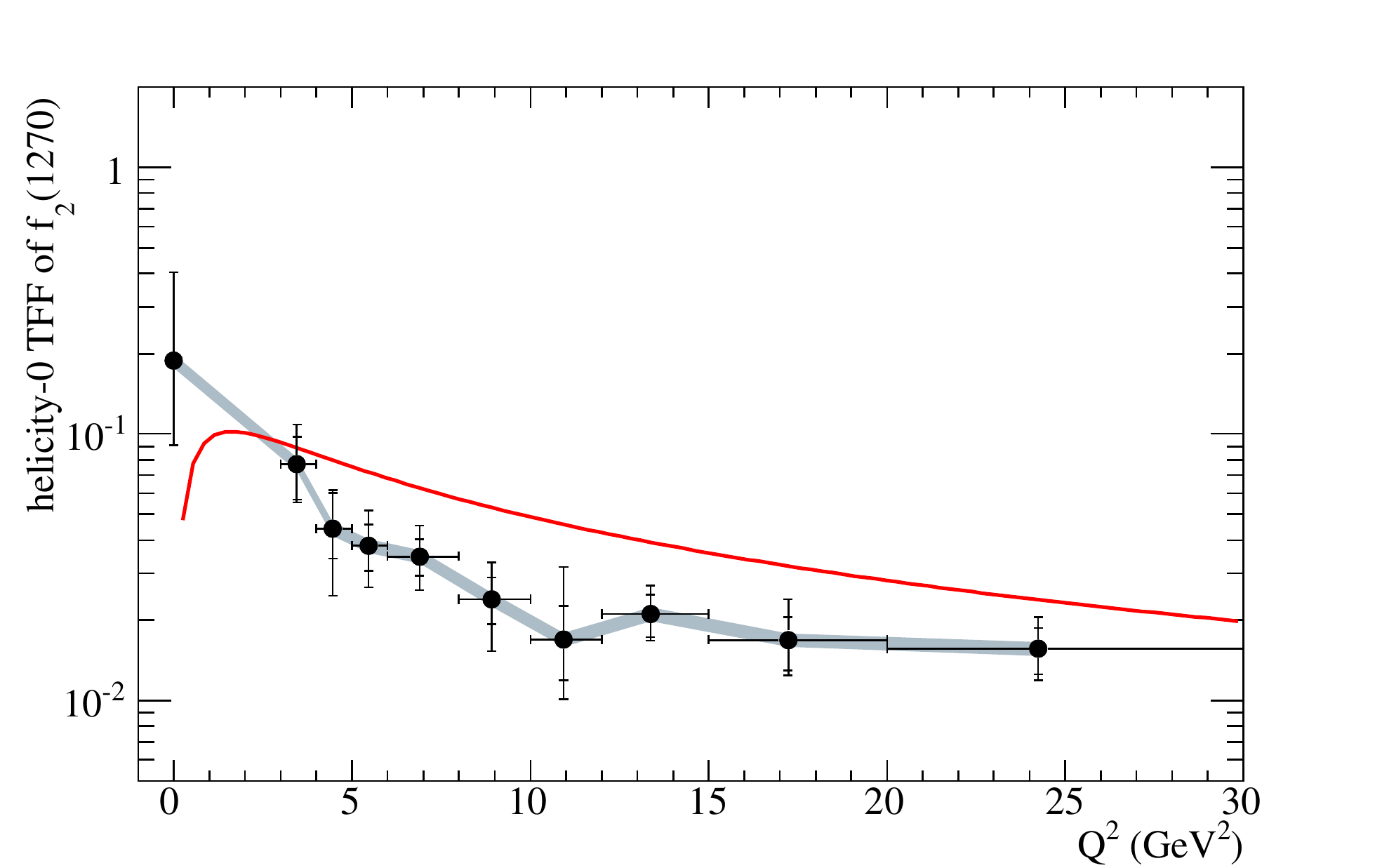,width=80mm}}
 \caption{
The measured helicity-0 TFF of the $f_2(1270)$
as a function of $Q^2$.
Short (long) vertical bars indicate statistical (statistical and systematic
combined) errors.
The shaded area corresponds to the overall systematic uncertainty
arising from that of $\Gamma_{\gamma \gamma}$.
Correlations of uncertainties between neighboring bins exist
and are included in the long vertical bars. 
The solid line shows the predicted $Q^2$ dependence in
Table~\ref{tab:pred} by Ref.~\cite{schuler}.}
\label{fig:f20q2}
\end{figure}

\begin{figure}
 \centering
  {\epsfig{file=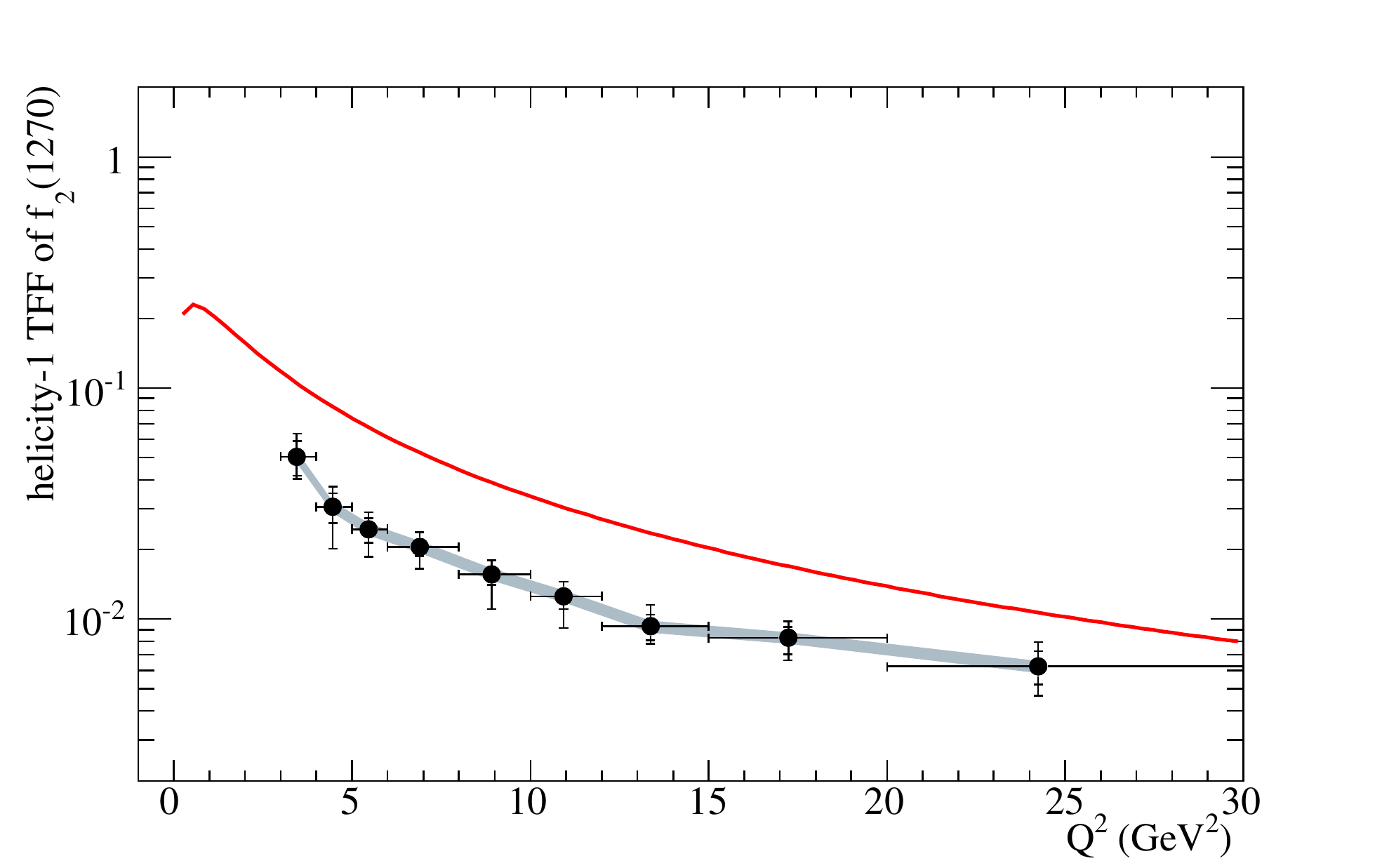,width=80mm}}
 \caption{
The measured helicity-1 TFF of the $f_2(1270)$
as a function of $Q^2$.
Short (long) vertical bars indicate statistical (statistical and systematic
combined) errors.
The shaded area corresponds to the overall systematic uncertainty
arising from that of $\Gamma_{\gamma \gamma}$.
Correlations of uncertainties between neighboring bins exist
and are included in the long vertical bars. 
Note that the helicity-1 TFF is calculated from $\sqrt{r_1(Q^2)} F_{f2}(Q^2)$,
where $r_1(Q^2)$ is given by Eq.~(\ref{eqn;r1q2}).
The solid line shows the predicted $Q^2$ dependence in
Table~\ref{tab:pred} by Ref.~\cite{schuler}.}
\label{fig:f21q2}
\end{figure}

Figure \ref{fig:f20q2} (\ref{fig:f21q2})
shows the helicity-0 (helicity-1) TFF obtained
for the $f_2(1270)$ together with the prediction~\cite{schuler}.
Note that the helicity-1 TFF is calculated from $\sqrt{r_1(Q^2)} F_{f2}(Q^2)$,
where $r_1(Q^2)$ is given by Eq.~(\ref{eqn;r1q2}).
The measured value calculated from $r_0$ at $Q^2=0$, 
$r_0 (0) = (3.56^{+0.22}_{-0.27}$$^{+12.81}_{-2.74})$\%~\cite{pi0pi0}, 
is also plotted in Fig.~\ref{fig:f20q2}, 
while Ref.~\cite{schuler} predicts zero at $Q^2=0$ (Table~\ref{tab:pred}).
The first and second errors of $r_0 (0) $ are statistical and systematic, respectively.
The prediction is 
a factor of $1.5-2$ larger 
than the measured values.


\begin{figure}
 \centering
  {\epsfig{file=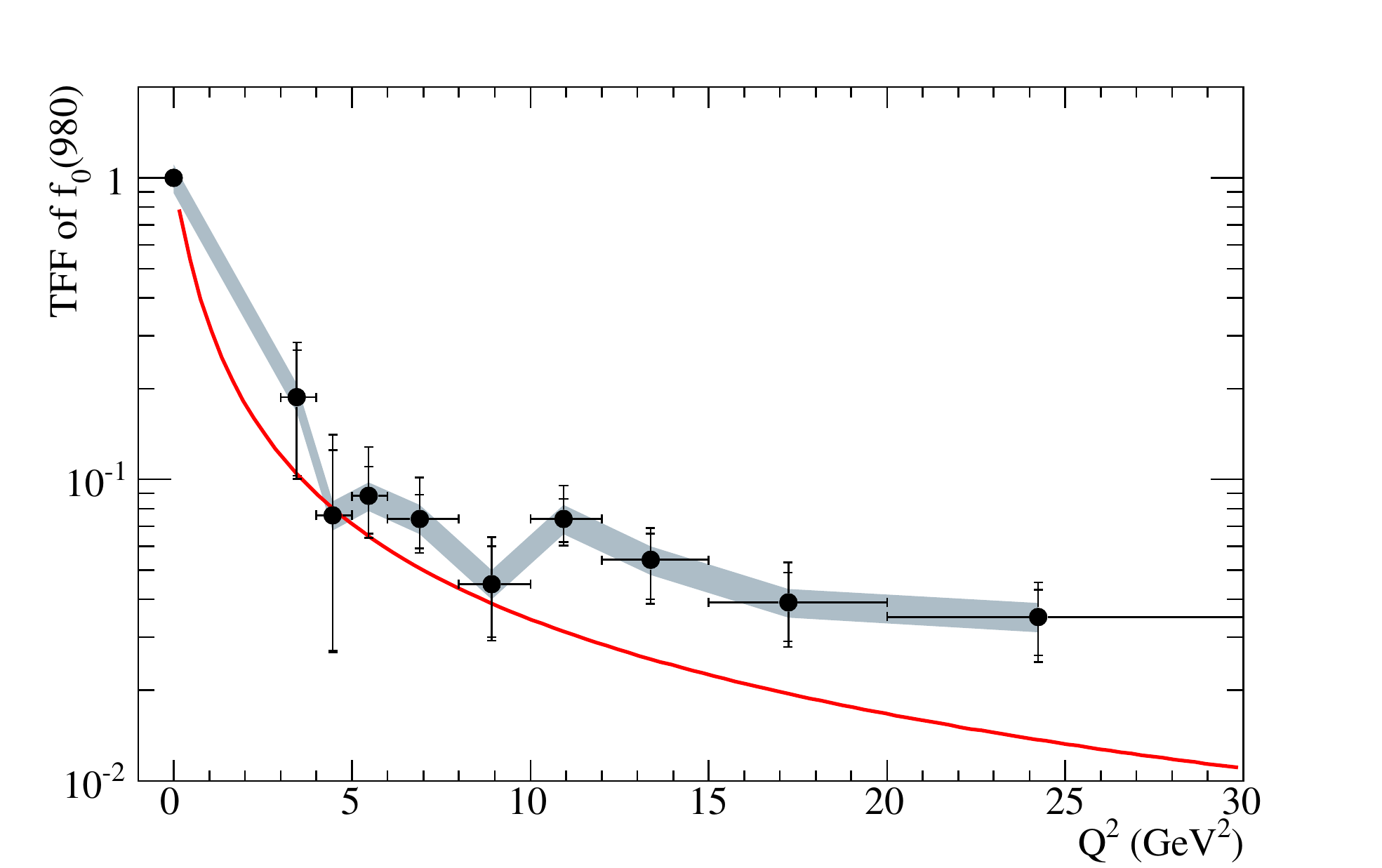,width=80mm}}
 \caption{
The measured $Q^2$ dependence of the TFF of the $f_0(980)$.
Short (long) vertical bars indicate statistical (statistical and systematic
combined) errors.
The shaded area corresponds to the overall systematic uncertainty
arising from that of $\Gamma_{\gamma \gamma}$.
Correlations of uncertainties between neighboring bins exist
and are included in the long vertical bars. 
The solid curve is the prediction for a scalar particle given
in Table~\ref{tab:pred} by Ref.~\cite{schuler}.
}
 \label{fig:f0q2}
\end{figure}

Figure \ref{fig:f0q2} shows the obtained $Q^2$ dependence of the TFF of 
the $f_0(980)$ for the $r_1$ nominal fit.
Here, the theoretical prediction for a scalar TFF in 
Ref.~\cite{schuler} agrees well with the measured ones
up to $Q^2 \simeq 10~\GeV^2$ but has steeper 
$Q^2$ dependence for $Q^2 > 10~\GeV^2$.

\section{Summary and Conclusions}
\label{sec:summary}
We have measured for the first time
the differential cross section of $\pi^0$ 
pair production in single-tag two-photon collisions,
$\gamma^* \gamma \to \pi^0 \pi^0$
up to $Q^2=30~\GeV^2$
based on a data sample of 759~fb$^{-1}$ collected
with the Belle detector~\cite{belle,belleptep} at the KEKB asymmetric-energy 
$e^+ e^-$ collider~\cite{kekb}.
The kinematical range of the data is
$0.5~\GeV < W < 2.1~\GeV$ and $|\cos \theta^*| < 1.0$ 
in the $\gamma^* \gamma$ center-of-mass system. 

The azimuthal angle dependence shows that the contribution of the
helicity-0 (helicity-1) component of the $f_2(1270)$ is large
(small but non-zero).
The differential cross section is fitted by parameterizing 
partial-wave amplitudes.
The transition form factors of the $f_2(1270)$ and $f_0(980)$ 
are measured for $Q^2$ up to $30~\GeV^2$ 
and compared with theoretical predictions.
For the $f_2(1270)$, the helicity-0, -1, and -2 TFFs are measured.
The measured helicity-2 TFF of the $f_2(1270)$ agrees well with
the theory prediction of Ref.~\cite{schuler} and with one of the two 
predictions in Ref.~\cite{ppv}.
The helicity-0 and -1 TFF are about 
a factor of $1.5-2$ smaller than
the prediction of Ref.~\cite{schuler}.

The TFF of the $f_0(980)$ is also extracted; the resulting $Q^2$ 
dependence agrees fairly well with the prediction 
of Ref.~\cite{schuler} for $Q^2 \le 10~\GeV^2$ but has less steeper 
$Q^2$ dependence for $Q^2 > 10~\GeV^2$.

\section*{Acknowledgments}
Special thanks are due to V.G.~Serbo for various useful discussions
and for kindly providing
the cross section formula that includes the $\varphi^*$ dependence 
for single-tag two-photon production of $\pi^0 \pi^0$,
without which we could not perform this analysis.

We thank the KEKB group for the excellent operation of the
accelerator; the KEK cryogenics group for the efficient
operation of the solenoid; and the KEK computer group,
the National Institute of Informatics, and the 
PNNL/EMSL computing group for valuable computing
and SINET4 network support.  We acknowledge support from
the Ministry of Education, Culture, Sports, Science, and
Technology (MEXT) of Japan, the Japan Society for the 
Promotion of Science (JSPS), and the Tau-Lepton Physics 
Research Center of Nagoya University; 
the Australian Research Council and the Australian 
Department of Industry, Innovation, Science and Research;
Austrian Science Fund under Grant No.~P 22742-N16 and P 26794-N20;
the National Natural Science Foundation of China under Contracts 
No.~10575109, No.~10775142, No.~10875115, No.~11175187, and  No.~11475187;
the Chinese Academy of Science Center for Excellence in Particle Physics; 
the Ministry of Education, Youth and Sports of the Czech
Republic under Contract No.~LG14034;
the Carl Zeiss Foundation, the Deutsche Forschungsgemeinschaft
and the VolkswagenStiftung;
the Department of Science and Technology of India; 
the Istituto Nazionale di Fisica Nucleare of Italy; 
National Research Foundation (NRF) of Korea Grants
No.~2011-0029457, No.~2012-0008143, No.~2012R1A1A2008330, 
No.~2013R1A1A3007772, No.~2014R1A2A2A01005286, No.~2014R1A2A2A01002734, 
No.~2014R1A1A2006456;
the Basic Research Lab program under NRF Grant No.~KRF-2011-0020333, 
No.~KRF-2011-0021196, Center for Korean J-PARC Users, No.~NRF-2013K1A3A7A06056592; 
the Brain Korea 21-Plus program and the Global Science Experimental Data 
Hub Center of the Korea Institute of Science and Technology Information;
the Polish Ministry of Science and Higher Education and 
the National Science Center;
the Ministry of Education and Science of the Russian Federation and
the Russian Foundation for Basic Research;
the Slovenian Research Agency;
the Basque Foundation for Science (IKERBASQUE) and 
the Euskal Herriko Unibertsitatea (UPV/EHU) under program UFI 11/55 (Spain);
the Swiss National Science Foundation; the National Science Council
and the Ministry of Education of Taiwan; and the U.S.\
Department of Energy and the National Science Foundation.
This work is supported by a Grant-in-Aid from MEXT for 
Science Research in a Priority Area (``New Development of 
Flavor Physics'') and from JSPS for Creative Scientific 
Research (``Evolution of Tau-lepton Physics'').



\begin{thebibliography}{99}
\bibitem{pw} We denote individual partial
waves by roman letters and parameterized waves by italic.
\end{thebibliography}
\end{document}